\documentclass[aps,twocolumn,prc,superscriptaddress,showpacs,nofootinbib,floatfix,amssymb,amsfonts,amsmath]{revtex4-1}
\pdfoutput=1
\UseRawInputEncoding
\usepackage{graphicx}% Include figure files
\usepackage{dcolumn}% Align table columns on decimal point
\usepackage{bm}% bold math
\usepackage{xcolor}
\usepackage{amsmath}    % need for subequations
\usepackage{amsfonts}   %note how statements can be commented out
\usepackage{amssymb}
\usepackage{graphicx}   % for figuresManMan//

\begin{document}

\title{Bubble nuclei: single-particle versus Coulomb interaction effects}

\author{U. C. Perera}
\affiliation{Department of Physics and Astronomy, Mississippi
State University, MS 39762}

\author{A. V. Afanasjev}
\affiliation{Department of Physics and Astronomy, Mississippi
State University, MS 39762}

\date{\today}

\begin{abstract}

   The detailed investigation of microscopic mechanisms leading to the formation of 
bubble structures in the nuclei  has been performed in the  framework of covariant 
density functional theory. The main emphasis of this study is on the role of 
single-particle degrees of freedom and Coulomb  interaction.  In general, the 
formation of bubbles lowers the Coulomb energy.  However, in nuclei this trend is 
counteracted by the quantum nature of the single-particle states: only specific 
single-particle states with specific density profiles can be occupied with increasing 
proton and neutron numbers. A significant role of
central classically forbidden region at the bottom of the wine bottle potentials 
in the formation of nuclear bubbles (via primarily the reduction of the densities of 
the $s$  states at $r=0$) has been revealed for the first time.  Their formation also 
depends on  the availability of low-$l$ single-particle states for occupation since 
single-particle densities  represent the basic building blocks of total densities. Nucleonic
potentials disfavor the occupation of such states in hyperheavy nuclei and this
contributes to the formation of bubbles in such nuclei.
Existing bubble indicators are  strongly affected by single-particle properties and 
thus  they cannot be reliable measures of  bulk properties (such as the Coulomb 
interaction). Additivity rule for densities has been proposed for the first time. It was 
shown  that the differences in the densities of bubble and flat density nuclei  follow 
this rule in the $A\approx 40$ mass region and in superheavy nuclei  with comparable accuracy.
This strongly suggests the same mechanism of the formation of central depression
in bubble nuclei of these two mass regions. Nuclear saturation mechanisms and
self-consistency effects also affect the formation of bubble structures. The detailed 
analysis of different aspects of bubble physics strongly suggests that the formation 
of bubble structures in superheavy nuclei is dominated by  single-particle  effects. The 
role of the Coulomb interaction increases in hyperheavy nuclei but even for such 
systems we do not find strong arguments that the formation of  bubble structures 
is dominated by it.

\end{abstract}

\maketitle

%%%%%%%%%%%%%%%%%%%%%%%%%%%%%%%%%%%%%%%%
\section{Introduction}
\label{Introduc}
%%%%%%%%%%%%%%%%%%%%%%%%%%%%%%%%%%%%%%%%

   The basic approximation which appears in many nuclear models is that 
the nuclear density  is constant in subsurface region. The simplest example is the 
Fermi function which is frequently used for the description of the density of the nuclei 
in phenomenological models (see, for example, Sec. 2 of Ref.\ \cite{NilRag-book}). 
However, theoretical investigations reveal that there is a density depletion in the 
central region in a number of  the nuclei. Such nuclei are typically called as bubble 
nuclei.

    The physics of bubble nuclei has first been studied by Wilson in 1946 
\cite{W.46} and the number of investigations of such nuclei in different theoretical  
frameworks have been carried out later.  The energies of spherical 
bubble nuclei have been studied using liquid drop model in Ref.\ \cite{SB.67}.
The investigation of bubble structures in 
$^{36}$Ar and $^{200}$Hg has been performed in non-relativistic Hartree-Fock 
approach in Ref.\ \cite{DWK.72,CS.73}. Additional nuclei (such as $^{68}$Se, 
$^{68}$Se, $^{100}$Sn, $^{138}$Ce) and the details of bubble formation mechanism 
have been studied using the same formalism in Ref.\ \cite{DKW.73}. The detailed
investigation of spherical bubble nuclei in the liquid drop and spherical shell models
has been performed in Ref.\ \cite{Wong.73}. However, this approach is too simplistic 
since it assumes zero density inside the bubble.  The shell structure of spherical 
nuclear bubbles has been investigated in simple phenomenological shell model
potentials allowing partial filling of the bubble in Ref.\ \cite{DP.97,DP.98}.

  More sophisticated and realistic models which take self-consistency 
effects into account have been used in the detailed studies
of bubble nuclei starting from 1990's. The bubble structure in $^{34}$Si 
and the density profiles of neighboring nuclei have been extensively 
studied in the non-relativistic  and relativistic density functional theories (DFTs) 
\cite{GGKNPSGV.09,KLRL.17,YML.13}, {\it ab initio} approaches \cite{DSLBN.17} 
and beyond mean field  approaches \cite{KLRL.17,YBBH.12,YML.13,WYL.14}. The 
possibility of the existence of deformed bubbles in light nuclei has been investigated
within the relativistic mean field (RMF) approach in Ref.\ \cite{SA.14} and the $^{24}$Ne, 
$^{32}$Si and $^{32}$Ar nuclei are found to be the best candidates. Bubble 
structures in very neutron-rich $^{68}$Ar nucleus have been investigated in Ref.\ 
\cite{KGMG.08} and in $^{22}$O in Refs.\ \cite{GABLNWY.07,GGKNPSGV.09}.
The impact of tensor force on the formation of bubble structures in light nuclei 
$Z=20$ or $N=20$ nuclei has been investigated in Ref.\ \cite{NSM.13}.  The 
bubble structures in superheavy nuclei have been studied in non-relativistic and 
relativistic DFTs in Refs.\ \cite{BRRMG.99,DBGD.03,AF.05-dep,SNR.17}. A 
systematic survey of bubble structures in spherical nuclei with N(Z)=8, 20, 28, 
40, 50, 82 and N=126 has been performed in the RMF framework in Ref.\ 
\cite{SKKJA.19}. These approaches have also been used in the studies of bubble 
structures in hyperheavy ($Z>126$) nuclei (see Refs.\ \cite{DBDW.99,DBGD.03,AATG.19,AA.21}). 
It is necessary to mention that the investigations of bubble structures in hyperheavy nuclei  
performed under restriction to spherical symmetry \cite{DP.97,DBDW.99,DBGD.03} 
ignore two facts \cite{AAG.18,AATG.19,AA.21}, namely,  (1) that the toroidal shapes 
are energetically more favored in such nuclei,  and (2) that the most of such nuclei 
cannot be stabilized  because of the absence of the local  
minimum in total energy at spherical shape.

\clearpage
   These investigations significantly advanced our understanding of the 
mechanisms of the formation of the bubble structures in nuclei. They also found the 
counteracting mechanisms: pairing correlations \cite{KLRL.17,YBBH.12}, beyond mean 
field effects \cite{YBBH.12,YML.13,WYL.14,WX.18} and deformation  
\cite{AF.05-dep,PXS.05,SA.14} 
soften fluctuations in the densities (as a function of radial coordinate in spherical 
nuclei) and somewhat reduce the bubble structures in the nuclei. In addition, some 
dependence of the predictions for the depletion in the central density on the model 
and employed functional has been found (see, for example, Refs.\ 
\cite{AF.05-dep,YBBH.12,DSLBN.17,SKKJA.19}).

   The most of predicted bubble structures are located in exotic nuclei  which 
either have not  been  measured so far or which are produced in very small amounts with 
very short lifetimes.  So far, only in $^{34}$Si the formation of proton bubble has been indirectly 
confirmed in experiment \cite{34Si-bubble-exp}. Direct measurements of charge density 
distributions  via electron scattering on unstable nuclei with sufficient luminosity are not 
possible today.  However, such experiments can be feasible in light bubble nuclei in near 
future at the FRIB, FAIR and RIKEN facilities.

 However, not in all respects of the physics of bubble nuclei the consensus has been reached.  For example, the 
analysis of bubble structures in $^{34}$Si, $^{48}$Ca and $N=82$, 126 and 184  isotopic 
chains based on the correlation analysis performed in Ref.\ \cite{SNR.17} suggests that the 
central depression in medium-mass nuclei is very sensitive to shell effects, whereas for 
superheavy nuclei it is firmly driven by the electrostatic repulsion. The later result is in 
contradiction with the conclusions of Ref.\ \cite{AF.05-dep} which clearly illustrated that 
the formation of central depression in the density distribution is driven by the filling of specific 
spherical subshells and shell structure of superheavy nuclei.  It also contradicts the 
observation that spherical superheavy nuclei with $Z=126$ have either no or significantly 
smaller depletion of the density in the central region as compared with the  $Z=120$ 
isotopes (see Fig.\ 2 in Ref.\ \cite{AF.05-dep}). 

    The main goal of the present paper is to perform a detailed microscopic 
analysis of the mechanisms which lead to the formation of central depression in nucleonic 
densities of atomic nuclei. To achieve that the pairs of light and superheavy nuclei with 
and without central depression in the densities will be compared. The detailed comparison 
of the single-particle and Coulomb interaction contributions into the proton and neutron 
densities of the nuclei in these pairs  allows to discriminate their role in the formation of 
central depression in nucleonic densities. This analysis will be further collaborated by the 
analysis of hyperheavy nuclei which possess pronounced bubble structure.

     The paper is organized as follows. A brief outline of the theory and the selection of the 
nuclei under study is given is Sec.\ \ref{theory}.   The role of the Coulomb interaction in the 
formation of bubble structure of superheavy nuclei is discussed in Sec.\ \ref{SHE-Coulomb}. 
Sec.\ \ref{single-particle-degrees} is dedicated to the discussion  of the role of the 
single-particle degrees of freedom in the formation of central depression in the density 
distributions. The mechanisms of the formation of the wine bottle potentials are
analyzed in Sec.\ \ref{wine-potentiall}. The additivity rule for the  densities of the pairs of 
the nuclei with and without  central depression is considered in Sec.\ \ref{additivity}.  
Other  general observations obtained in the present study are discussed in Sec.\ 
\ref{Gen-observations}.   Sec.\ \ref{bubble-indicators} critically analyzes existing bubble 
indicators and their physical content.  The factors affecting 
the availability of the low-$l$ states for occupation are analyzed in Sec.\ \ref{factors_1}.
Potential impact of deformation on the balance of the single-particle and 
Coulomb interaction contributions to the bubble structures is discussed in Sec.\ 
\ref{factors_2}.  Finally, Sec.\ \ref{concl}  summarizes the results of our paper.

%%%%%%%%%%%%%%%%%%%%%%%%%%%%%%%%%%%%%%%%%%%%%%%%
\begin{figure}[htb]
\centering
\includegraphics[width=8.4cm]{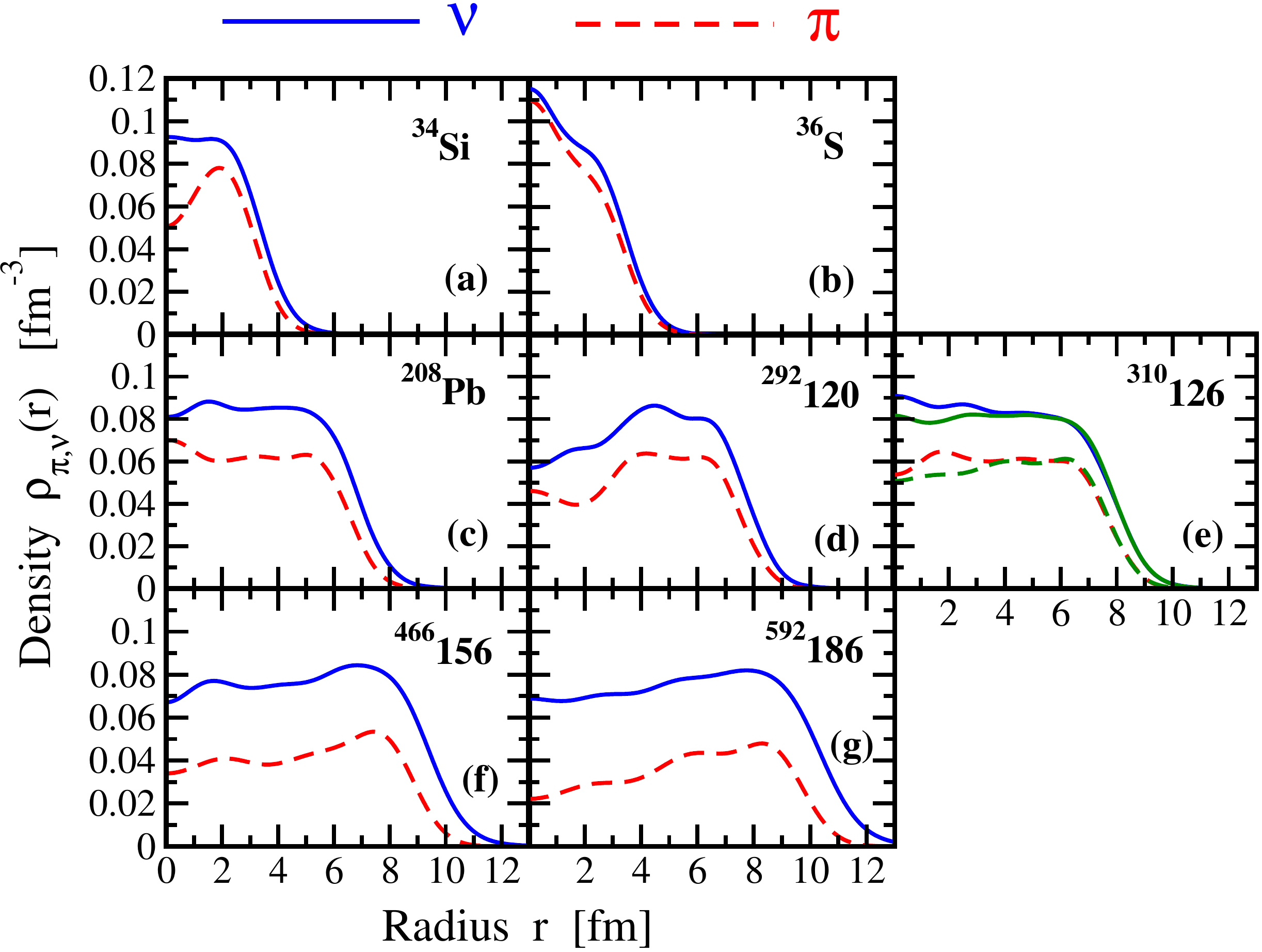}
\caption{The proton and neutron densities as a function of radial coordinate $r$
for indicated spherical nuclei. Light, superheavy and hyperheavy nuclei are shown 
in the top, middle and bottom rows, respectively. The results of the calculations without 
pairing are shown by blue solid and red dashed lines for neutrons and protons, respectively. 
Note that the pairing 
collapses in all nuclei with exception of the $^{320}$126 nucleus. The RHB results for 
this nucleus are shown by green solid (neutrons) and dashed (protons) lines in panel
(f).
\label{Total_den}
}
\end{figure}
%%%%%%%%%%%%%%%%%%%%%%%%%%%%%%%%%%%%%%%%%%%%%%%%%%

%%%%%%%%%%%%%%%%%%%%%%%%%%%%%
\section{Theoretical method and the selection of the nuclei}
\label{theory} 
%%%%%%%%%%%%%%%%%%%%%%%%%%%%%%

  Theoretical calculations have been performed within the framework of 
covariant density functional theory (CDFT) \cite{VALR.05} employing the modified 
version of the computer code restricted to spherical  symmetry used in Ref.\ \cite{AF.05-dep}. 
The pairing correlations are neglected in the calculations in order to better understand 
the underlying physical mechanisms. In reality, the pairing collapses in all nuclei 
considered in the present paper with the exception of the $^{310}$126 one (see 
Fig.\ \ref{Total_den} and further comments on this nucleus below) in relativistic 
Hartree-Bogoliubov (RHB) calculations with separable pairing interaction of two types (one 
from Ref.\ \cite{AARR.14} and another one [isospin dependent] from Ref.\ 
\cite{TA.21}). 

%%%%%%%%%%%%%%%%%%%%%%%%%%%%%%%%%%%%%%%%%%%%% 
\begin{table}[!ht]
\begin{center}
\caption{Rms radii of proton and neutron matter distributions in the nuclei 
under study. The radii in the  $^{34}$Si/$^{36}$S and $^{292}$120/$^{310}$126
pairs of nuclei are shown in bold.}
\begin{tabular}{l|c|c|} \hline \hline 
 Nuclei & Proton  $ r_{rms}$ [fm]& Neutron  $r_{rms}$ [fm] \\ 
  \hline 
$^{34}$Si             & {\bf 3.046} & {\bf 3.304} \\
$^{36}$S   & {\bf 3.171} & {\bf 3.297} \\
$^{208}$Pb          & 5.450 & 5.738 \\
$^{292}$120       & {\bf 6.223} & {\bf 6.386} \\
$^{310}$126           &  {\bf 6.302} & {\bf 6.519} \\
$^{466}$156           &  7.352 & 7.775 \\
$^{592}$186           &  8.048 & 8.569 \\
			\hline	\hline
\end{tabular}
\label{Table-rms-radii} 
\end{center}
\end{table}
%%%%%%%%%%%%%%%%%%%%%%%%%%%%%%%%%%%%%%%%%%%%% 
   Since the details of the CDFT framework are widely available (see, for 
example, Ref.\  \cite{VALR.05}), we focus on the physical quantities of the interest.
The proton ($i=\pi$) and neutron ($i=\nu$) nucleonic potentials are defined as follows:
\begin{eqnarray}
V_\pi & = & V + S + V_{Coul} , \\
V_\nu & = & V + S ,
\end{eqnarray} 
where scalar potential is given by
\begin{eqnarray} 
S(r) = g_{\sigma} \sigma(r),
\end{eqnarray}
meson defined part of the vector potential is written as
\begin{eqnarray} 
V(r) = g_{\omega} \omega_0 (r) + g_{\rho} \tau_3 \rho_0(r) ,  
\end{eqnarray}
and 
\begin{eqnarray} 
V_{Coul}(r) = e A_0(r)
\end{eqnarray}
is the Coulomb potential. Note that for the sake of discussion we split vector
potential (see, for example, Eq. (9) in Ref.\ \cite{VALR.05}) into meson defined 
and Coulomb parts. In addition, we consider only time-like components of
vector mesons since only even-even nuclei are studied in the present paper.

  The calculations are performed with the NL3* covariant energy density functional 
(CEDF) \cite{NL3*}.  This functional has a lot of similarities with the NL3 one used
earlier in the study of bubble structures in superheavy nuclei (see Ref.\ \cite{AF.05-dep})
but provides improved description of the masses and charge radii on the global scale (see 
Refs.\ \cite{AARR.14,PAR.21}). It was verified that main conclusions obtained in the
present paper do not depend on the selection of the functional.

   In the present study the pair of  light nuclei $^{34}$Si/$^{36}$S and the pair of 
superheavy nuclei $^{292}$120/$^{310}$126 nuclei are considered. The first nucleus 
in these pairs ($^{34}$Si and $^{292}$120) is characterized by substantial central 
depression, while such depression is either absent or almost suppressed in the second 
nucleus of the pair (see Fig.\ \ref{Total_den} and Refs.\
\cite{DBDW.99,DBGD.03,AF.05-dep,KLRL.17,GGKNPSGV.09}).  Moreover, this 
feature exists in different theoretical frameworks. The detailed comparison of the 
single-particle and Coulomb interaction contributions into the differences of the 
proton and neutron densities of the nuclei in above mentioned pairs allows to 
discriminate their contribution into the formation of central depression in nucleonic 
densities.  Note that significant central depression in the density distribution of 
the $^{292}$120 nucleus and flat density in the $^{310}$126 nucleus have been 
found both in relativistic and non-relativistic DFTs (see Refs.\ 
\cite{AF.05-dep,BRRMG.99,DBGD.03}).  The fact that rms radii of proton/neutron
matter distribution of the nuclei in these pairs are very similar (see Table  \ref{Table-rms-radii})
also simplifies the analysis of the additivity of the single-particle densities (see Sec.\ 
\ref{additivity}).  Note that in the $^{34}$Si, $^{36}$S and $^{292}$120 nuclei the 
spherical minimum is the lowest one corresponding to the ground state in the RHB 
calculations (see Refs.\ \cite{AARR.14, Mass-Explorer,AANR.15}). In contrast, the 
same calculations bring oblate ground state for the $^{310}$126 nucleus (see Ref.\ 
\cite{AANR.15}). However, the spherical solution in this nucleus is considered here 
in order to have a benchmark theoretical solution with near flat density distribution 
in the region of superheavy nuclei.

   The $^{208}$Pb nucleus is also analyzed for the sake of comparison with superheavy 
nuclei.  In addition, hyperheavy $^{466}$156 and $^{592}$186 nuclei are investigated in 
detail  in order to get a better understanding of the factors affecting the profiles of density 
distributions with increasing proton number $Z$.  These nuclei are located in the centers 
of the islands of potentially relatively stable spherical hyperheavy nuclei (see Refs.\ 
\cite{AAG.18,AATG.19,AA.21}). Note, however,  that they correspond to highly excited 
local spherical minima and the lowest in energy solutions  in axial RHB calculations 
have toroidal shapes.

%%%%%%%%%%%%%%%%%%%%%%%%%%%%%%%%%%%%%%
\section{Superheavy bubble nuclei and the role of Coulomb interaction}
\label{SHE-Coulomb} 
%%%%%%%%%%%%%%%%%%%%%%%%%%%%%%%%%%%%%%

%%%%%%%%%%%%%%%%%%%%%%%%%%%%%%%%%%%%%%%%%%%%
\begin{figure}[htb]
\centering
\includegraphics[width=8.4cm]{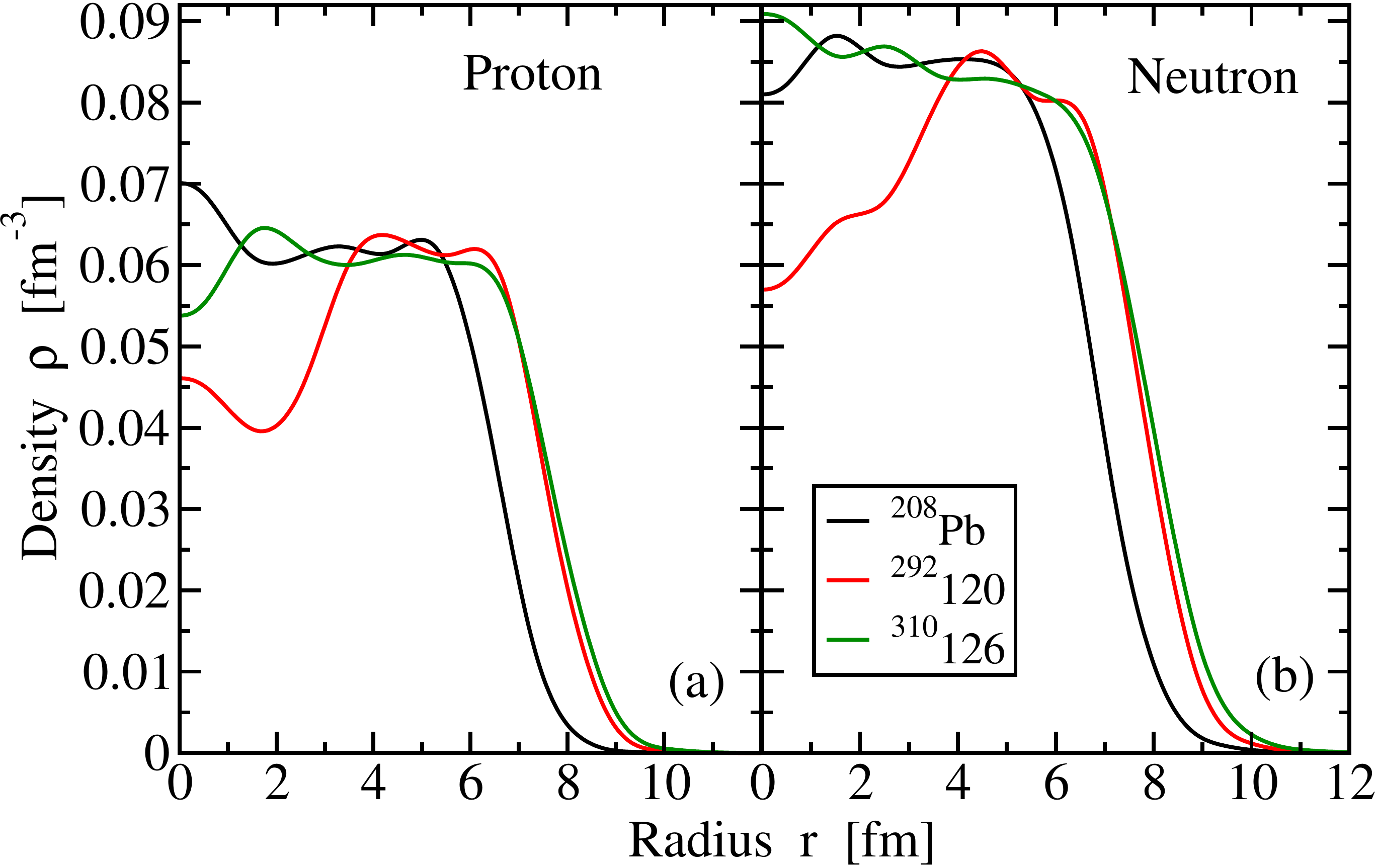}
\caption{Proton and neutron density distributions of indicated nuclei.
\label{Total_density_3_nuclei}
}
\end{figure}
%%%%%%%%%%%%%%%%%%%%%%%%%%%%%%%%%%%%%%%%%%%%

    Proton and neutron density distributions of doubly magic $^{208}$Pb nucleus 
and superheavy  $^{292}$120 and $^{310}$126 nuclei are shown in Fig.\ \ref{Total_density_3_nuclei}.
The $^{292}$120 nucleus shows very pronounced depression in central densities.
\clearpage
  In contrast, such depression is absent in the  $^{208}$Pb and $^{310}$126 nuclei. Thus,
the increase of proton number on going from $^{292}$120 to $^{310}$126 nucleus
does not trigger the enhancement of the central depression as it would be 
expected in the case when central depression is firmly defined by electrostatic 
repulsion (as suggested by Ref.\ \cite{SNR.17}).

%%%%%%%%%%%%%%%%%%%%%%%%%%%%%%%%%%%%%%%%%%%%
\begin{figure}[htb]
\centering
\includegraphics[width=8.4cm]{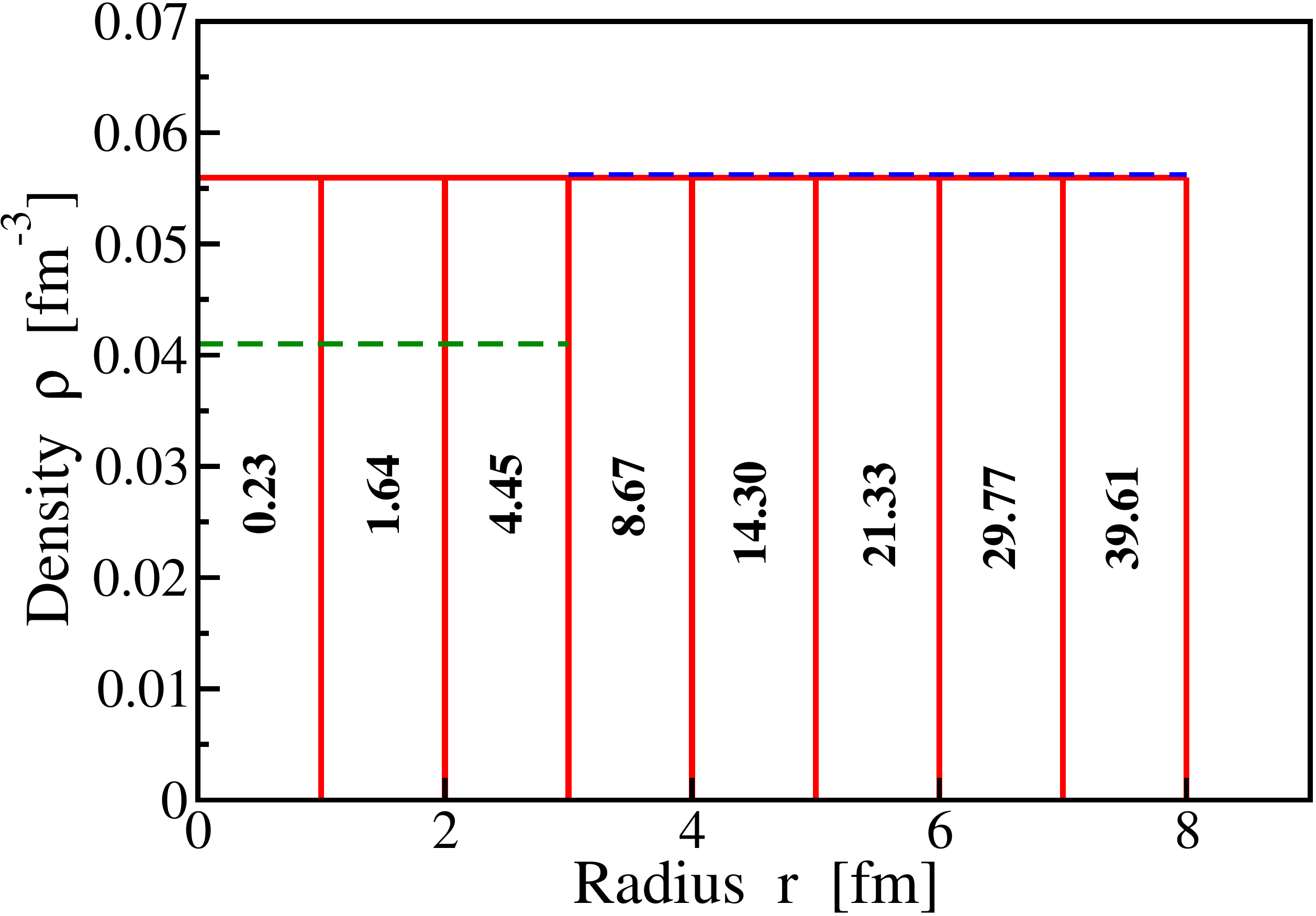}
\caption{The numbers of the particles in spherical shells between $r_{min} = j$ fm 
to $r_{max} = j+1$ fm ($j=0,1, ...7$) of the sphere with  $R=8.0$ fm with uniform 
density $\rho$. They are shown inside the bins. The total number of particles is 120. 
Dashed lines show resulting density distributions after creation of central depression
(see text for more detail).
\label{particle_filling}
}
\end{figure}
%%%%%%%%%%%%%%%%%%%%%%%%%%%%%%%%%%%%%%%%%%%%

    When considering central depression in medium mass to superheavy nuclei, one should
keep in mind that they are created by the transfer of relatively small number of particles from
the central region to near surface region.  As a consequence, the bubble should not be
considered as a bulk property \cite{CS.73}. The density plots as a function of radial 
coordinate tend to overemphasize the importance of central region since they ignore
the fact that the number of particles $dn$ in spherical shell of thickness $dr$ is given
by $4\pi  r^2 \rho(r) dr$. To illustrate that we simplify the case of the proton subsystem of 
the $^{292}$120 nucleus [see Fig.\ \ref{Total_density_3_nuclei}(a)]  to the sphere of radius 
$R=8.0$ fm and uniform density distribution $\rho$. Then the number of particles $n$ in 
spherical shell with inner radius $r_{min}$ and outer radius $r_{max}$ is given by
\begin{eqnarray}
n= 4\pi \rho \int_{r_{min}}^{r_{max}} r^2dr .
\end{eqnarray}
The distribution of particles over spherical shells is shown in Fig.\ \ref{particle_filling}.
There are only 0.23 particles in the inner sphere of radius 1.0 fm and 
1.64 and 4.45 particles in the first and second spherical shells with outer radii 2.00 and 
3.00 fm, respectively. Based on Fig.\ \ref{Total_density_3_nuclei}(a) one can assume 
that central depression with average density $\rho_{dep} = 0.041$ fm$^{-3}$ is formed 
up to radius $r=3.0$ fm (see green dashed line in Fig.\ \ref{particle_filling}). To create such 
central depression one should move 0.061 particles from inner sphere, 0.44 particles 
from first inner shell, and 1.19 particles from second inner shell into outer shells located 
between 3.0 and 8.0 fm. If these particles are redistributed uniformly among  outer shells 
this would lead only to a marginal increase of densities (see blue dashed line in Fig.\ 
\ref{particle_filling}).

%%%%%%%%%%%%%%%%%%%%%%%%%%%%%%%%%%%%%%%%%%%%%%%%
\begin{figure}[htb]
\centering
\includegraphics[width=8.4cm]{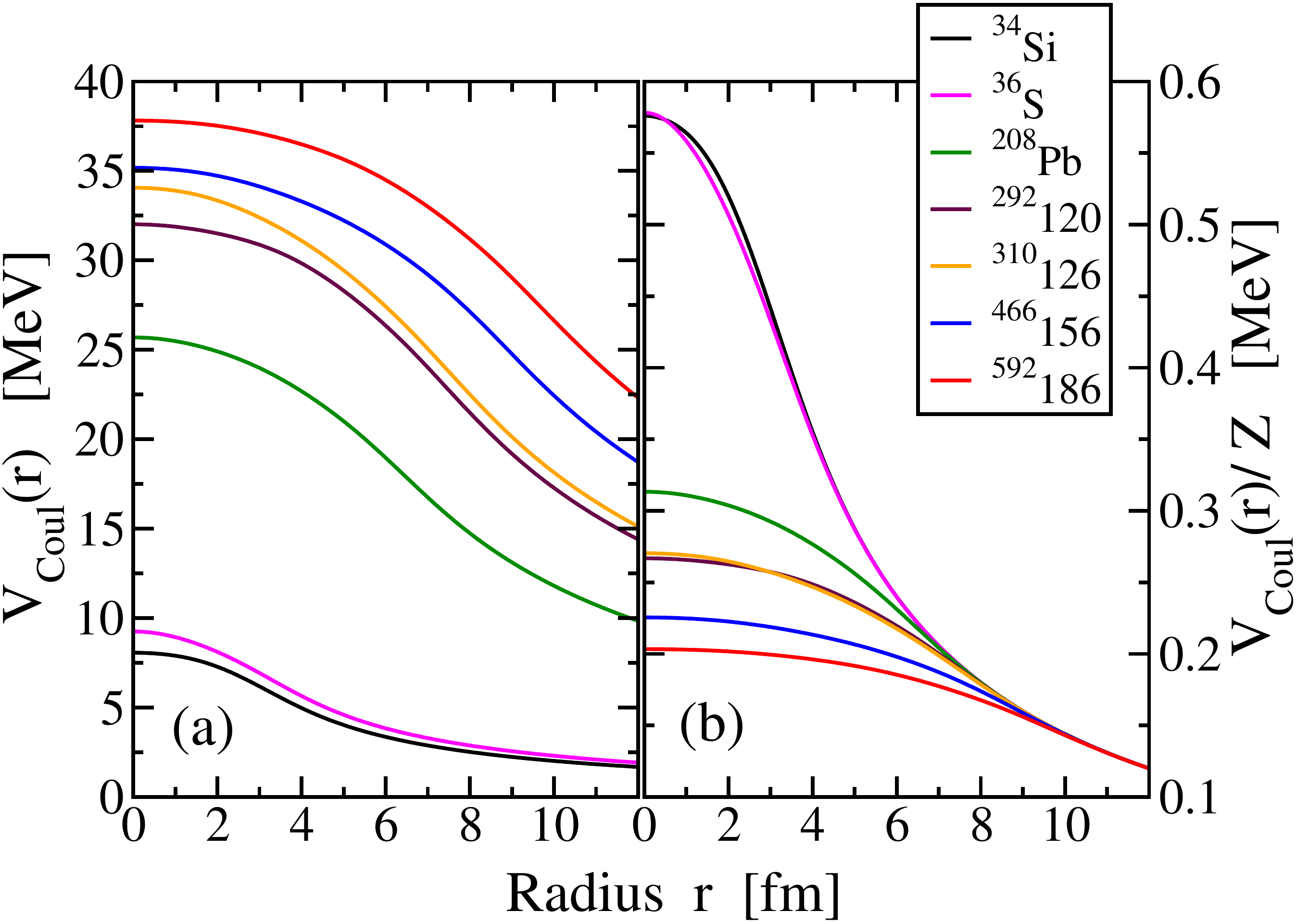}
\caption{Coulomb potentials of indicated nuclei as a function of radial coordinate $r$. Panels (a) 
and (b) show absolute ($V_{Coul}$) and normalized ($V_{Coul}/Z$) (to the proton number $Z$) 
Coulomb  potentials. 
\label{Coulomb_pot}
}
\end{figure}
%%%%%%%%%%%%%%%%%%%%%%%%%%%%%%%%%%%%%%%%%%%%%%%%

   The Coulomb potentials $V_{Coul}(r)$ are shown  as a function of radial 
coordinate  for the nuclei under study  in Fig.\ \ref{Coulomb_pot}. Their absolute values 
and evolution with radial coordinate $r$ are very similar for the 
$^{292}$120 and $^{310}$126 nuclei (see Fig.\ \ref{Coulomb_pot}(a). This similarity  
becomes even more pronounced when normalized values $V_{Coul}(r)/Z$ of the 
Coulomb potential per number of protons are compared in Fig.\ \ref{Coulomb_pot}(b).  
These results strongly suggest that the formation of the bubble structure in the 
$^{292}$120 nucleus is not driven predominantly by the electrostatic repulsion
since such bubble structure is absent in the $^{310}$126 nucleus for which $V_{Coul}$
is larger. Note that  the situation in this pair of the nuclei is very similar to the one seen 
in the pair of the nuclei $^{34}$Si and $^{36}$S (see Fig.\ \ref{Coulomb_pot})  in which the 
formation of the proton bubble in $^{34}$Si is attributed solely to  the single-particle 
effects (see Ref.\ \cite{SNR.17}). 

    It is also interesting to compare normalized values $V_{Coul}(r)/Z$ of the Coulomb 
potential per number of protons for all nuclei under study (see Fig.\ \ref{Coulomb_pot}(b)).
At low radial coordinate $V_{Coul}(r)/Z$ has highest value  in light nuclei 
and then it gradually decreases with increasing proton number. This correlates with the 
evolution of the proton density with proton number (see Fig.\ \ref{Total_den}). Note that in 
all nuclei the asymptotic behavior of the $V_{Coul}(r)/Z$ is the same at $r>10 $ fm.

    The Coulomb potential alone in all these systems favors the arrangement of the 
protons into bubble like structures since $V_{Coul}(r=0) - V_{Coul}(r_{surf})>0$, where $r_{surf}$ 
is the  radial coordinate at which the density is maximal in near surface region. 
$V_{Coul}(r) - V_{Coul}(r_{surf})$  is equal approximately to 0.78, 1.3, 5.7, 6.8 and 7.8 MeV 
in the $^{34}$Si, $^{36}$S, $^{208}$Pb, $^{292}$120 and $^{310}$126 nuclei, respectively 
(see Fig.\ \ref{Coulomb_pot}).  However, even in superheavy nuclei these contributions to the 
building of the wine bottle proton potential  are smaller than those coming from nuclear 
interactions. For example, in the wine bottle potential of the $^{292}$120 nucleus the difference 
$V_{\pi}(r=0) - V_{\pi}(r_{surf}) \approx 21$ MeV [see Fig.\ \ref{Single-particle-120-126}(a)] and 
$V_{Coul}(r=0) - V_{Coul}(r_{surf})\approx 6.8$ MeV accounts for less than one third of this 
value. 
 
    Although the role of the Coulomb potential in the formation of wine bottle
potential increases in hyperheavy nuclei, even in those systems it does not become 
dominant. Indeed, in the $^{592}$186 nucleus $V_{\pi}(r=1.7\,{\rm fm}) - V_{\pi}(r=8.03\,{\rm fm}) = 13.79$ 
MeV and $V_{Coul}(r=1.7\,{\rm fm}) - V_{Coul}(r=8.03\,{\rm fm}) = 6.55$ MeV.  Similar situation 
exist in the $^{466}$156 nucleus in which $V_{\pi}(r=1.81\,{\rm fm}) - V_{\pi}(r=6.84\,{\rm fm}) = 11.22$ 
MeV and $V_{Coul}(r=1.81\,{\rm fm}) - V_{Coul}(r=6.84\,{\rm fm}) = 5.37$ MeV. The values of the proton 
($V_{\pi}$) and Coulomb ($V_{Coul}$) potentials in these differences are defined  at radial
coordinates corresponding to minimum and maximum points of the wine bottle part of the proton 
potentials shown in Figs.\ \ref{Single-particle-156-186}(a) and (b) below.

%%%%%%%%%%%%%%%%%%%%%%%%%%%%
\section{The role of single-particle degrees of freedom}
\label{single-particle-degrees}
%%%%%%%%%%%%%%%%%%%%%%%%%%%%

%%%%%%%%%%%%%%%%%%%%%%%%%%%%%%%%%%%%%%%%%%%%%%%%  
\begin{figure*}[htb] 
\centering
\includegraphics[width=6.64cm]{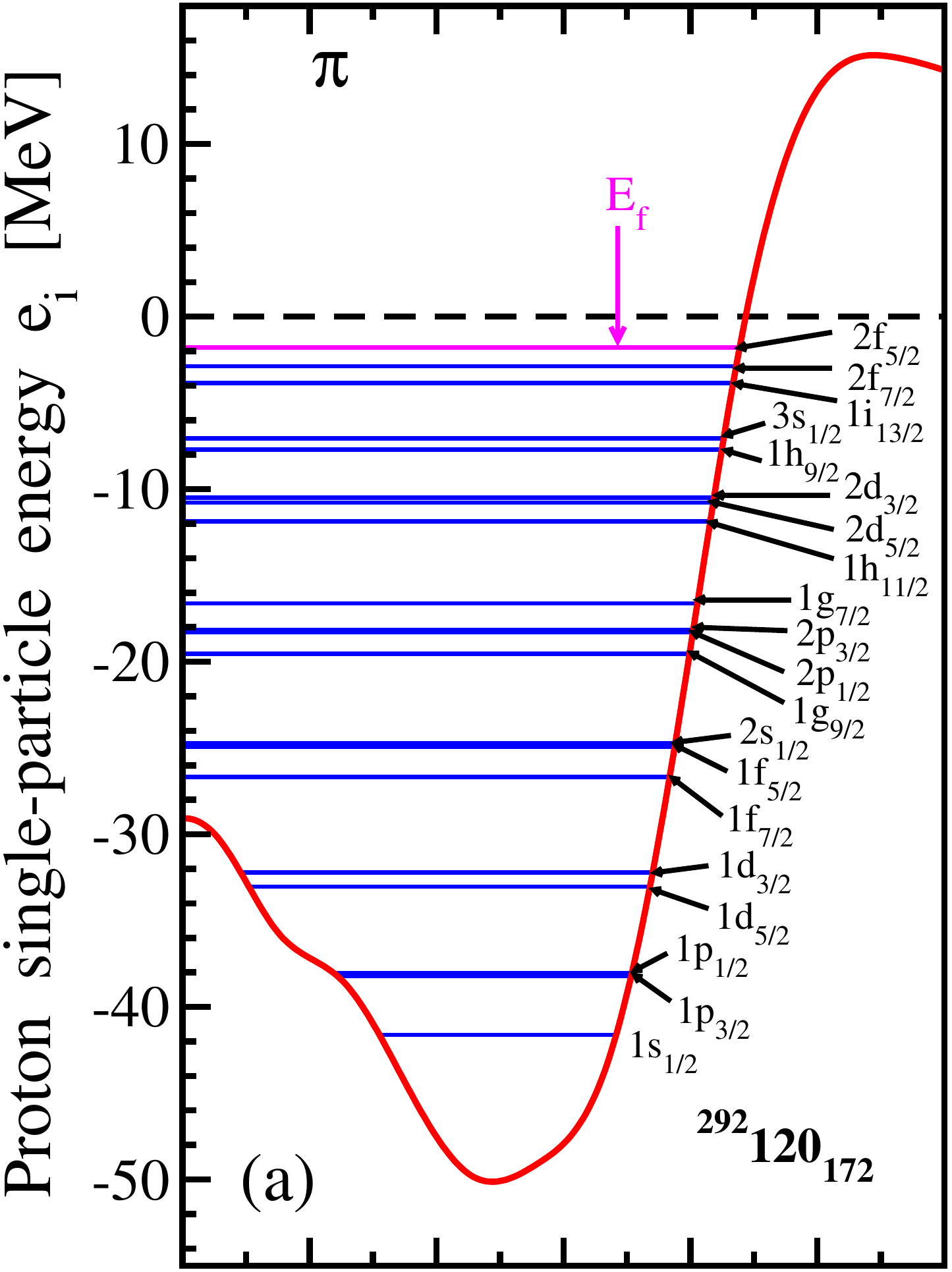}
\includegraphics[width=5.4cm]{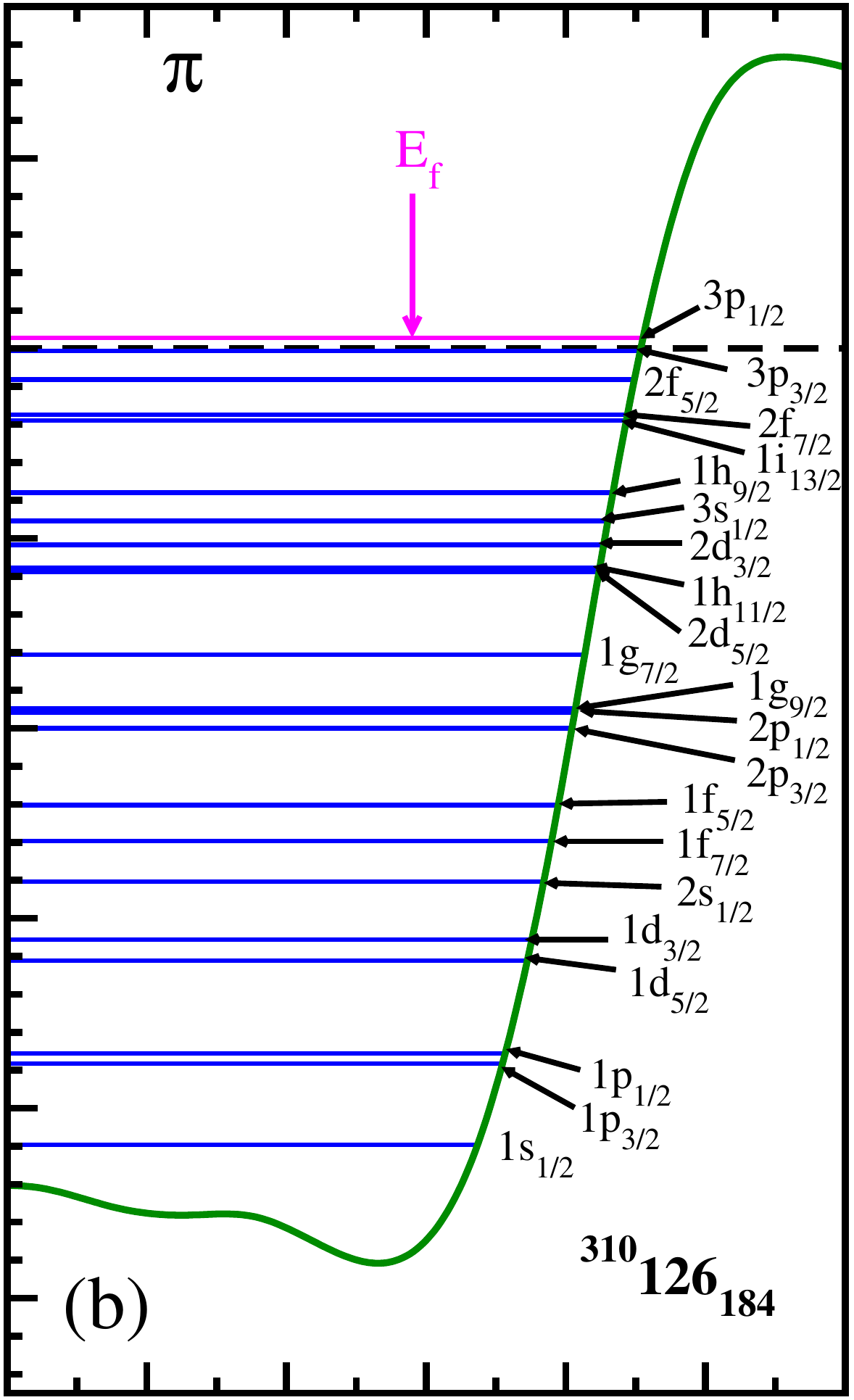}
\includegraphics[width=6.64cm]{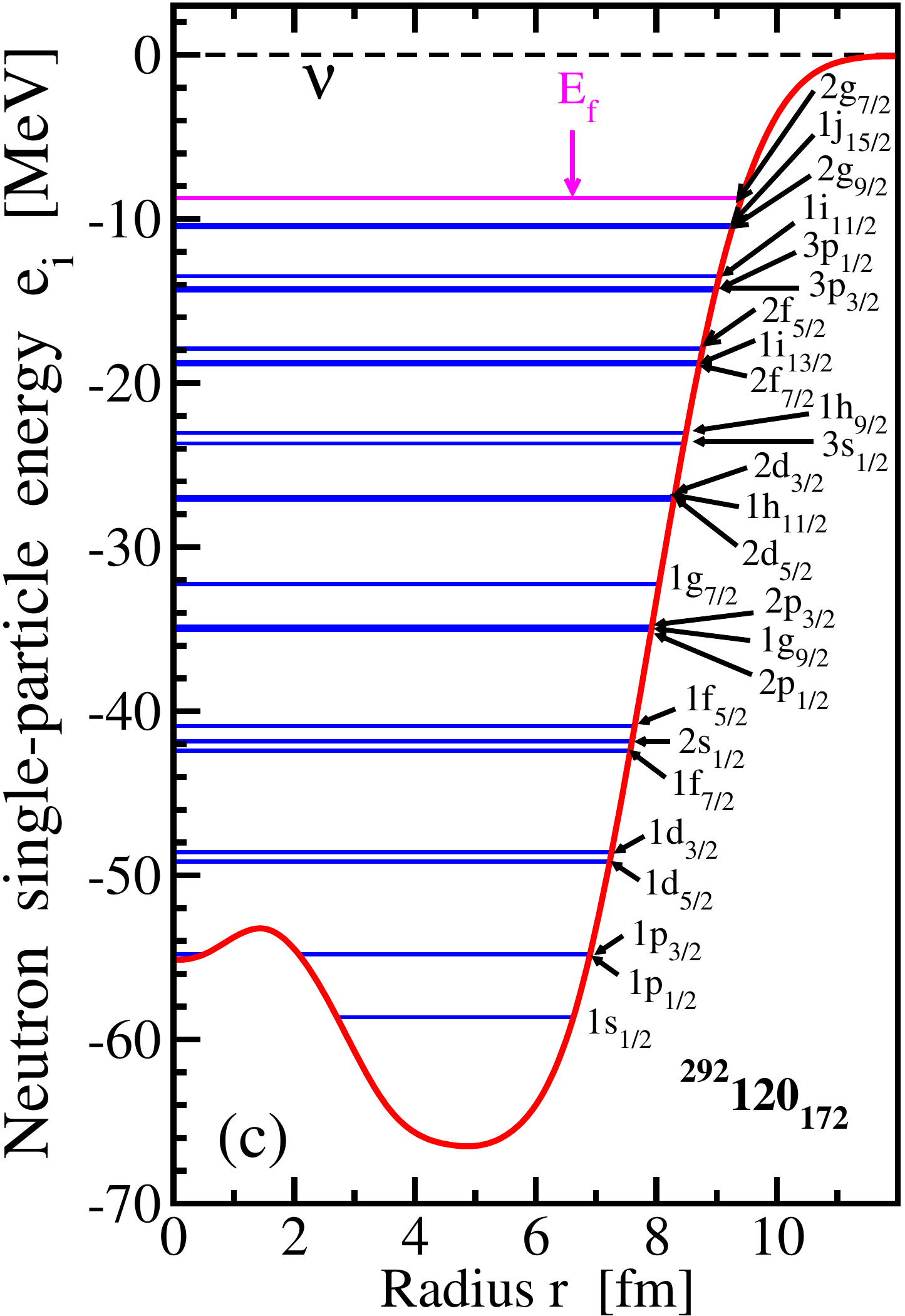}
\includegraphics[width=5.4cm]{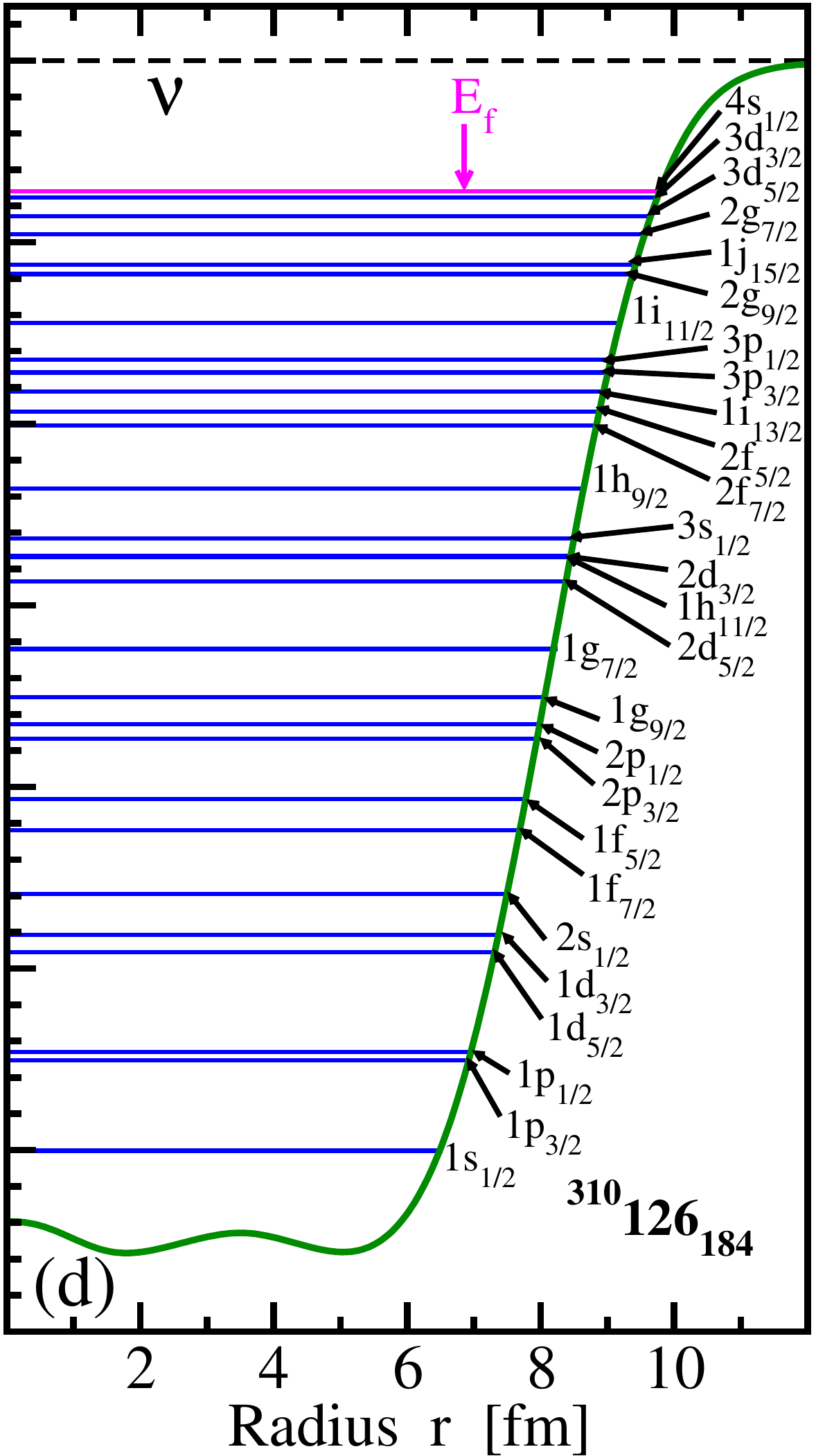}
\caption{Nucleonic potentials and occupied single-particle states of the
ground state configurations in the
$^{292}120_{172}$ and  $^{310}126_{184}$ nuclei. The Fermi level
E$_{\rm f}$ in the calculations without pairing coincides with the last
occupied state: it is shown by pink arrow.  
\label{Single-particle-120-126}
}
\end{figure*}
%%%%%%%%%%%%%%%%%%%%%%%%%%%%%%%%%%%%%%%%%%%%%%%%  
  
    In order to obtain better microscopic understanding of the origin of the 
bubble nuclei  and the role of the single-particle structure in their formation we carry 
out a detailed investigation of the single-particle properties in the pair of superheavy 
nuclei $^{292}$120 and $^{310}$126 and in the pair of the $N=20$ isotones  
$^{34}$Si and $^{36}$S.

    We start from the analysis of the first pair. The nucleonic potentials of  these nuclei 
and their occupied states are shown in Fig.\ \ref{Single-particle-120-126}.  The nucleonic 
potentials of the $^{310}$126 nucleus are similar to those of $^{208}$Pb (compare Fig. 
3(c),(d) in Ref.\ \cite{AF.05-dep} with Fig.\ \ref{Single-particle-120-126}(b) and (d) in the
present paper): they have flat bottom potentials. In contrast, the nucleonic potentials 
of the  $^{292}$120 nucleus are wine bottle shaped and this is especially pronounced 
for the proton subsystem (see Figs.\ \ref{Single-particle-120-126}(a) and (c)).

 Total nucleonic density $\rho_{tot}(r)$ in a given subsystem (proton or neutron) is built 
from the contributions of individual  particles as follow
\begin{eqnarray} 
\rho_{tot}(r)  =  \sum_{i} (2j_i+1) \rho^{sp}_i (r).
\end{eqnarray} 
Here we consider only the nuclei in which full spherical subshells (indicated by
subscript $i$) are occupied. Thus, the sum runs over spherical subshells $i$
with multiplicity $(2j_i+1)$ and $\rho^{sp}_i (r)$ is the density of the single-particle
state belonging to the $i$-th subshell with the normalization
\begin{eqnarray}
\int \rho^{sp}_i (r) d^3r = 1.0 .
\label{norm-cond}
\end{eqnarray}

\begin{figure*}[htb]
\centering
\includegraphics[width=16.0cm]{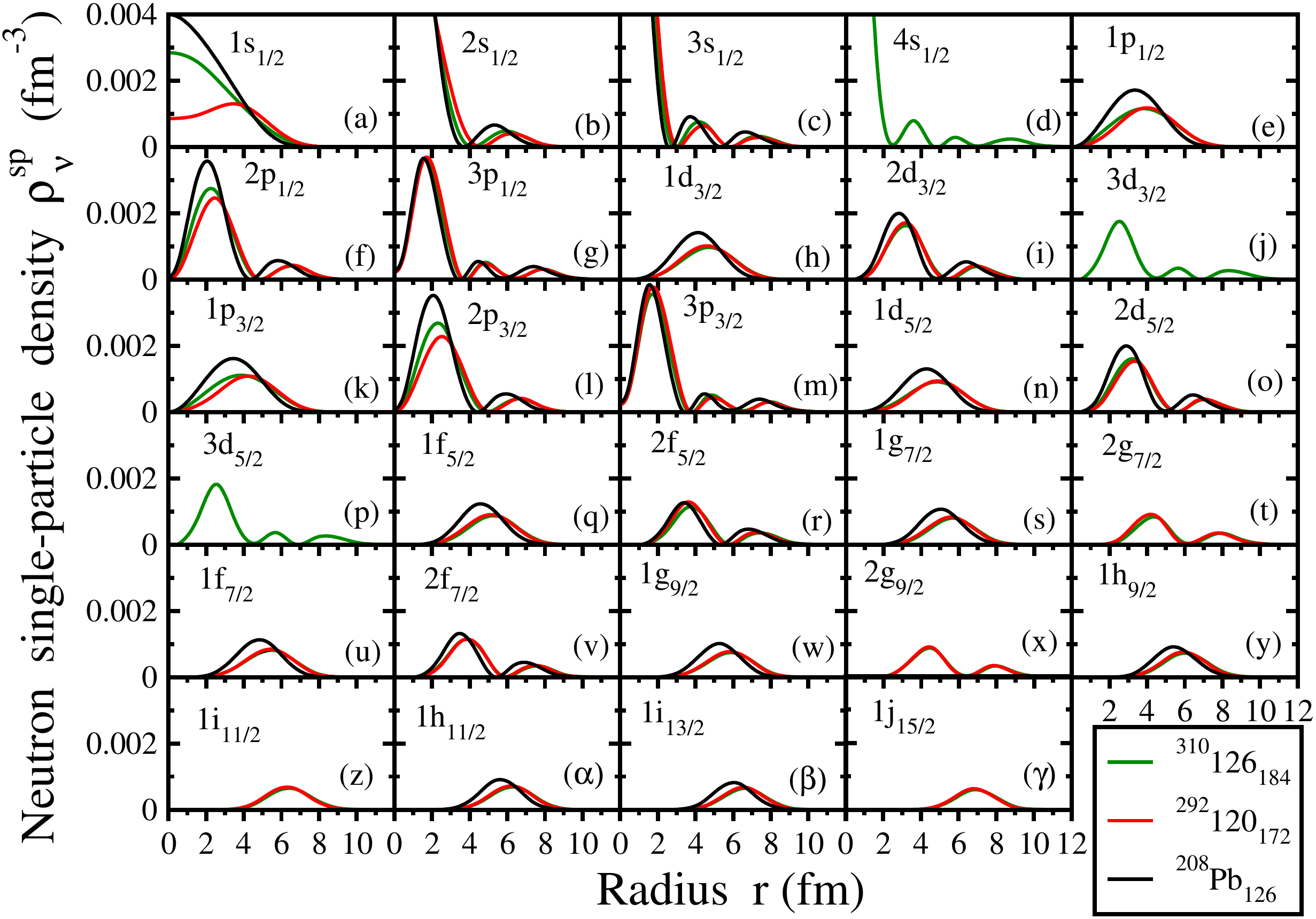}
\caption{Single-particle neutron density distributions
$\rho^{sp}_i (r)$ of the 
states in the occupied spherical subshells of indicated nuclei.
\label{sp-neutron-SHE}
}
\end{figure*}

\begin{figure}[ht]
\centering
\includegraphics[width=8.4cm]{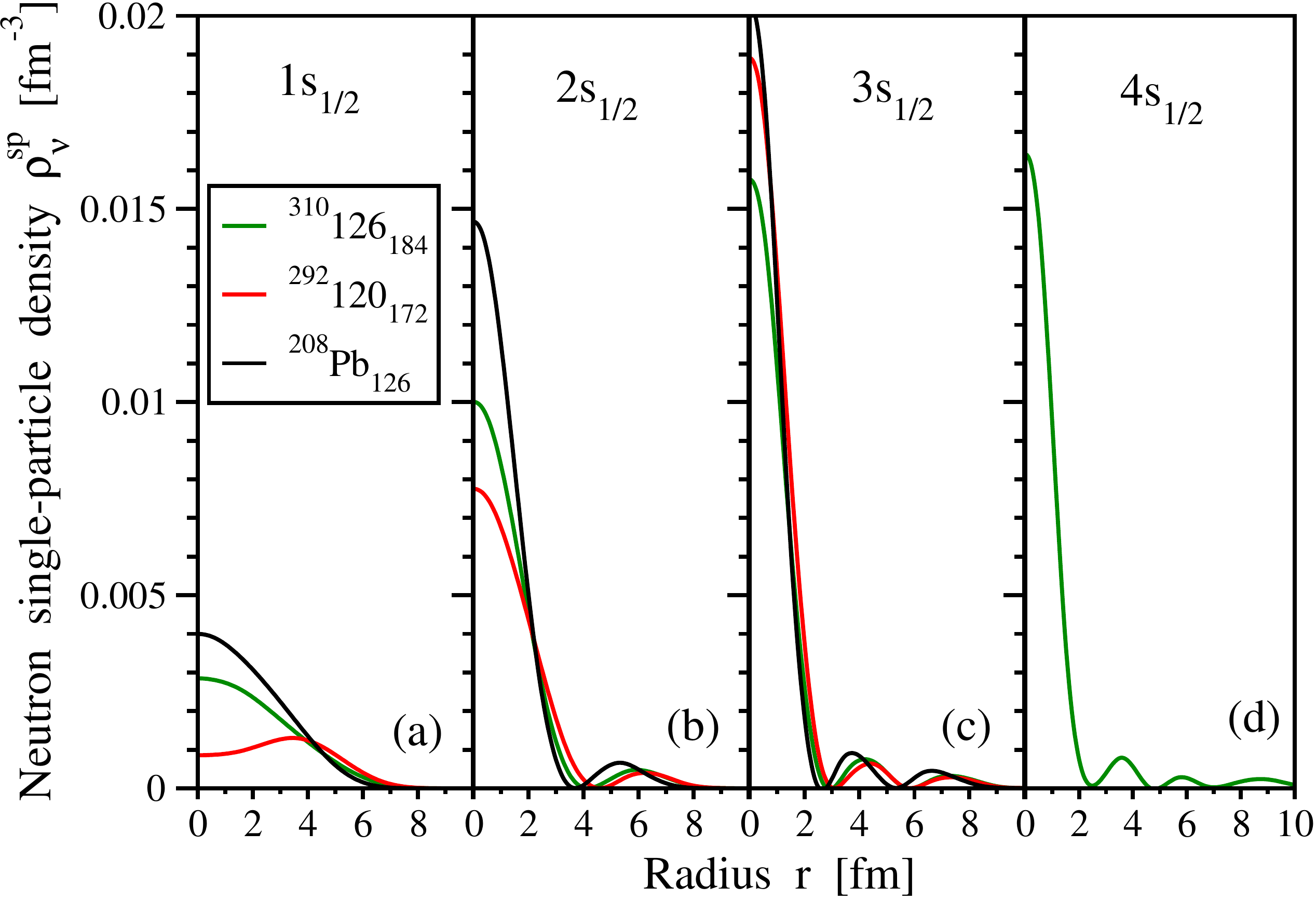}
\includegraphics[width=7.6cm]{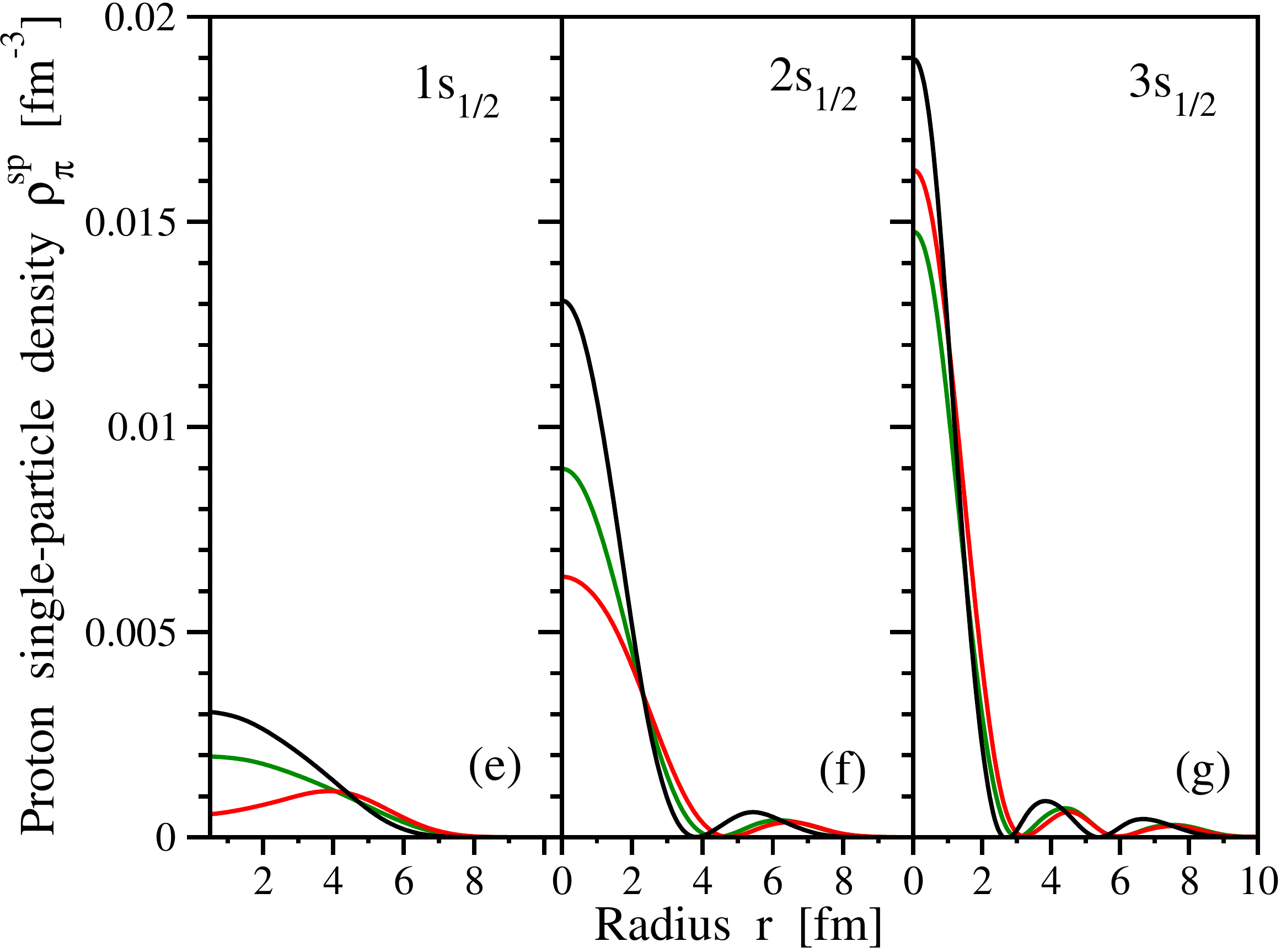}
\caption{The same as in Fig.\ \ref{sp-neutron-SHE} but only for the neutron (top panels) and 
proton (bottom panels) $s$ states. Note that the range on the vertical axis is increased as 
compared with  Fig.\ \ref{sp-neutron-SHE}. 
\label{sp-den-s-states}
}
\end{figure}

    The calculated neutron single-particle densities of the $^{208}$Pb, $^{292}$120 
and $^{310}$126 nuclei are shown in Fig.\ \ref{sp-neutron-SHE}. For the $l\geq 1$ subshells, 
proton single-particle densities are very similar to the neutron ones. Thus, they are not shown. 
The single-particle densities for the neutron and proton $s$ states are shown in greater detail 
in Fig.\ \ref{sp-den-s-states}.

   The following general features emerge from the analysis of these densities. First, 
the density at the center is built almost entirely by the $s$ states because centrifugal interaction 
does not allow the buildup of the density at $r=0$ for the $l\geq 1$ states (see discussion in 
Sec. 6 of Ref.\ \cite{NilRag-book}). In the relativistic framework, there is some contribution to 
the density at $r=0$ coming from the $p$ states which is especially pronounced for the  
$3p_{1/2}$ and $3p_{3/2}$ states [see Figs.\ \ref{sp-neutron-SHE}(g) and (m)]. It originates from the 
fact that small components of the Dirac spinor have opposite parity to the large component.
As a consequence, the $p$ state have the part of small component in the $s$  state which
builds the density at $r=0$. Note that in non-relativistic framework this mechanism is absent 
and the density at $r=0$ is built solely by the $s$ states (see Ref.\ \cite{NilRag-book}).

  Second, the single-particle densities of the $l\geq 1$ states in the  
$^{292}$120 and $^{310}$126 nuclei are very similar: this is a consequence of similar rms radii 
in respective subsystems of these nuclei (see Table \ref{Table-rms-radii}). The densities of the
single-particle states in the $^{208}$Pb nucleus have similar radial dependences as those in
superheavy nuclei but they are somewhat compressed in radial direction because of smaller
rms radii (see Table \ref{Table-rms-radii}). 

Third, the peaks of the single-particle density of the 
states with principal quantum number $n=1$ move to higher radial coordinate 
$r$ with increasing $l$.  The analysis of the $n=2$ and $n=3$ states is complicated by the 
presence of two and three peaks in density distribution, respectively.  However, these densities
also move to higher radial coordinate with increasing $l$.

Fourth, for the majority of the states
located substantially above the bottom of nucleonic potential the densities of the spin-orbit
partner orbitals are very similar [compare, for example, the $2g_{9/2}$ and $2g_{7/2}$ states
in Fig.\ \ref{sp-neutron-SHE}(t) and (x)]. The densities of the $j=l+1/2$ states of the spin-orbit 
doublets are only slightly compressed in radial coordinate  as compared with the ones of  their  
$j=l-1/2$ partners since these states are located deeper in the nucleonic potential due to
spin-orbit interaction.
%%%%%%%%%%%%%%%%%%%%%%%%%%%%%%%%%%%%%%%%%%%%%%%%
%%%%%%%%%%%%%%%%%%%%%%%%%%%%%%%%%%%%%%%%%%%%%%%% 

\begin{figure}[htb] 
\centering
\includegraphics[width=4.41cm]{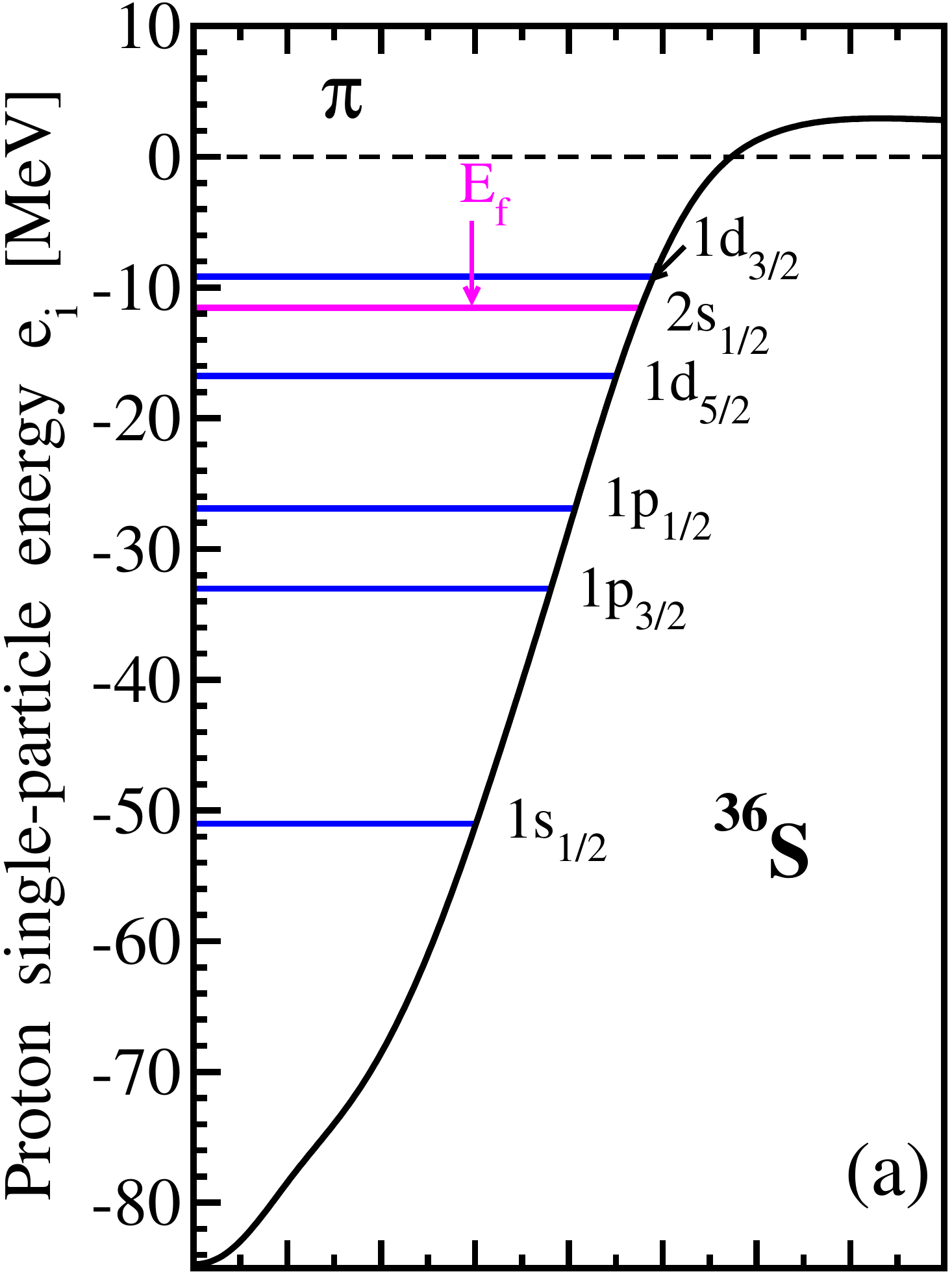}
\includegraphics[width=3.54cm]{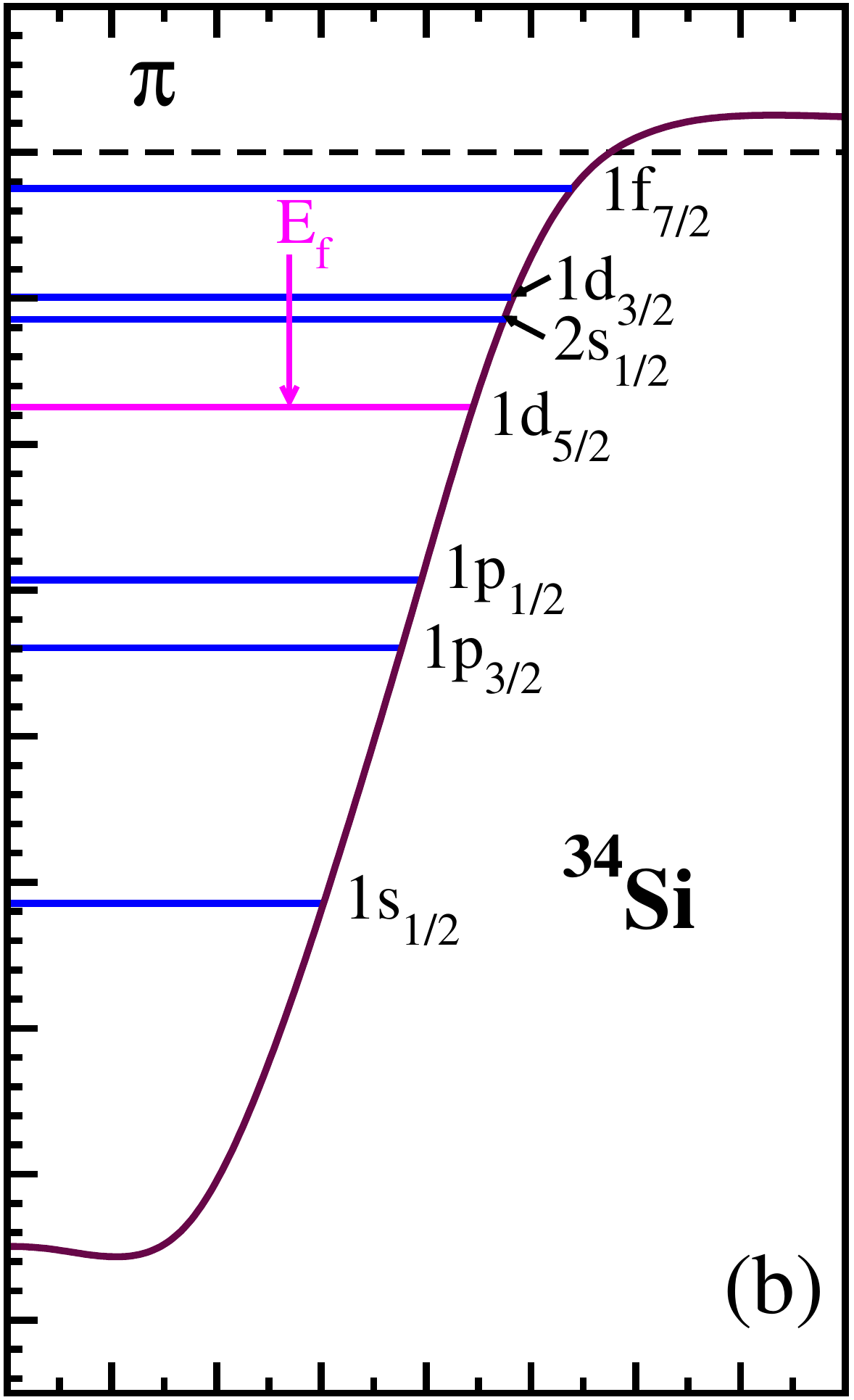}
\includegraphics[width=4.41cm]{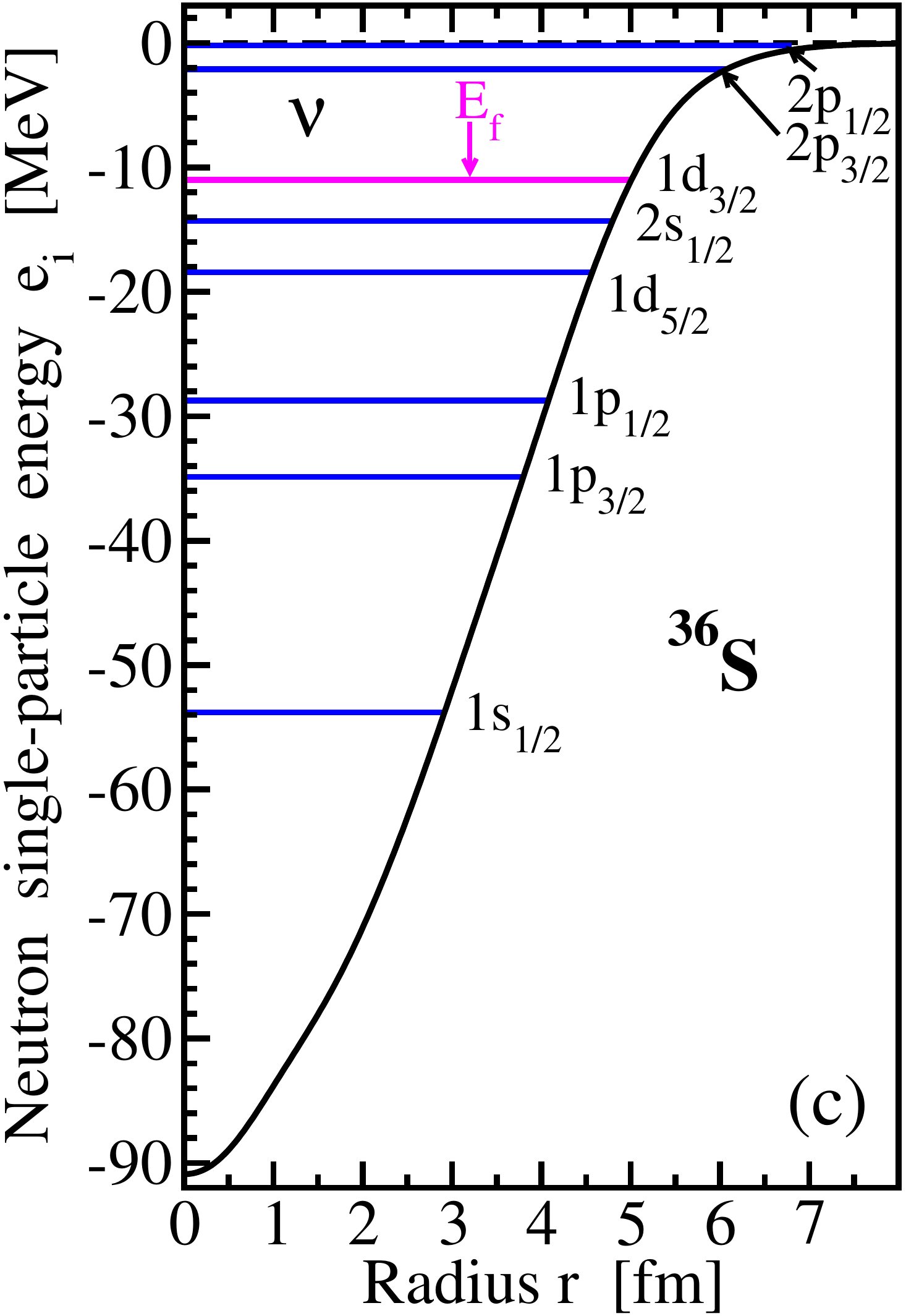}
\includegraphics[width=3.54cm]{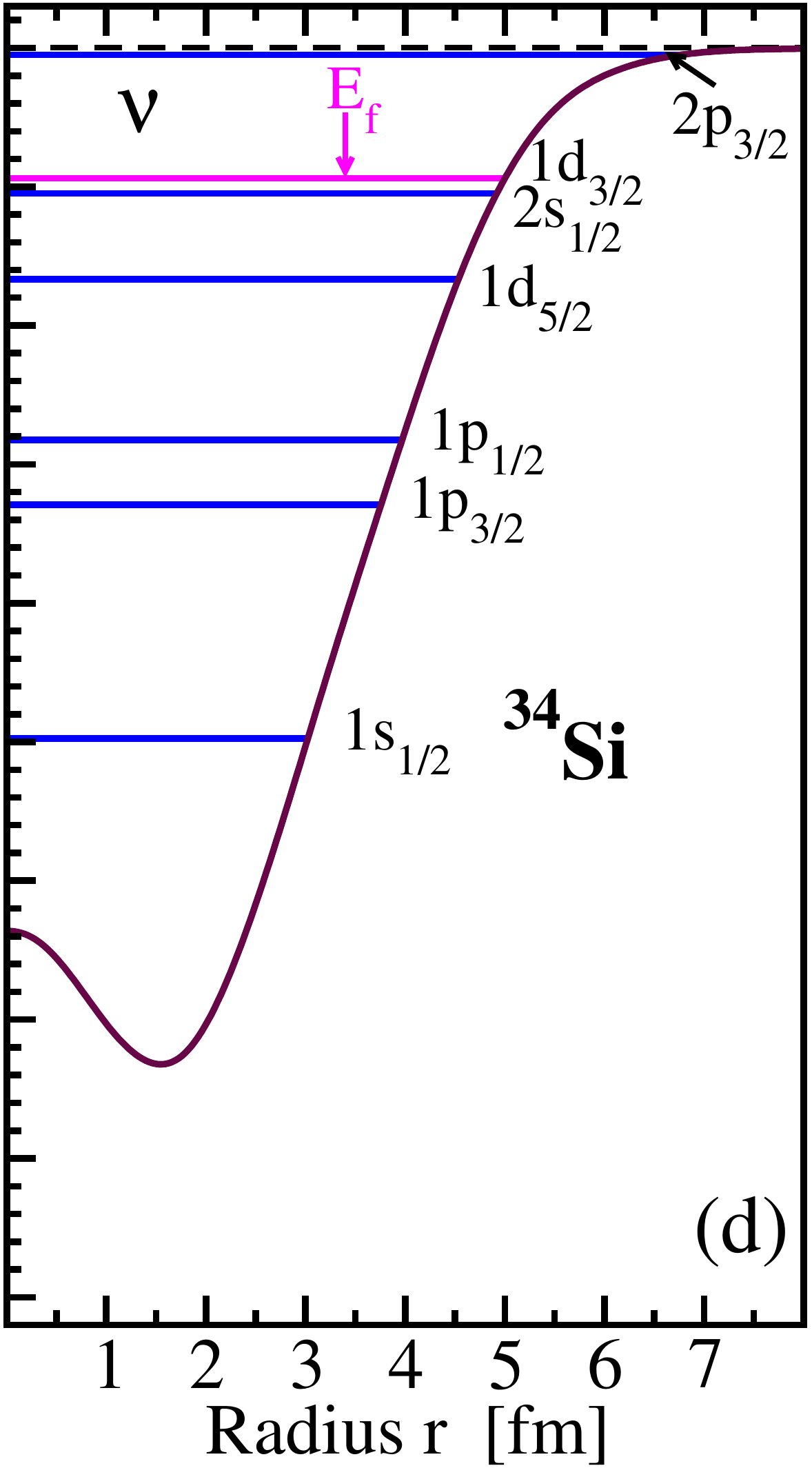}
\caption{The same as in Fig.\ \ref{Single-particle-120-126} but for the $^{36}$S 
             and $^{34}$Si nuclei.
\label{Potentials_Si_S}
}
\end{figure}

   A specific feature of the bubble nuclei is the formation of wine bottle shaped 
potentials (see Ref.\ \cite{AF.05-dep} and Figs.\ \ref{Single-particle-120-126} and 
\ref{Potentials_Si_S}). The transition from the $^{310}$126 nucleus, characterized by the 
flat bottom potentials, to the $^{292}$120 one, characterized by wine bottle potentials, is 
done by removing the protons from the $3p_{1/2}$ and $3p_{3/2}$ spherical subshells and 
the neutrons from the $3d_{5/2}$, $3d_{3/2}$ and $4s_{1/2}$ spherical subshells (see Fig. 2 
in Ref.\ \cite{AF.05-dep} and Fig.\ \ref{Single-particle-120-126}). These orbitals built the density 
in central and near-central regions of the nuclei and their removal leads to the depletion of 
central density and, as a consequence, to the formation of wine bottle proton and neutron potentials.  
In a similar fashion, the removal of two neutrons from the  $2s_{1/2}$ subshell in $^{36}$S leads 
to the formation of wine bottle neutron potential  in $^{34}$Si [compare Figs.\ \ref{Potentials_Si_S}(c) and (d)] and
flattening of proton potential in $^{34}$Si  [compare Figs.\ \ref{Potentials_Si_S}(a) and (b)].

However, the  impact of wine bottle nucleonic potentials on the single-particle  states and 
on their  densities has not been studied so far.  The analysis of Figs.\ \ref{Single-particle-120-126}(a) 
and (c) reveals that for some single-particle states located near the bottom of potential there is 
a classically forbidden region at radial coordinate $r<3.0$ fm.  These are proton $1s_{1/2}$, 
$1p_{3/2}$, $1p_{1/2}$, $1d_{5/2}$ and $1d_{3/2}$ states [see Fig.\ \ref{Single-particle-120-126}
(a)] and neutron $1s_{1/2}$, $1p_{3/2}$ and $1p_{1/2}$ states [see Fig.\ \ref{Single-particle-120-126}
(c)].  The presence of this classically forbidden region leads to a substantial reduction of the densities
of the proton and neutron $1s_{1/2}$ states in the $^{292}$120 nucleus for radial coordinate $r=0$ 
and near it as compared with the ones in the $^{208}$Pb and $^{310}$126 nuclei, which are 
characterized by near flat bottom potential
[see Figs.\ \ref{sp-den-s-states}(a) and (e)]. In addition, the profiles of the density distributions of 
the $1s_{1/2}$ states in the $^{292}$120 nucleus as a function of radial coordinate change drastically: the 
peak of the density is localized at $r\approx 4.0$ fm in the $^{292}$120 nucleus while in the
$^{208}$Pb and $^{310}$126 nuclei it is located at $r=0$ fm
[see Figs.\ \ref{sp-den-s-states}(a) and (e)].

    Classically forbidden regions for the proton and neutron $1p_{3/2}$ and $1p_{1/2}$ 
states are located for radial coordinate $r$ which is smaller than approximately 2.0 fm (see Figs.\ 
\ref{Single-particle-120-126}(a) and (c)). However, its impact on the densities of these states
is small since in the $^{292}$120 and $^{310}$126 nuclei the peak of their density distributions 
is located at $r\approx 4.0$ fm and the difference between their densities in these two nuclei 
is small  [see Figs.\ \ref{sp-neutron-SHE}(e) and (k)]. The impact of classically forbidden regions 
of the proton potential on the densities of the proton $1d_{5/2}$ and $1d_{3/2}$ orbitals is even 
smaller.  This is because the peak of their density distributions [at $r\approx 5.0$ fm, see Figs.\ 
\ref{sp-neutron-SHE}(n) and (h)] is located far away from the boundary of the classically forbidden 
region [at $r\approx 1.0$ fm, see Fig.\  \ref{Single-particle-120-126}(a)].

\begin{figure}[htb]
\centering
\includegraphics[width=8.4cm]{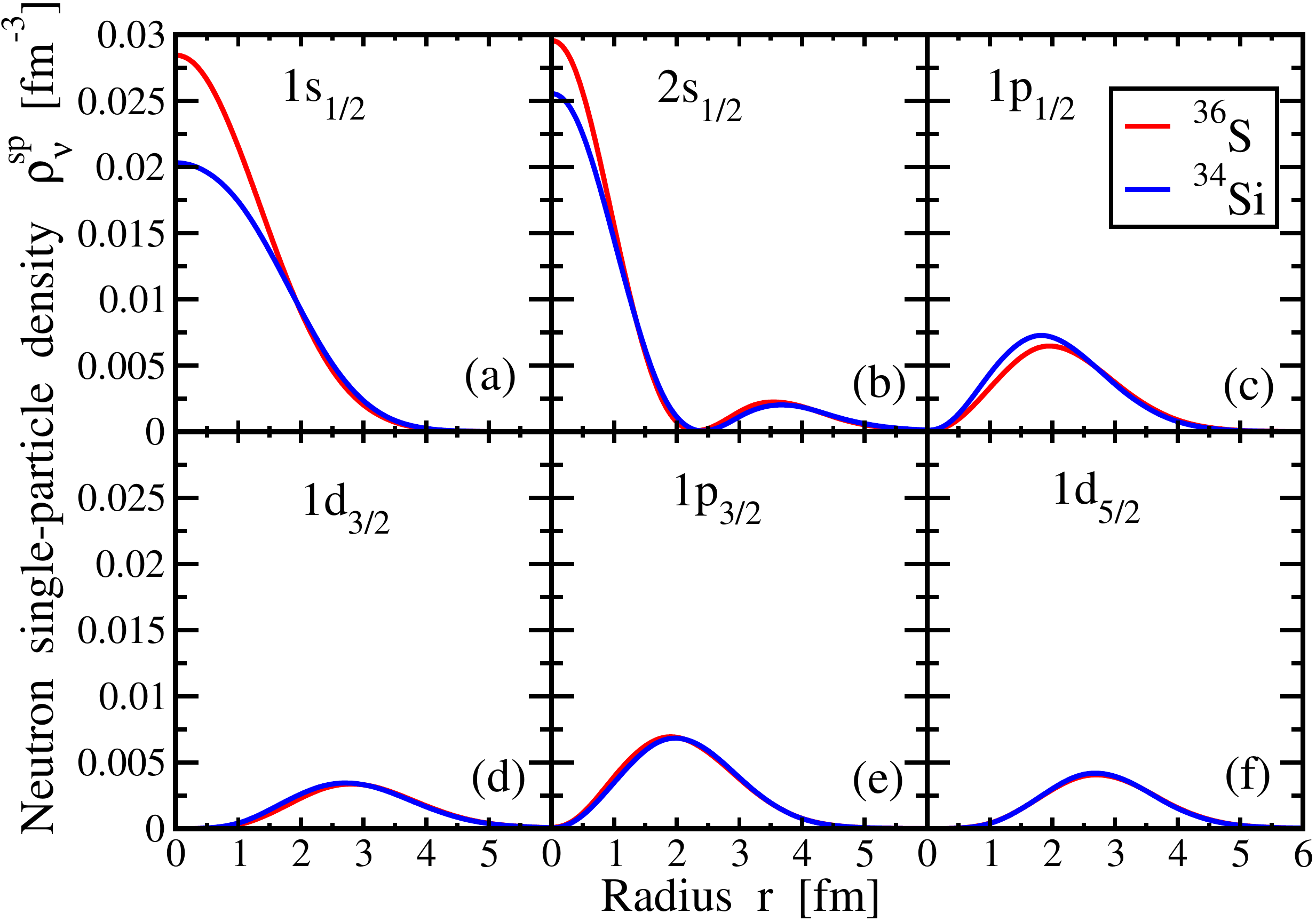}
\includegraphics[width=8.4cm]{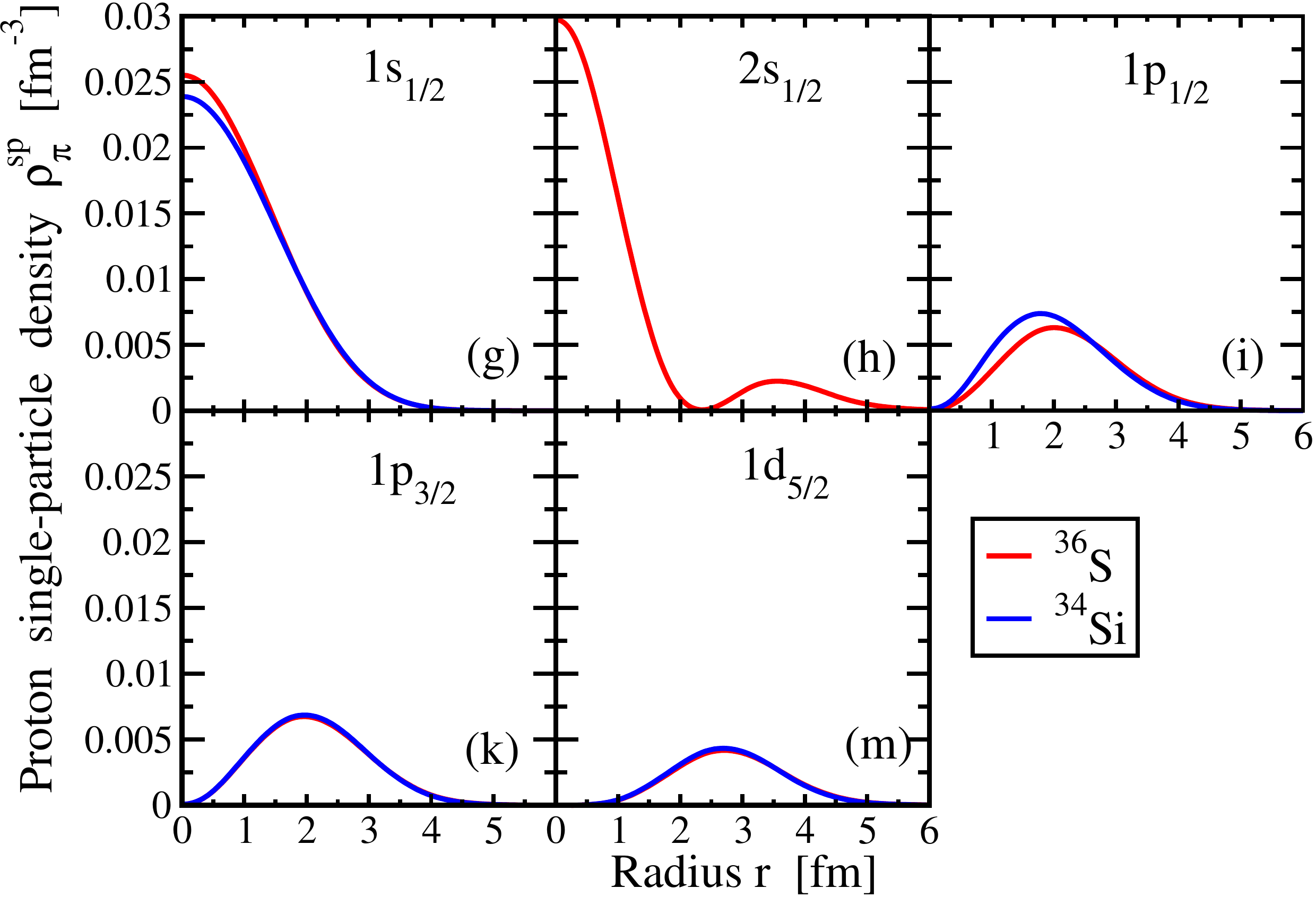}
\caption{Single-particle proton and neutron density distributions  of the occupied 
states in the $^{34}$Si and $^{36}$S nuclei as a function of radial coordinate $r$. 
\label{S_Si_SPD}
}
\end{figure}

   In addition, wine bottle potential affects the density distributions of other states which 
are located above its bottom and this effect is especially  pronounced for the $l=0$ $s$ states.  
For example, it  has substantial impact on the densities of the proton and neutron $2s_{1/2}$ states 
which for $r <1.0$ fm are substantially smaller  in the $^{292}$120 nucleus than those in the $^{208}$Pb 
and  $^{310}$126 nuclei [see Figs.\ \ref{sp-den-s-states}(b) and (f)]. Note that the total density of the 
nucleus at $r=0$ fm is built almost entirely by the $s$ states.  As a consequence, the differences in the proton 
and neutron densities at $r=0$ seen in the pairs of nuclei $^{208}$Pb/$^{292}$120 and  
$^{292}$120/$^{310}$126 are predominantly due to the impact of the change of the occupation of the 
$s$ states  and the impact of wine bottle nucleonic potentials of the $^{292}$120 nucleus on the density 
distributions of  these states. A similar impact is also seen in the $^{36}$S/$^{34}$Si pair of the nuclei for 
which the  removal of two protons from the $2s_{1/2}$ states in $^{36}$S leads to the formation of 
wine bottle neutron potential in $^{34}$Si (see Fig.\ \ref{Potentials_Si_S}). The consequence of this 
process is a substantial decrease of the single-particle densities of the neutron $1s_{1/2}$ and $2s_{1/2}$ 
states in the $^{34}$Si nucleus at low radial coordinate $r$ as compared with those in $^{36}$S 
[see Figs.\ \ref{S_Si_SPD}(a) and (b)]. Note that this reduction is almost absent in proton
subsystem since proton potential of $^{34}$Si has flat bottom [see Fig.\ \ref{Potentials_Si_S}(b)]. 
Note that spherical hyperheavy $^{466}$156 and  $^{592}$186 nuclei are characterized by 
wine bottle nucleonic potentials (see Fig.\ \ref{Single-particle-156-186} below) and the densities of  
low-lying states are affected by their presence in a similar way to that discussed above for the $^{292}$120 nucleus.

%%%%%%%%%%%%%%%%%%%%%%%%%%%%%%%%%%%%%%%%%%%%%%%%
%%%%%%%%%%%%%%%%%%%%%%%%%%%%%%%%%%%%%%%%%%%%%%%% 

%%%%%%%%%%%%%%%%%%%%%%%%%%%%%%%%%%%%%%
\section{The mechanisms of the formation of the wine bottle potentials}
\label{wine-potentiall}
%%%%%%%%%%%%%%%%%%%%%%%%%%%%%%%%%%%%%%

   In order to better understand the mechanisms of the formation of 
wine bottle potentials we consider the evolution of nucleonic potentials and densities 
along the isotopic and isotonic chains in Fig.\ 
\ref{density-potentials}.  We start  from  $^{208}$Pb and then sequentially occupy 
spherical subshells in the order shown in Table \ref{Table-wine}. In this way, the 
densities and potentials are built first along $Z_{fix}=82$ (first column in Fig.\ 
\ref{density-potentials}), then along 
$N_{fix}=172$ (second column in Fig.\  \ref{density-potentials}) and $Z_{fix}=120$ (third 
column in Fig.\ \ref{density-potentials}) and finally along $N_{fix}=184$ (fourth column 
in Fig.\ \ref{density-potentials}). Note that not necessary the occupation of spherical
subshells in the order shown in  Table \ref{Table-wine} leads to the ground states in
the nuclei of interest. However, this is acceptable since we are interested in the 
understanding of the mechanisms leading to the formation of the wine bottle potentials 
and their dependence on the occupation of specific single-particle states and 
this is easier to achieve by considering the occupation of full spherical subshells.
Note that obtained solutions in $^{209}$Pb,
$^{292}$120 and $^{310}$126 nuclei correspond to the ground states.

%%%%%%%%%%%%%%%%%%%%%%%%%%%%%%%%%%%%%%%%%%%%%%%%  
%%%%%%%%%%%%%%%%%%%%%%%%%%%%%%%%%%%%%%%%%%%%%%%%  
 Let us start from the $^{208}$Pb nucleus and to see how the densities and potentials are 
affected by the addition of neutrons and protons. The neutron potential of this nucleus is flat 
bottom one [see Fig.\ \ref{density-potentials}(i)].  However,  proton potential shows some 
development of wine bottom features but the difference between the $(V+S)$ values at 
$r=0$ and $r=4.2$ fm is only around 6 MeV (see Fig.\  \ref{density-potentials}(e)). Although 
some fluctuations induced by the single-particle effects exist, the proton and neutron densities 
of this nucleus in the subsurface region are close to flat ones [see  Fig.\  \ref{density-potentials}(a) 
and (m)]. Note that when considering the addition of particle(s) one should take into account
the structure of their single-particle density distributions shown in Fig.\ \ref{sp-neutron-SHE} and 
the location of the maxima (peaks) of their density distributions in radial coordinate ($r_{peak(s)}$) 
(see Table \ref{Table-wine}). The single-particle density is typically localized within $r-2.0\,\,{\rm fm} < 
r_{peak(s)} < r+2.0$ fm region around the peak.
 \clearpage
%%%%%%%%%%%%%%%%%%%%%%%%%%%%%%%%%%%%%%%%%%%%%%%%
%%%%%%%%%%%%%%%%%%%%%%%%%%%%%%%%%%%%%%%%%%%%%%%%

  The Pb $(Z=82$) isotopic chain will be considered first. The surface region of these
nuclei are located at $r>5.5$ fm [see  Fig.\  \ref{density-potentials}(m)].  Thus,  the occupation 
of the $\nu 1i_{11/2}$ subshell, leading to the $N=138$ isotope, builds the neutron density
mostly in the near surface and surface regions.  As a consequence, the neutron density profile is
similar for the $N=126$ and $N=138$ isotopes in the subsurface region [see  Fig.\  
\ref{density-potentials}(m)]. The occupation of the $\nu 2g_{9/2}$ subshell, leading to the
$N=148$ isotope builds the density in the subsurface region around $r \approx 4.4$ (due to 
the first peak of the single-particle density distribution) and in the surface region (due to the second 
peak of the single-particle density at $r_{peak(s)} \approx 8.0$ fm). As a consequence, the neutron 
density of the $N=148$ isotope is larger than those of the $N=126$ and $N=138$ isotopes 
at $r\approx 4.4$ fm but  smaller\footnote{The density in the central ($r<2 $ fm) region 
of the nucleus typically decreases with increasing proton or neutron number if no new $s_{1/2}$ 
state(s) is(are) occupied [see Figs.\ \ref{density-potentials}(a), (m), (b), (n), and (c)].  This is due 
to the stretching out the radial profile of the density distribution of the single-particle states with 
increasing proton and neutron number (see Fig.\ \ref{sp-neutron-SHE}).}
 (and similar in radial profile) for $r<3$ fm [see  Fig.\  
\ref{density-potentials}(m)]. 
 The occupation of the $\nu 1j_{15/2}$ subshell, leading to the 
$N=164$ isotope, contributes 
density mostly in the surface region since the peak of its single-particle density 
is located at $r_{peak(s)}\approx 6.8$ fm (see Table \ref{Table-wine}). As a result, in the
subsurface region the density profiles as a function of radial coordinate are very 
similar for the $N=148$ and $N=168$ isotopes [see  Fig.\  \ref{density-potentials}(m)].
The effect of the occupation of the $\nu 2g_{7/2}$ subshell is very similar to that of the $\nu 2g_{9/2}$ 
subshell discussed above [see  Fig.\  \ref{density-potentials}(m)].

The final result of the  sequence of these occupations of the spherical subshells is the formation of 
the bubble structure in  the neutron density of the $N=172$ isotope [see  Fig.\  
\ref{density-potentials}(m)]. It is created by the combination of two factors, namely, (i) the buildup 
of the densities at $r\approx 4.4$ fm due to  the first peaks of the single-particle densities of the 
$\nu 2g_{9/2}$ and  $\nu 2g_{7/2}$  subshells and (ii) the reduction of the neutron densities in 
the central region (in particular, at $r=0$) due to a general stretching out of  the nucleus with 
increasing neutron number. The latter effect is even more pronounced 
in proton subsystem [compare Figs.\ \ref{density-potentials}(a) 
and (m)].  Note, however,  that in the subsurface region the radial profile of the proton 
densities remains more or less the same but its magnitude decreases drastically with 
increasing neutron number [see Figs.\ \ref{density-potentials}(a)].  

  The consequences of these density changes for the nucleonic potentials 
are somewhat counterintuitive. The
neutron potentials of the Pb isotopes remain close to the flat bottom ones [see Fig.\ 
\ref{density-potentials}(i)] despite the formation of the neutron bubble structures in 
the $N=164$ and $N=172$ isotopes [see Fig.\  \ref{density-potentials}(m)].  In
contrast, the wine bottom features become enhanced in the proton potentials
of the $N=164$ and, especially, $N=172$ isotopes as compared with those of the 
$N=126$ isotope [see Fig.\  \ref{density-potentials}(e)].

  Similar features are also seen in the $N=172$ isotopic chain. The occupation of  the 
$\pi 1h_{9/2}$ and $\pi 1i_{13/2}$ spherical subshells builds density near $r\approx 6$ 
fm and leads to the formation of pronounced proton bubble structures in the $Z=96$
and, especially, $Z=106$ isotones [see Table \ref{Table-wine} and  Fig.\ \ref{density-potentials}(b)].  
The subsequent occupation
of the $\pi 2f_{7/2}$ and $\pi 2f_{5/2}$ subshells leads to an additional buildup of the 
densities near $r= 3.8$ fm in the $Z=120$ isotone but this process still
preserves the proton bubble structure [see Fig.\ \ref{density-potentials}(b)].
These modifications of the proton densities feed back into proton potentials the wine
bottom features of which become more enhanced in the $Z=106$ and $Z=120$ 
isotones as compared with the $Z=82$ one [see Fig.\ \ref{density-potentials}(f)].  Because 
of the isovector force, which tries to keep the neutron and proton density profiles alike, neutron 
bubble structures are also somewhat enhanced in the $Z=106$ and $Z=120$ isotones as 
compared with the $Z=82$ one [see Fig.\ \ref{density-potentials}(n)]. This feeds back into the 
neutron potentials of these isotones which contrary to the $Z=82$ and $Z=96$ ones develop 
wine bottom features [see Fig.\ \ref{density-potentials}(n)].

%%%%%%%%%%%%%%%%%%%%%%%%%%%%%%%%%%%%%%%%%%%%%%%%  
\begin{figure*}[t] 
\centering
\includegraphics[width=6.64cm]{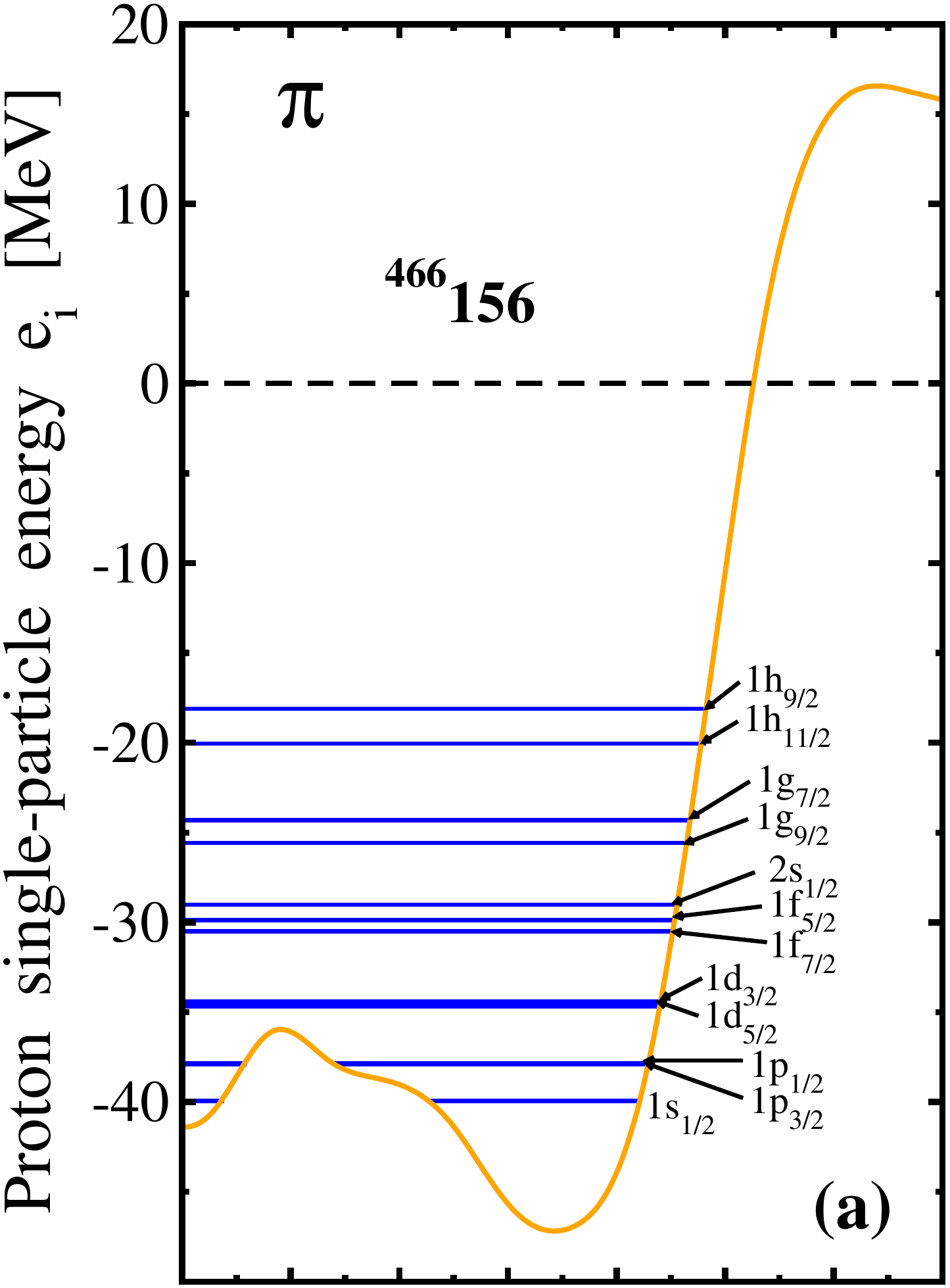}
\includegraphics[width=5.4cm]{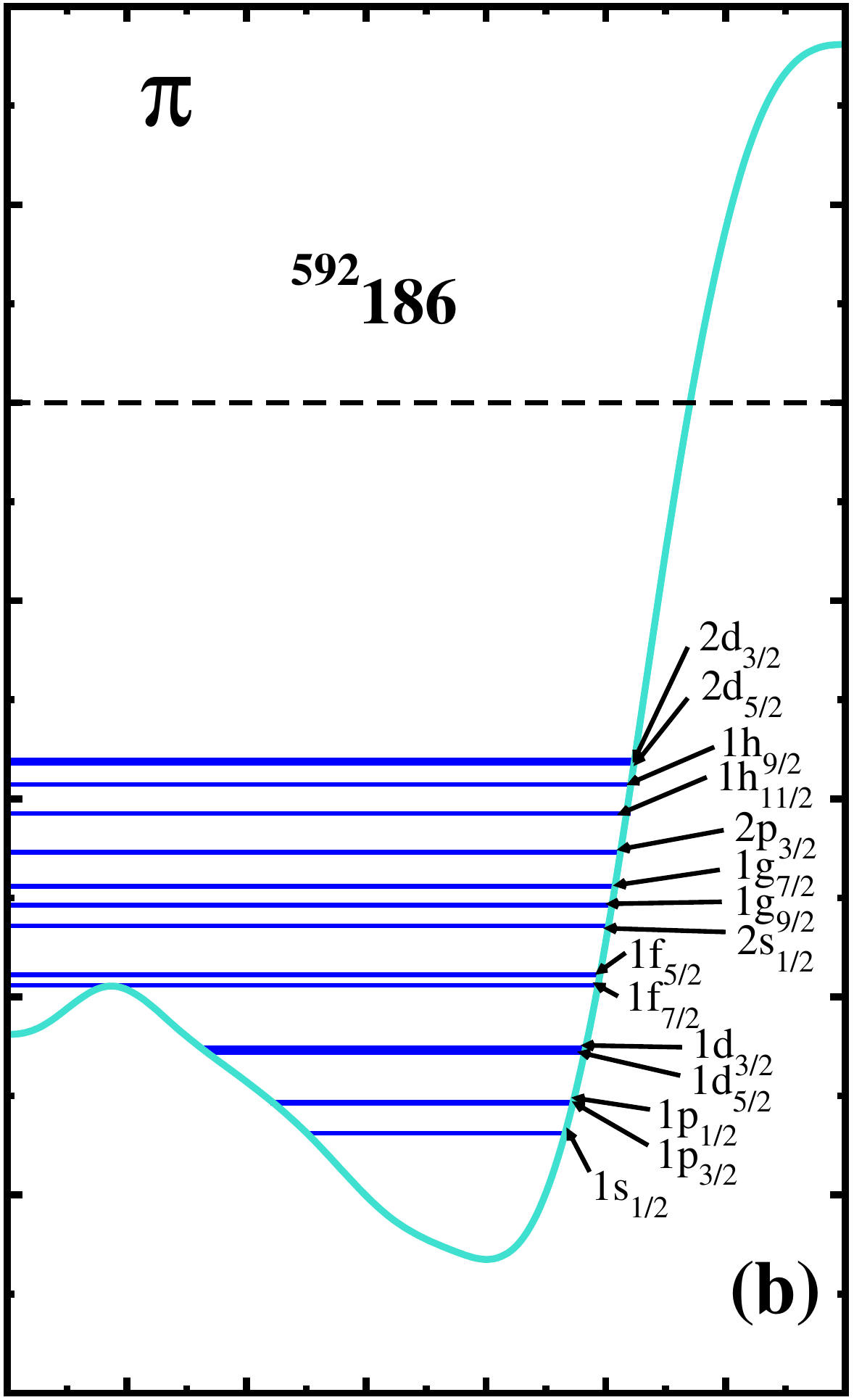}
\includegraphics[width=6.64cm]{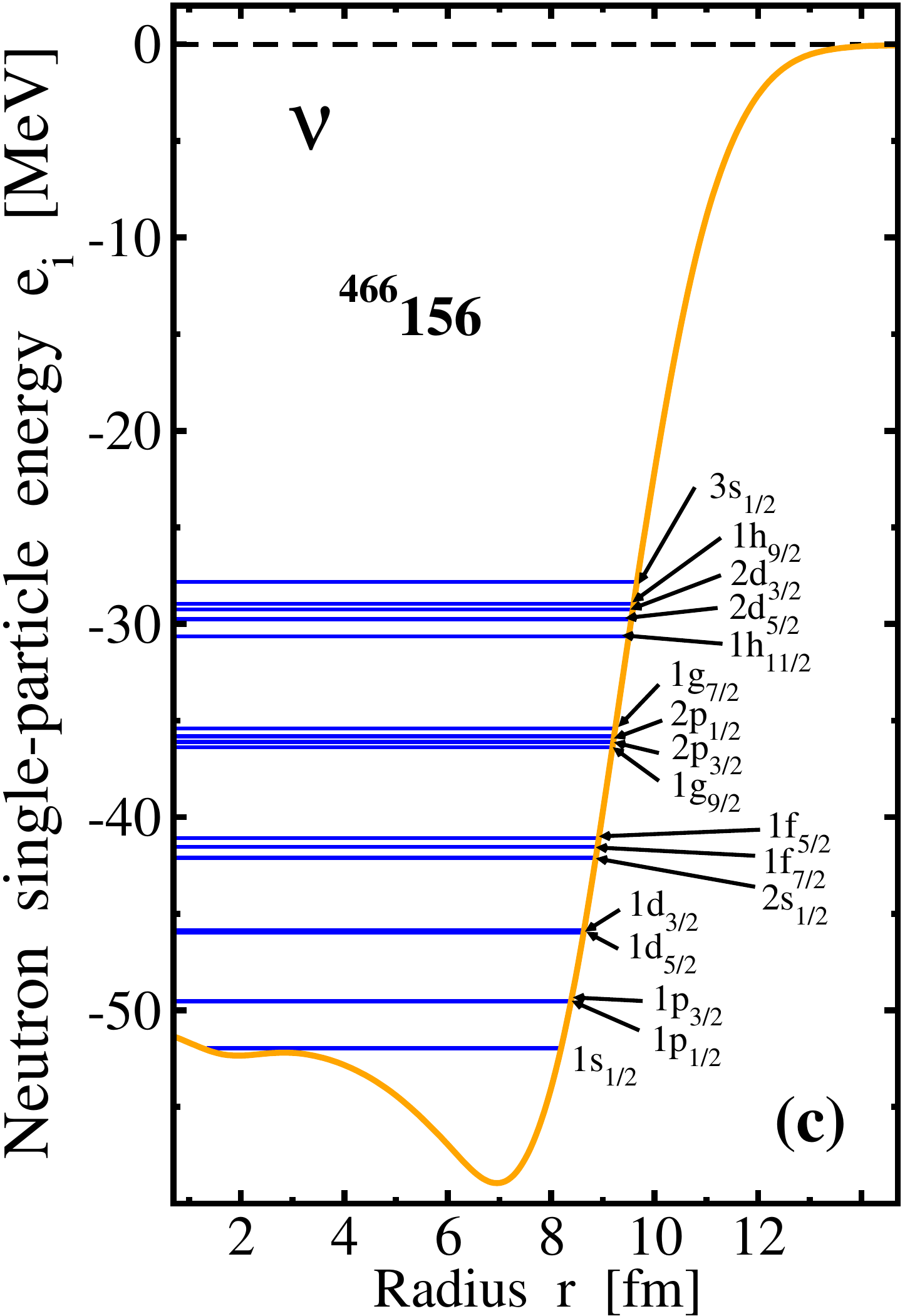}
\includegraphics[width=5.42cm]{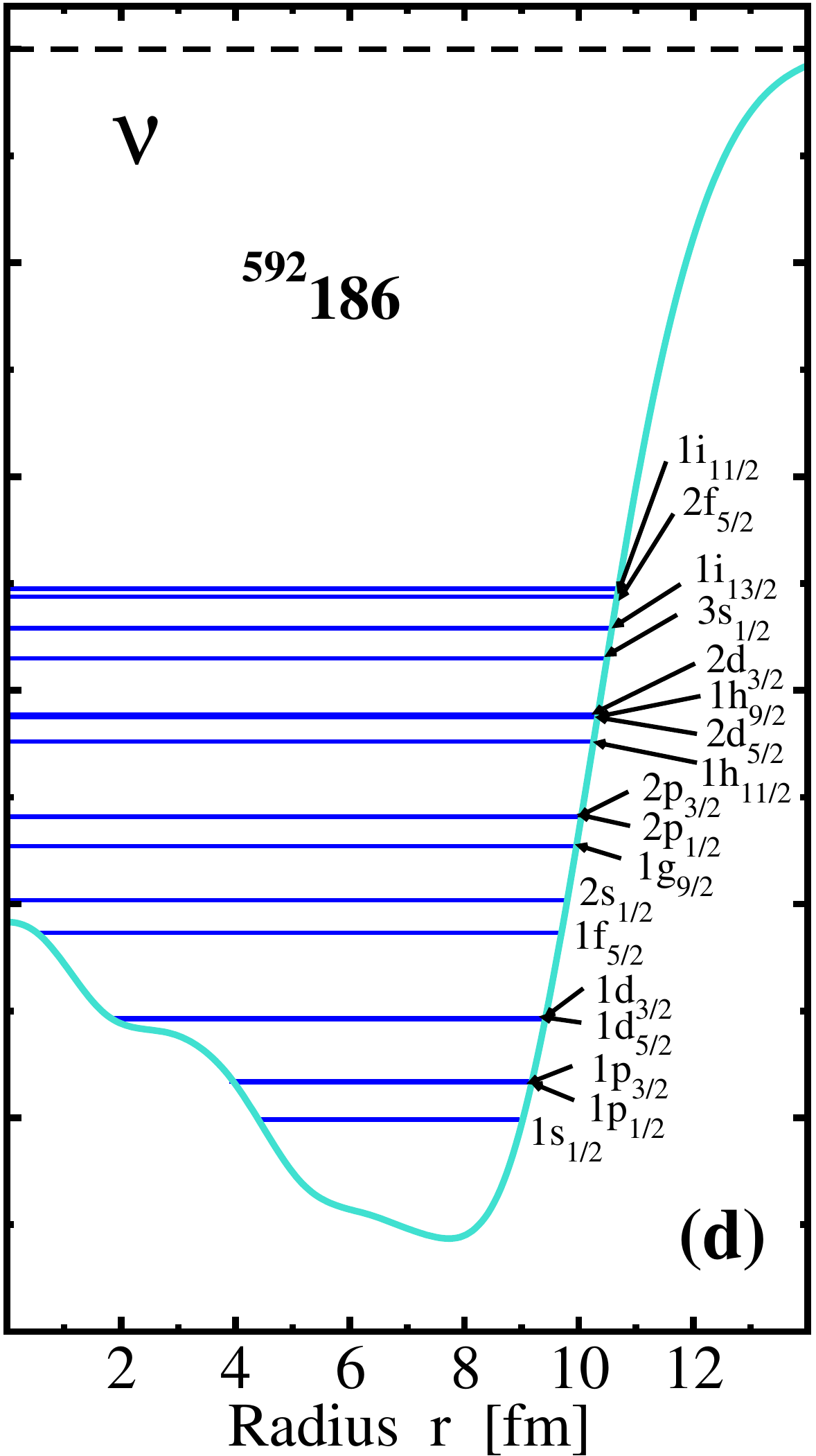}
\caption{The same as in Fig.\ \ref{Single-particle-120-126} but for spherical hyperheavy nuclei. 
For simplicity, only single-particle states within approximately 25 MeV from the bottom of 
potential are shown.
\label{Single-particle-156-186}
}
\end{figure*}
%%%%%%%%%%%%%%%%%%%%%%%%%%%%%%%%%%%%%%%%%%%%%%%%  

  The next step is along the $Z=120$ line. The occupation of the $\nu 3d_{5/2}$ and 
$\nu 3d_{3/2}$  subshells leading  to the $N=182$ isotope  builds density near $r \approx 2.5$ 
fm  making neutron density flat in the $2.5\,\,{\rm fm} < r < 6.5$ fm region [see Fig.\ \ref{density-potentials}(o)].
However, the neutron bubble still survives. Only the occupation of the $\nu 4s_{1/2}$ subshell,
leading to the $N=184$ isotope, eliminates this bubble [see Fig.\ \ref{density-potentials}(n)]. 
Note that proton bubble structures survive in these nuclei but they become less pronounced
[see Fig.\ \ref{density-potentials}(c)]. These neutron density changes somewhat reduce wine 
bottom features of the neutron potential but do not eliminate them completely [see Fig.\ 
\ref{density-potentials}(k)]. The situation is more drastic for the proton potential in which the 
occupation of the neutron $1s_{1/2}$ subshell significantly reduces wine bottom features of the 
potential [see Fig.\ \ref{density-potentials}(g)].
\clearpage
   Let us consider the latter case of the transition from $N=182$ to $N=184$
in detail. This is definitely fully self-consistent process in which the drastic increase of the 
neutron density at and near $r=0$ induced by the occupation of the $\nu 4s_{1/2}$ state 
[see Fig.\ \ref{density-potentials}(o)] leads to a moderate increase of the proton densities at 
and near $r=0$ due to isovector nature of nuclear force [see Fig.\  
\ref{density-potentials}(c)]. This in turn requires the increase of the single-particle
densities of the occupied proton $1s_{1/2}$, $2s_{1/2}$, and $3s_{1/2}$ states [which
is seen in detailed analysis]  that can be achieved only by the transition from
wine bottom  to near flat bottom proton potential  [see Fig.\  
\ref{density-potentials}(g)].

  Finally, the transition from the $Z=120$ to $Z=126$ isotone along the $N=184$ 
line is carried out by occupying the $\pi 3p_{3/2}$ and $\pi 3p_{1/2}$ subshells. This leads
to the flattening of the proton  density in the subsurface region [see Fig.\  \ref{density-potentials}(d)]  
because the major peak of single-particle density of these proton subshells is located at 
$r_{peak(s)} \approx 1.8$ fm  [see Table \ref{Table-wine} and Figs.\ \ref{sp-neutron-SHE}(g) and (m)].  
Similar effect is seen in the neutron densities because of the isovector character of nuclear force 
[see Fig.\  \ref{density-potentials}(p)]. However, the impact of this  process on the features of 
the proton potential is rather small [see Fig.\ \ref{density-potentials}(h)]. In contrast, it completely 
removes wine bottom features from the neutron potential which becomes flat bottom one 
[see Fig.\ \ref{density-potentials}(l)].

 The cases discussed above reveal that the formation/suppression of the bubble structure 
in the densities of one subsystem (let us call it as A) of the nucleus leads to a significant enhancement 
of wine  bottle (flat bottom) features of the potential in another subsystem (let us call it as B)\footnote{If 
A=proton then B=neutron and vise versa.}.  Note 
that the potentials of  the subsystem A are only moderately affected by this process. Similar features  have been seen earlier in the analysis of the ground state and excited 
configurations of the  $^{292}$120 nucleus (see the discussion of Fig. 3 in Ref.\ \cite{AF.05-dep}). The following 
explanation  is in place.  Let us consider the case of the formation of the bubble structure in the densities of
the subsystem A.  It proceeds by the occupation of the states in the vicinity of the Fermi level and 
it has only minor impact on the nucleonic potential of this subsystem. The isovector interaction tries 
to keep proton and neutron densities alike. For a  fixed number of the particles in the subsystem B, 
the formation of the bubble structure  in its densities can be achieved only by a significant 
enhancement of wine bottom features of its potential.

%%%%%%%%%%%%%%%%%%%%%%%%%%%%%%%%%%%%%%%
\section{Additivity rule for densities}
\label{additivity}
%%%%%%%%%%%%%%%%%%%%%%%%%%%%%%%%%%%%%%%

%%%%%%%%%%%%%%%%%%%%%%%%%%%%%%%%%%%%%%%%%%%%% 
\begin{table}[!ht]
\begin{center}
\caption{The occupation of  neutron/proton single-particle subshells in Fig.\ 
\ref{density-potentials}  for fixed proton (neutron) particle numbers $Z_{fix}(N_{fix})$ 
on going from the nucleus with $N_{in}(Z_{in})$ to the nucleus with $N_{fin}(Z_{fin})$.
The sequence of the states is defined by the general trends of the evolution of the
single-particles structure with proton and neutron numbers (see Fig. 1 in Ref.\ 
\cite{PAR.21} for $^{208}$Pb and Fig.\ \ref{Single-particle-120-126} in the present 
paper for superheavy nuclei). The approximate positions of the peaks of the
single-particle density of these states are shown in the last column. They are
taken from Fig.\ \ref{sp-neutron-SHE}. Note that the number of peaks is equal to the
principal quantum number $n$.  The states of the spin-orbit doublets emerging from 
the low-$l$ subshells (such as  the ($\nu 3d_{5/2}$, $\nu 3d_{3/2}$) states) are 
characterized by relatively small energy splitting (see Fig.\ \ref{Single-particle-120-126}), 
very similar single-particle densities (see Fig.\ \ref{sp-neutron-SHE}) and relatively 
moderate number of particles which can occupy them. Thus, for simplicity, such states 
are occupied together in Fig.\ \ref{density-potentials}.
}
\begin{tabular}{c|c|c|c} \hline \hline
$N_{in}(Z_{in})$ &  orbital & $N_{fin}(Z_{fin})$ & peak(s) [fm] \\ \hline 
      1       &        2   & 3   & 4            \\ \hline 
\multicolumn{4}{c}{$Z_{fix}=82$} \\ \hline 	 		
%         &                          &  \\ \hline
 126  & $\nu 1i_{11/2}$ & 138 & 6.2 \\
 138  & $\nu 2g_{9/2}$  & 148 & 4.4 and 8 \\
 148  & $\nu 1j_{15/2}$ & 164 & 6.8 \\
 164  & $\nu 2g_{7/2}$  &  172 & 4.3 and 8 \\ \hline 
\multicolumn{4}{c}{$N_{fix}=172$} \\ \hline 	 		
%  & $N_{fix}=172$  &  \\
  82  & $\pi 1h_{9/2}$ &  92 & 5.7 \\
  92  & $\pi 1i_{13/2}$ & 106 &  6.5 \\
 106 & $\pi 2f_{7/2}$ + $\pi 2f_{5/2}$ & 120 &  3.8 and 7.4 \\ \hline
\multicolumn{4}{c}{$Z_{fix}=120$} \\ \hline 	 		
%        & $Z_{fix}=120&  \\
  172   & $\nu 3d_{5/2}$ +$\nu 3d_{3/2}$  & 182 &  2.5, 5.8 and 8.5 \\
  182  & $\nu 4 s_{1/2}$ & 184 & 0 \\ \hline
%      & $N_{fix}=184 &  \\  
\multicolumn{4}{c}{$N_{fix}=184$} \\ \hline 	 		
   120  & $\pi 3p_{3/2}$ + $\pi 3p_{1/2}$ & 126 & 1.8, 5.0, and 8.0 \\
            \hline	\hline
\end{tabular}
\label{Table-wine} 
\end{center}
\end{table}
%%%%%%%%%%%%%%%%%%%%%%%%%%%%%%%%%%%%%%%%%%%%% 

The addition or removal of particle(s) to the nucleonic configuration modifies the total 
physical observables. But it also creates the polarization effects on the physical properties
(both in time-even and time-odd channels) of initial configuration. The comparison of
relative properties of two configurations can shed important light both on the impact of 
the added/removed particle(s) in specific orbital(s) on physical observable of interest 
and  on the related polarization effects. In this context the additivity rule of physical 
observables plays an extremely important role since it allows to verify whether
the independent particle motion is realized in finite nuclei \cite{SDDN.96,A.22}. This rule states that
physical observable $O^B$ in the configuration $B$ can be approximated as a sum 
of physical observable $O^A$ in reference configuration $A$ and single-particle
contributions $o_i$ of the states by which the configurations $A$ and $B$ differ
\begin{eqnarray} 
O(B) = O(A) + \sum_i o_i
\end{eqnarray} 
The additivity rule was successfully tested for the effective alignments and relative
quadrupole moments of the superdeformed rotational bands in unpaired regime
(see Refs.\ \cite{Rag.93,SDDN.96,ALR.98,MADLN.07}). This justifies the use of an 
extreme single-particle model in an unpaired regime typical of high angular momentum.  
Note that the basic idea behind the additivity rule for one-body operators is rooted 
in the independent particle model \cite{Rag.93,SDDN.96,A.22}.

%%%%%%%%%%%%%%%%%%%%%%%%%%%%%%%%%%%%%%%%%%%%%%%%
\begin{figure*}[h]
\centering
\includegraphics[width=5.3cm]{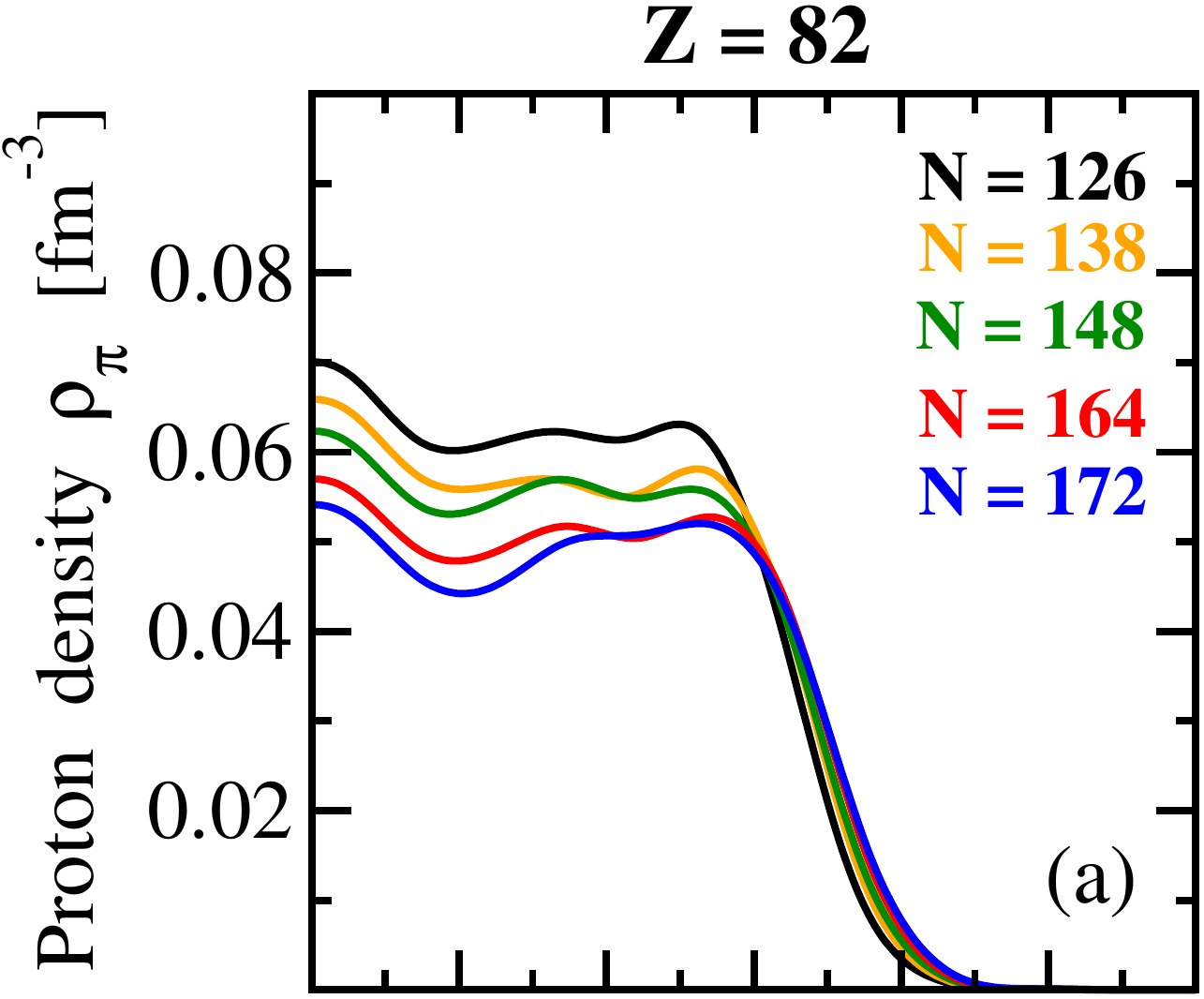}
\includegraphics[width=3.96cm]{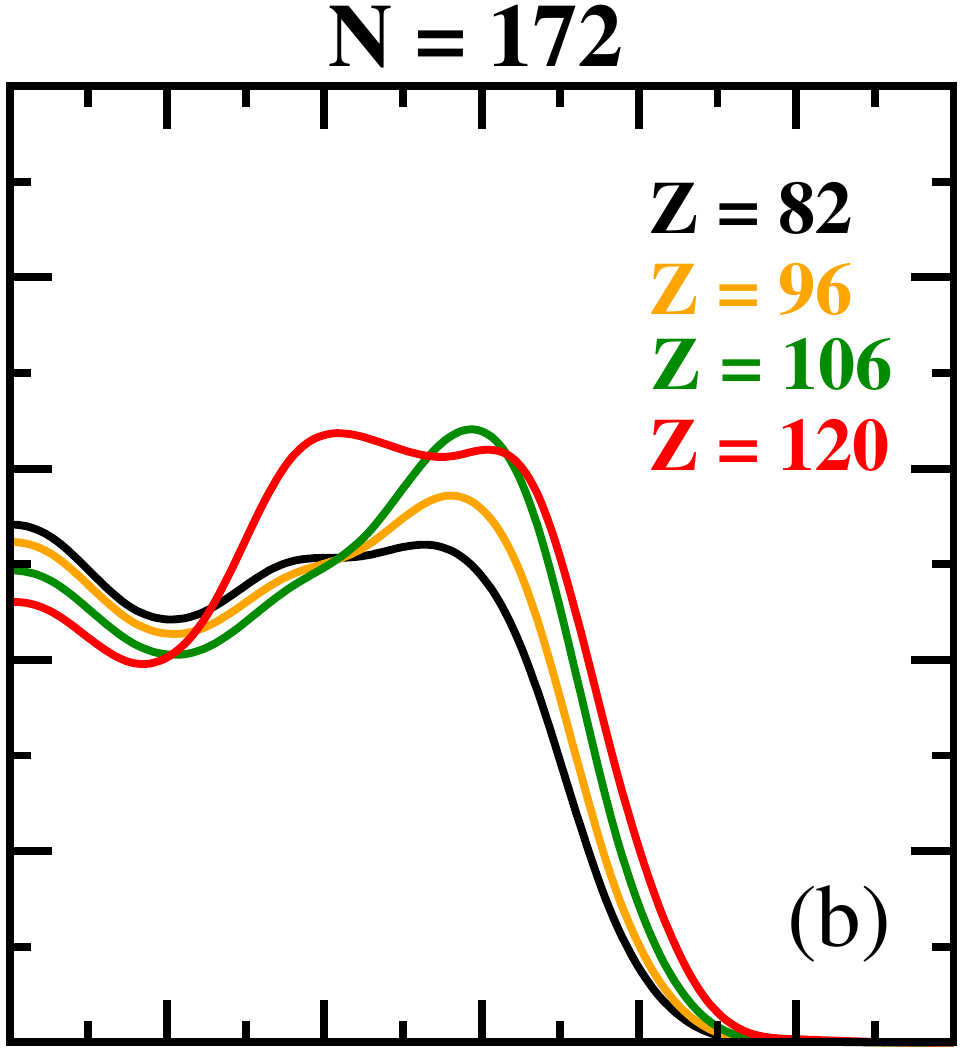}
\includegraphics[width=3.96cm]{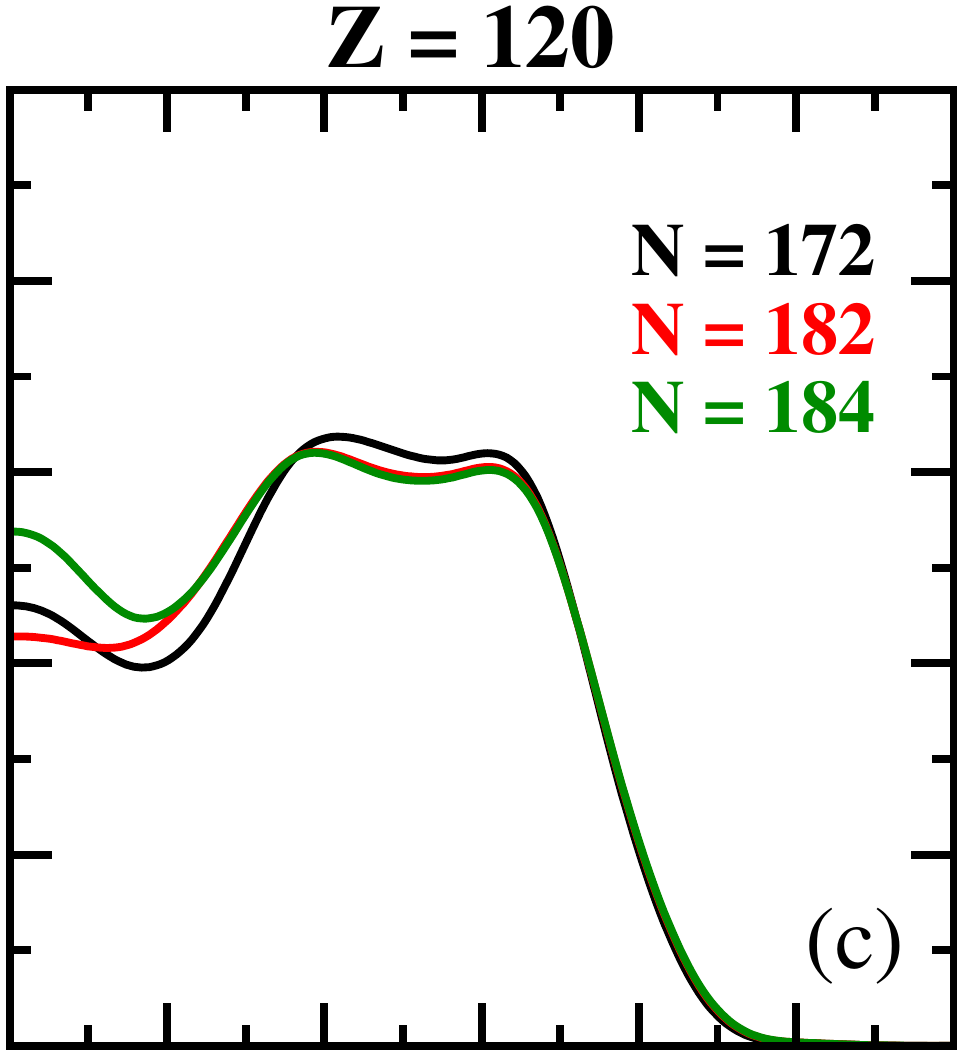}
\includegraphics[width=3.96cm]{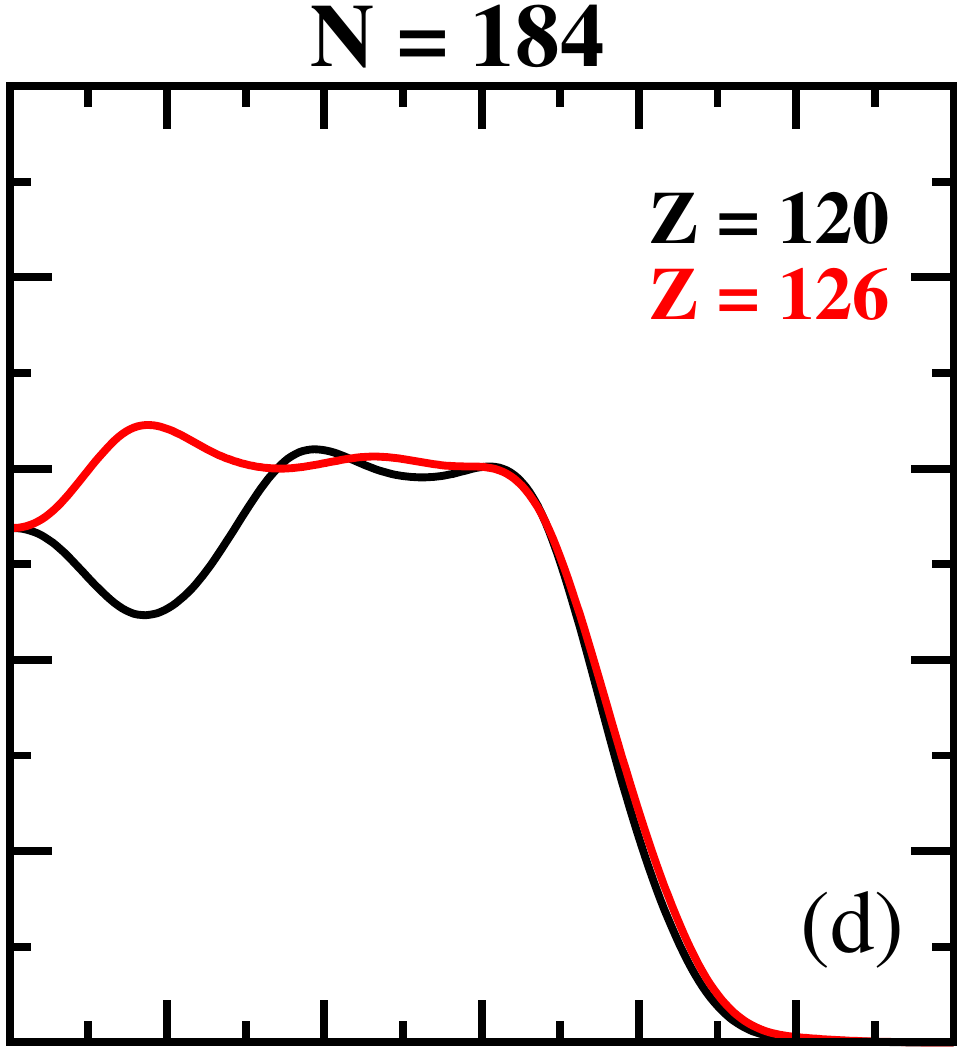}\\
%\hspace{0.38cm}\includegraphics[width=4.93cm]{Z_82_N_138_172_pro_VPS.eps}
\hspace{0.38cm}\includegraphics[width=4.93cm]{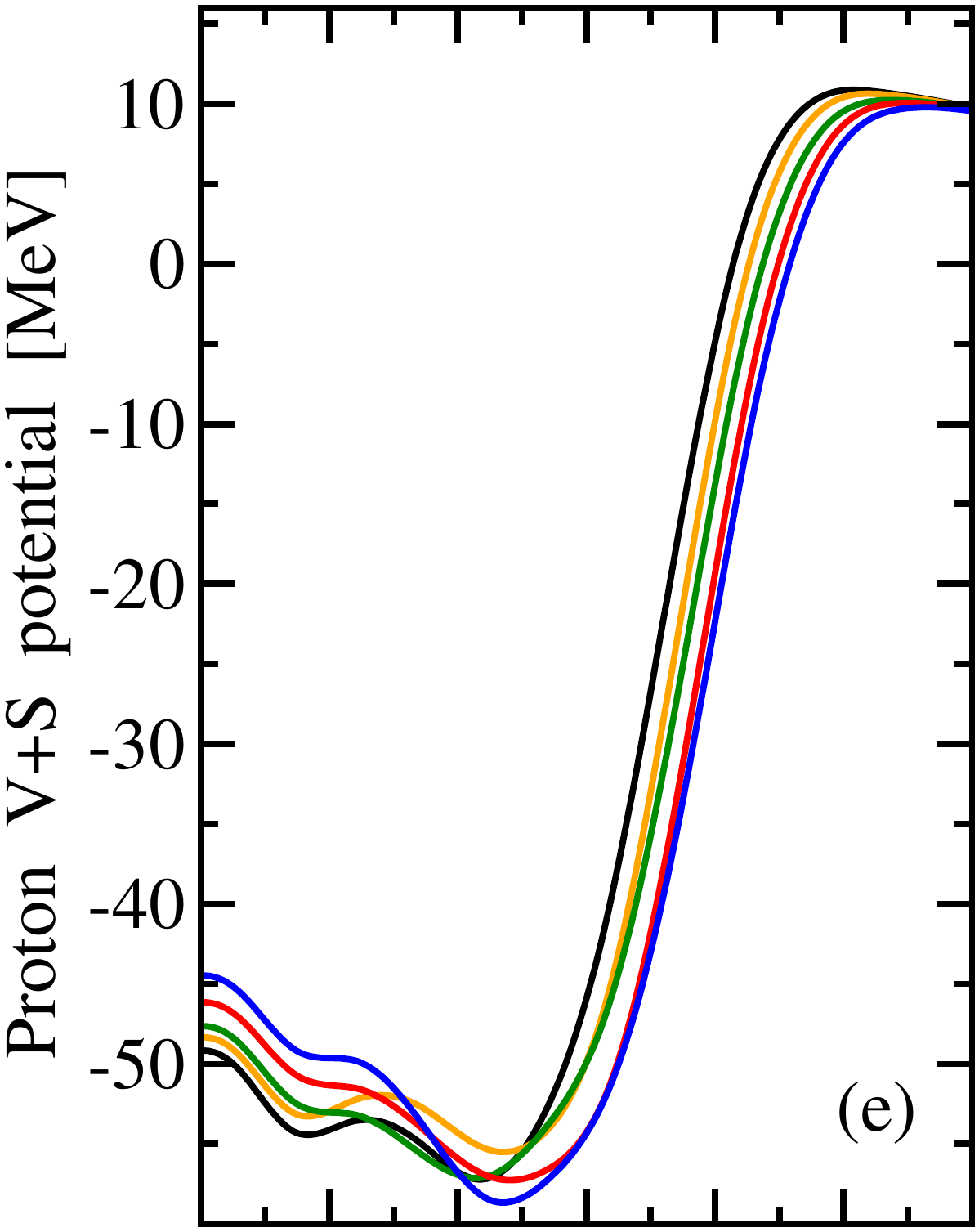}
\includegraphics[width=3.96cm]{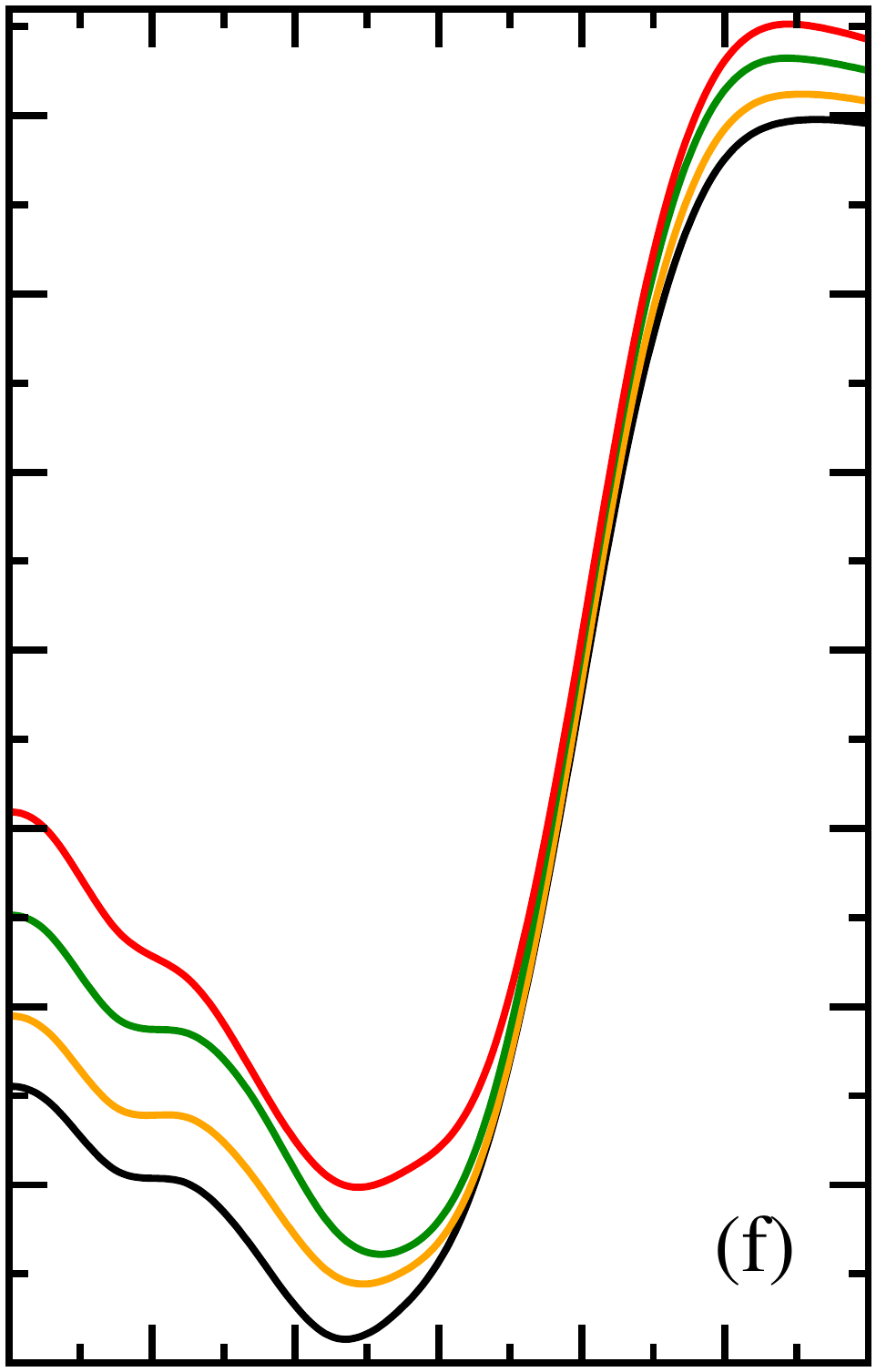}
\includegraphics[width=3.96cm]{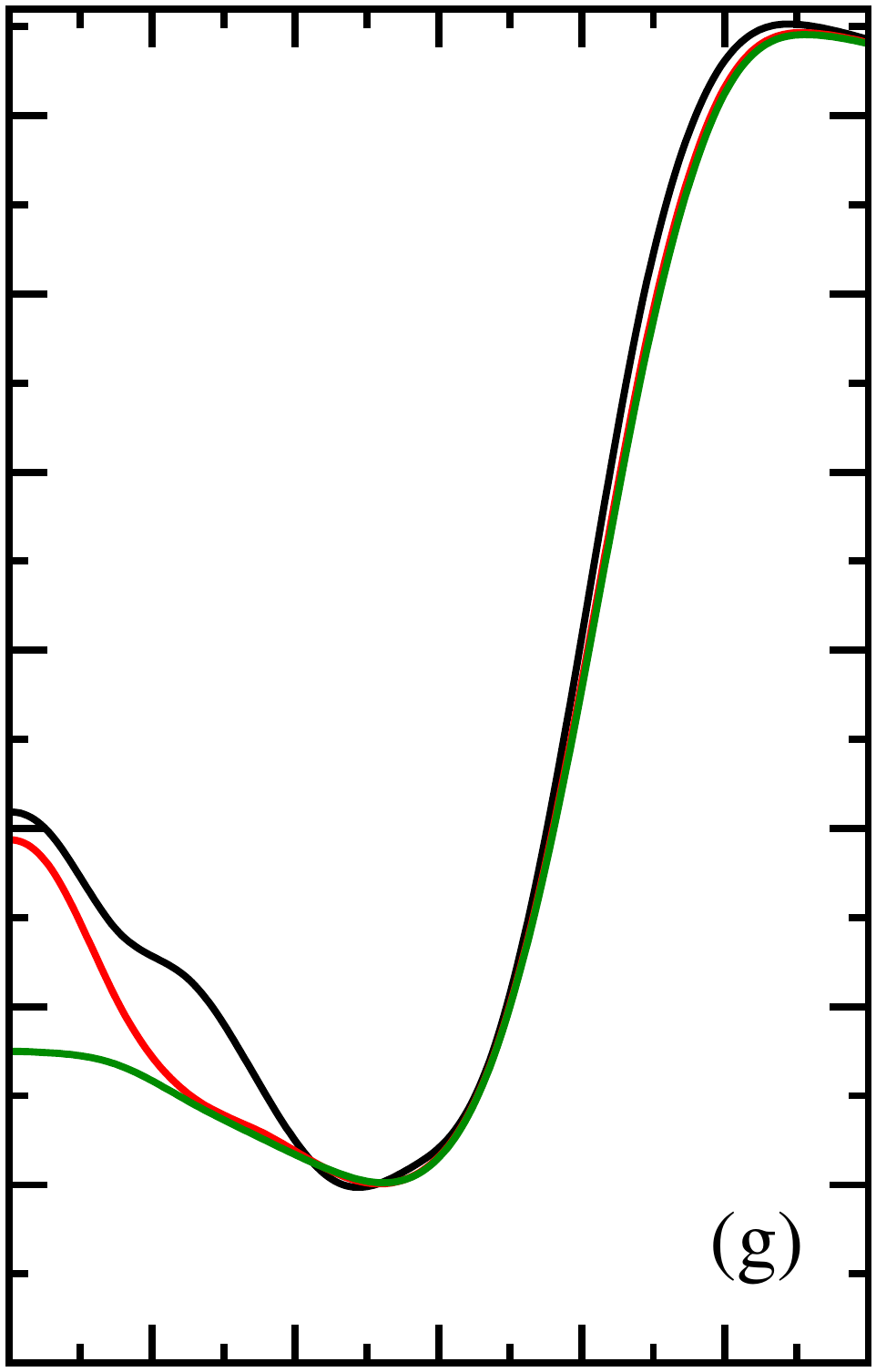}
\includegraphics[width=3.96cm]{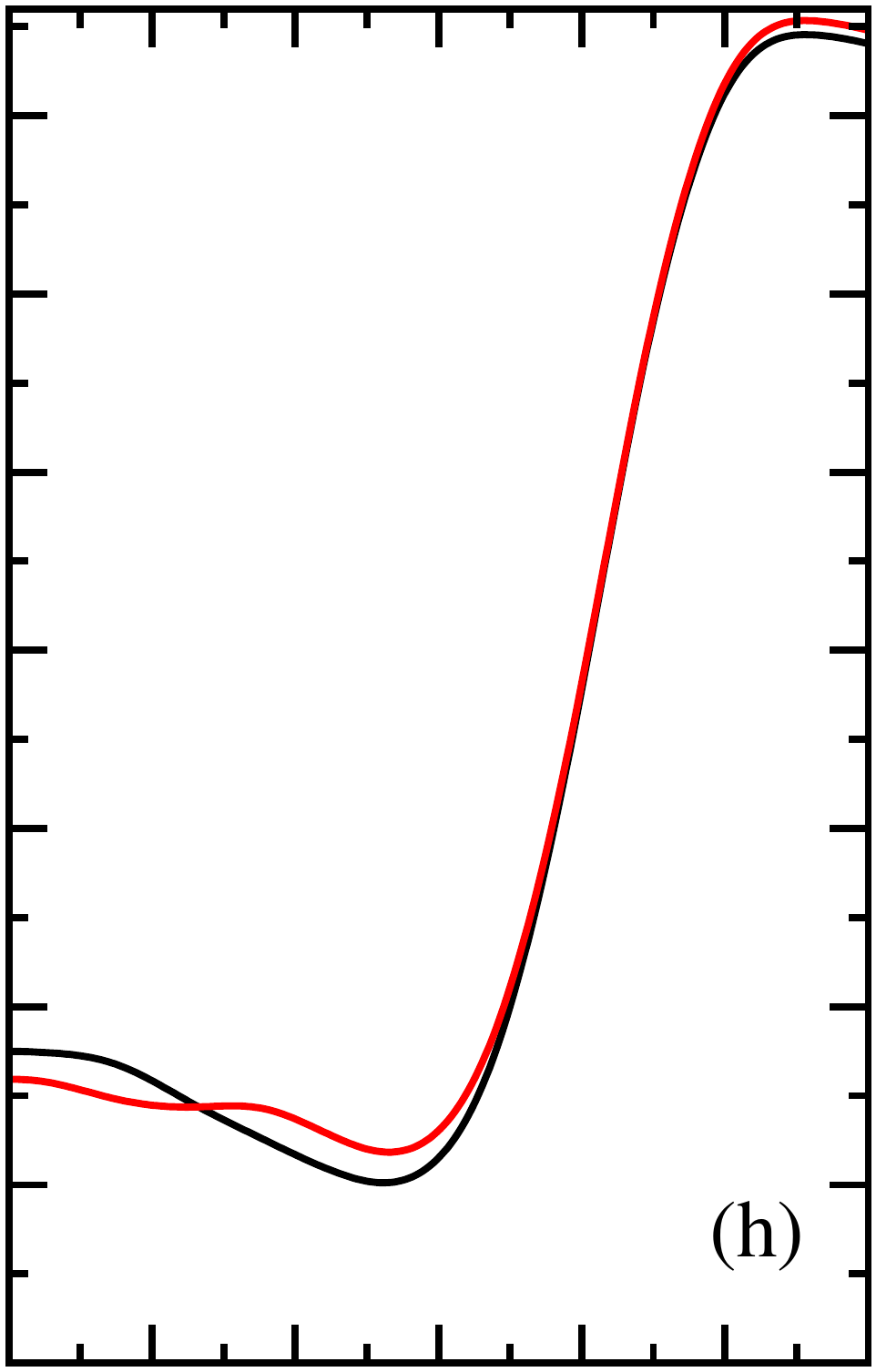}\\
%\hspace{0.38cm}\includegraphics[width=4.93cm]{Z_82_N_138_172_neu_VPS.eps}
\hspace{0.38cm}\includegraphics[width=4.93cm]{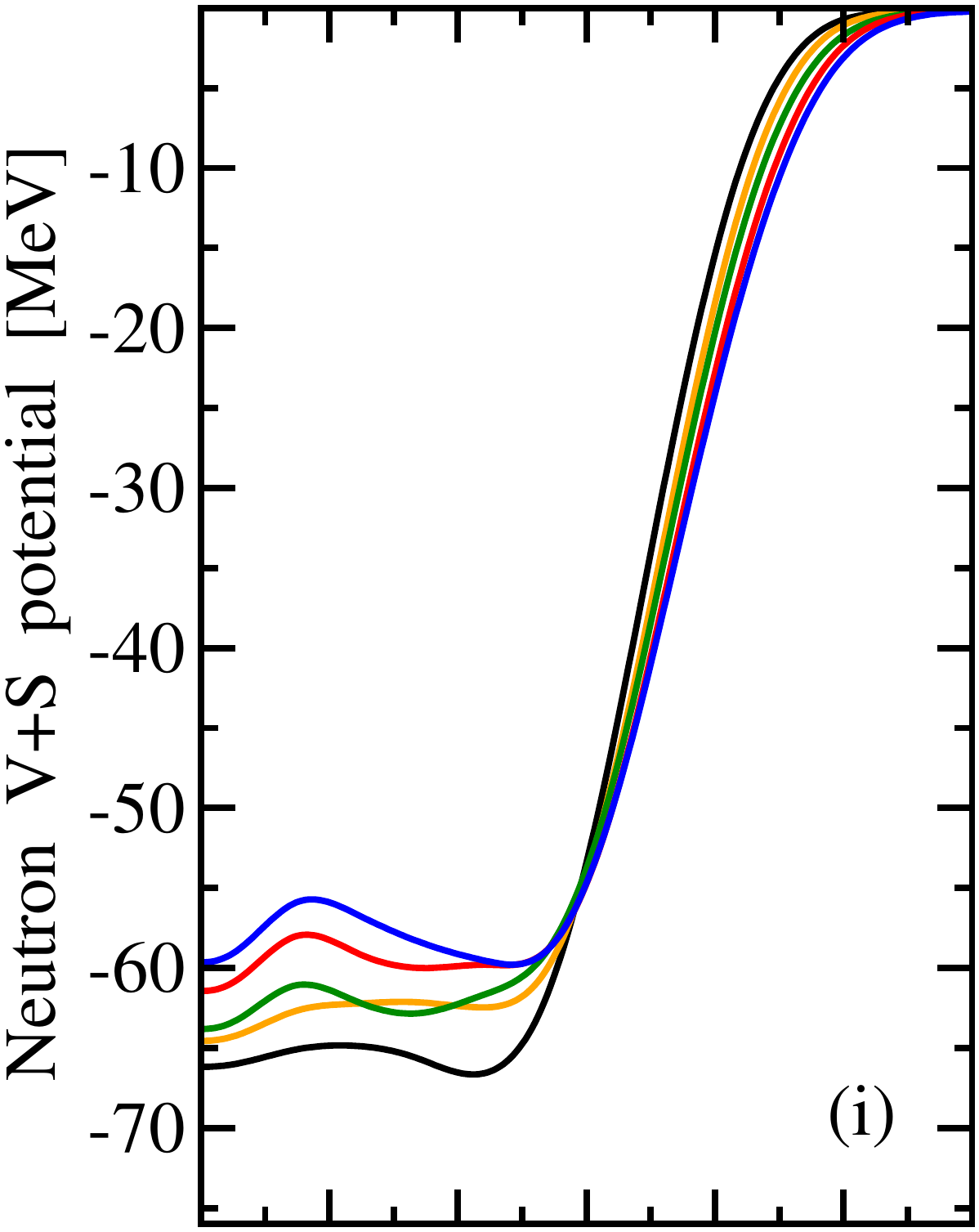}
\includegraphics[width=3.96cm]{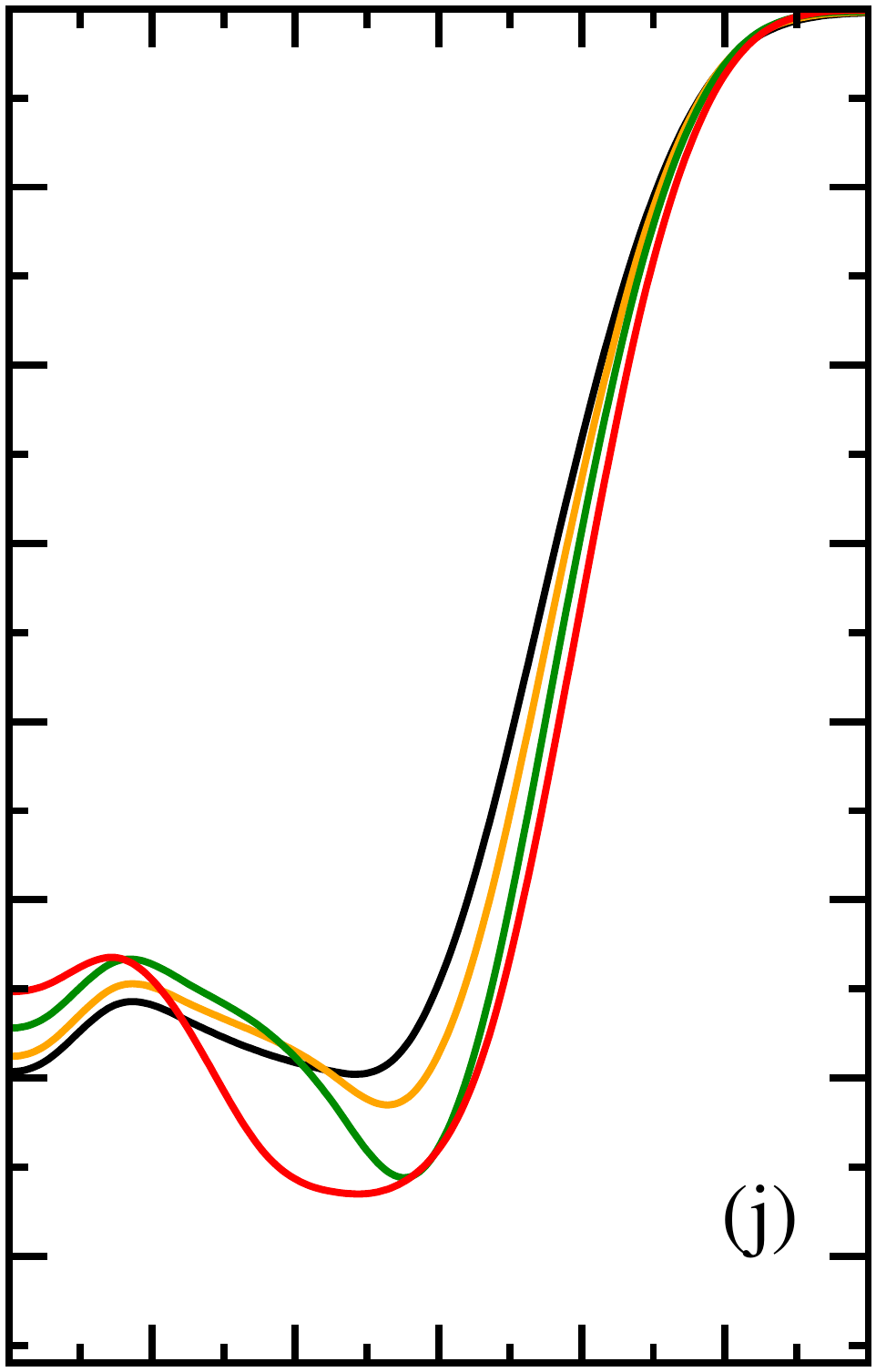}
\includegraphics[width=3.96cm]{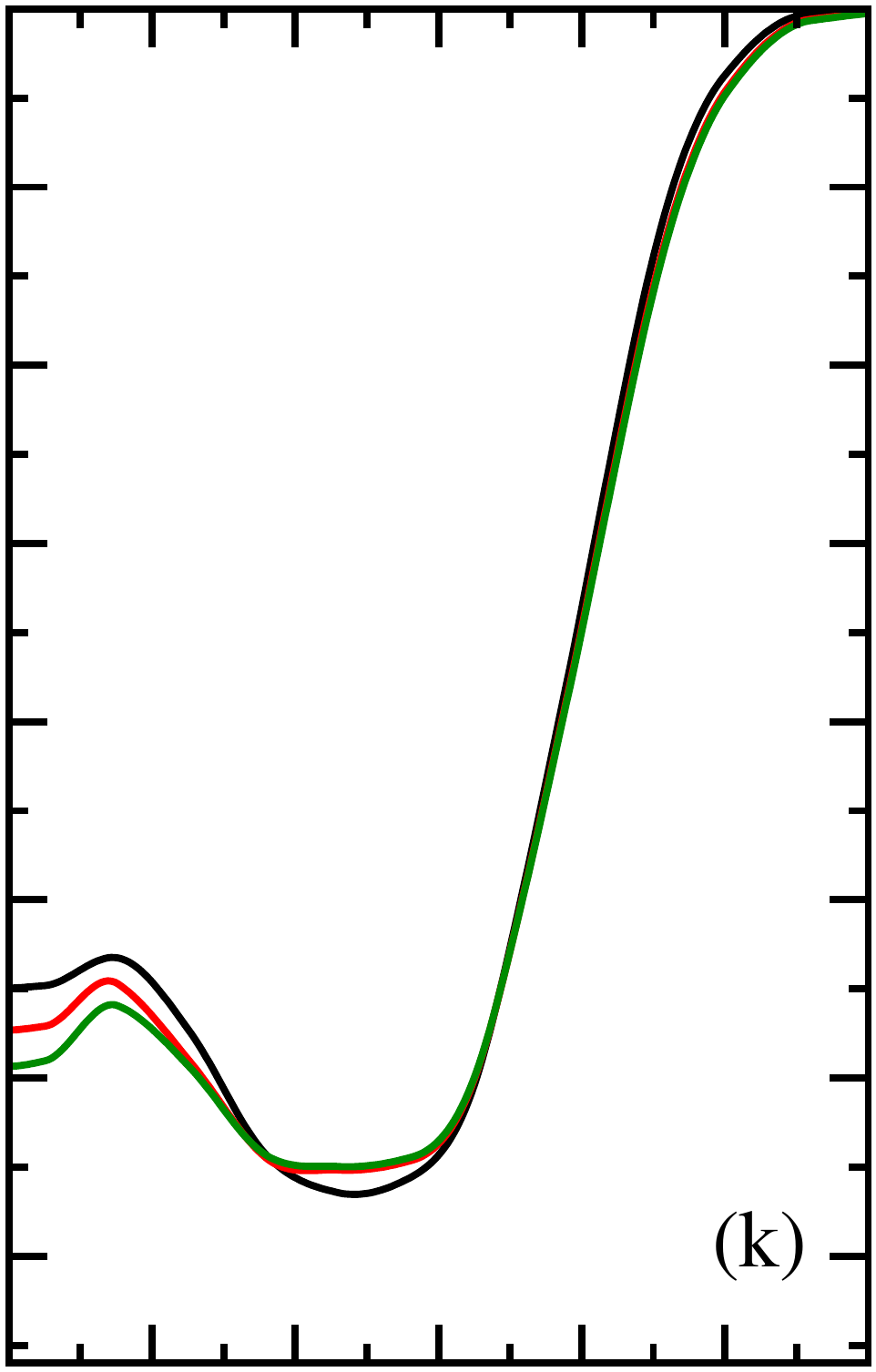}
\includegraphics[width=3.96cm]{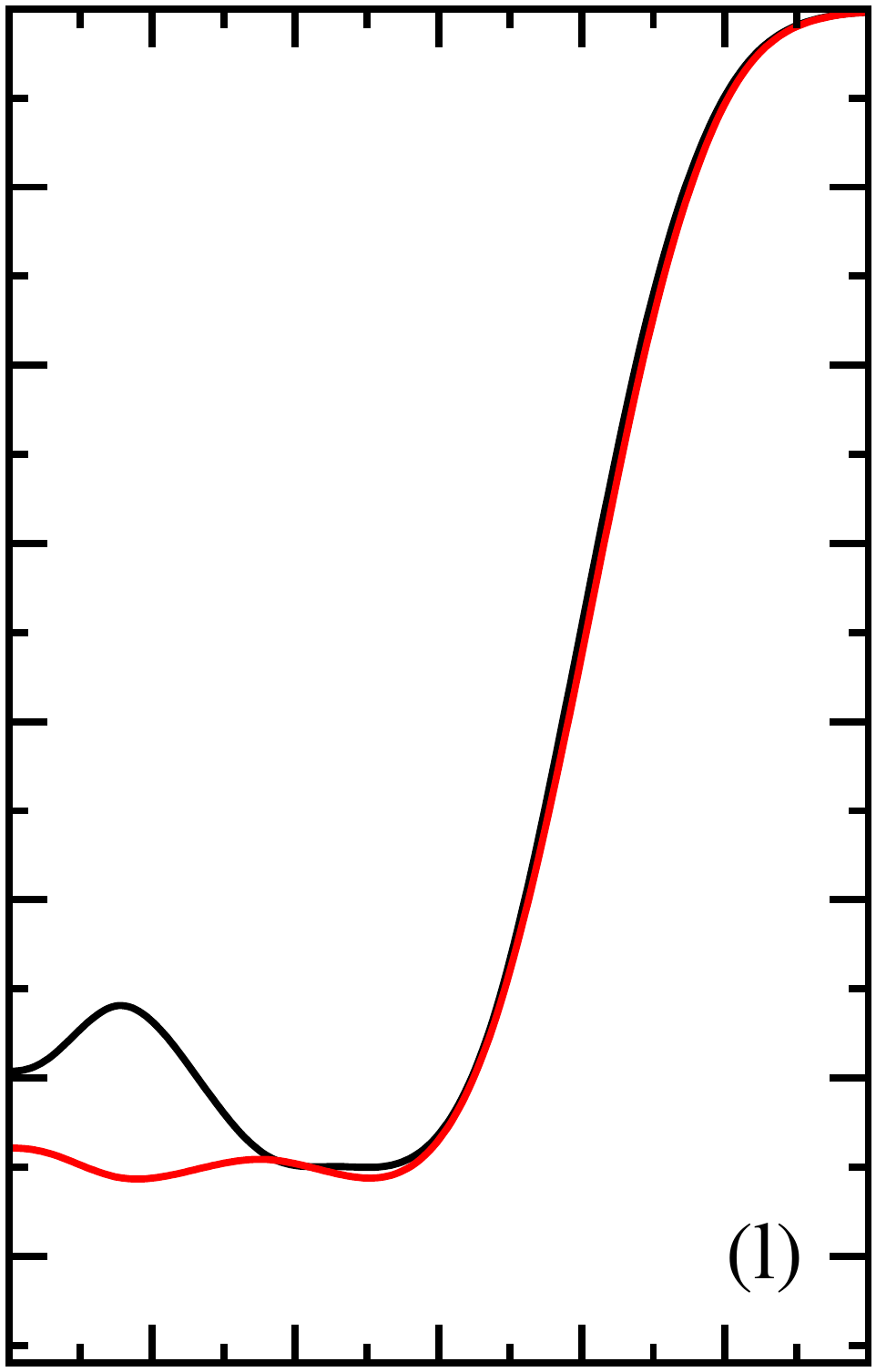}\\
\includegraphics[width=5.3cm]{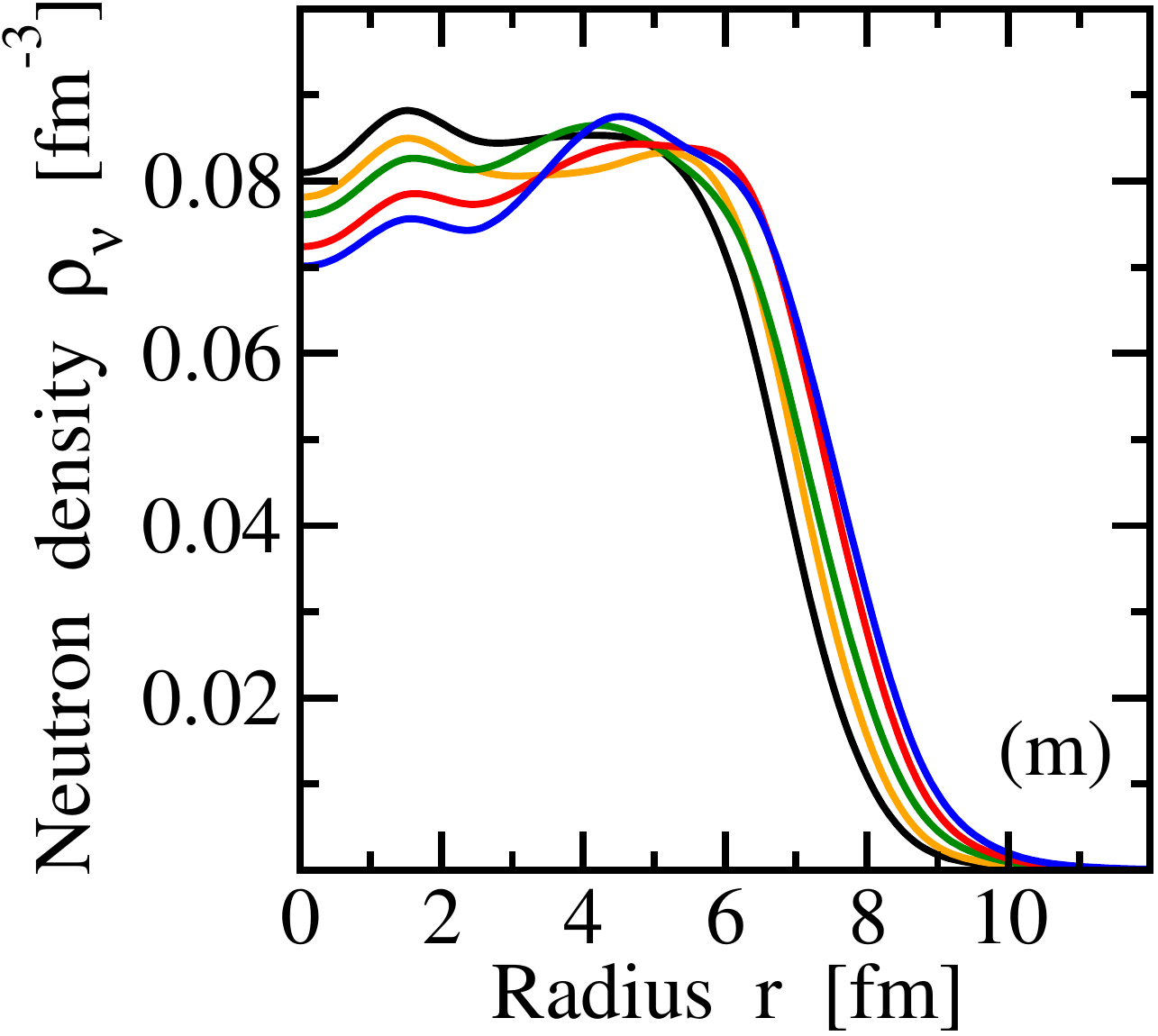}
\includegraphics[width=3.96cm]{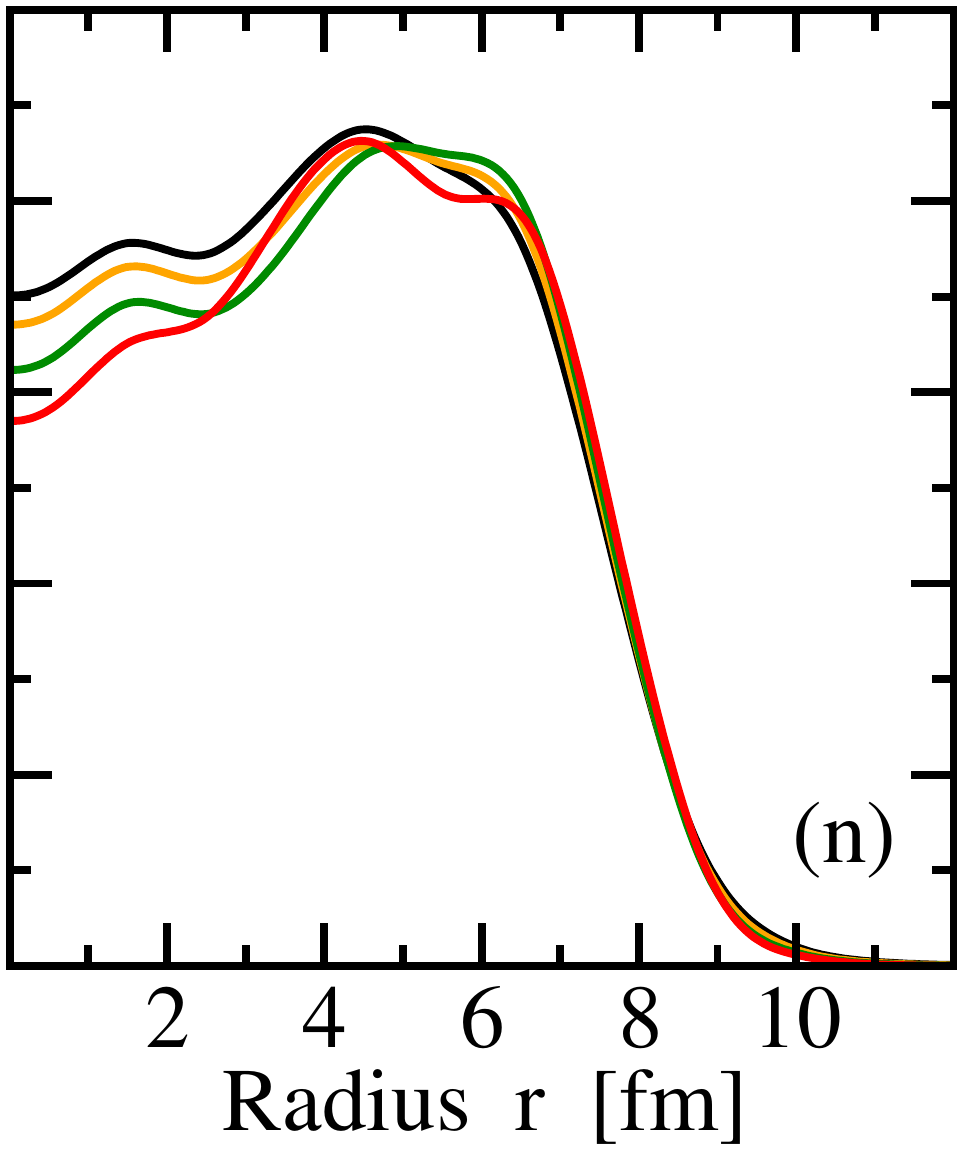}
\includegraphics[width=3.96cm]{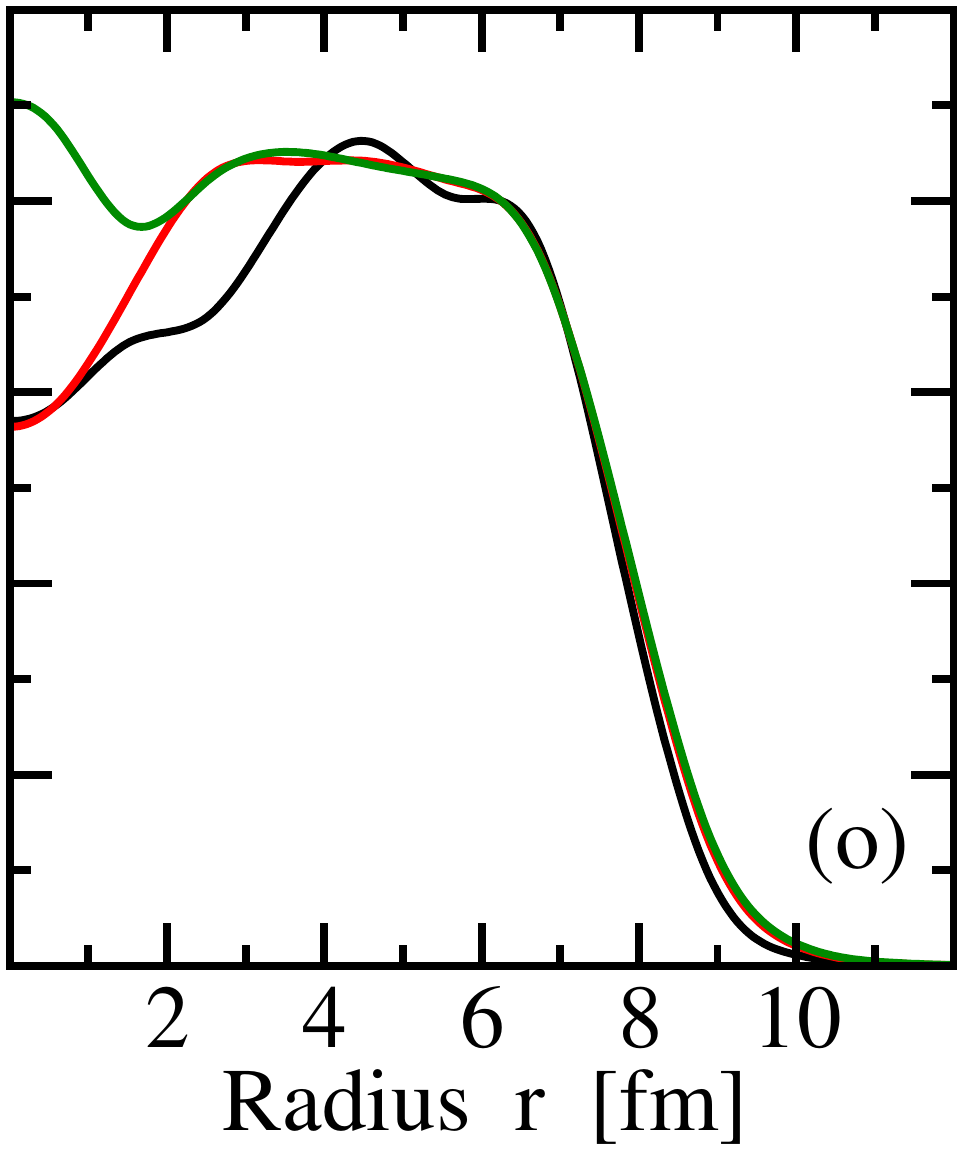}
\includegraphics[width=3.96cm]{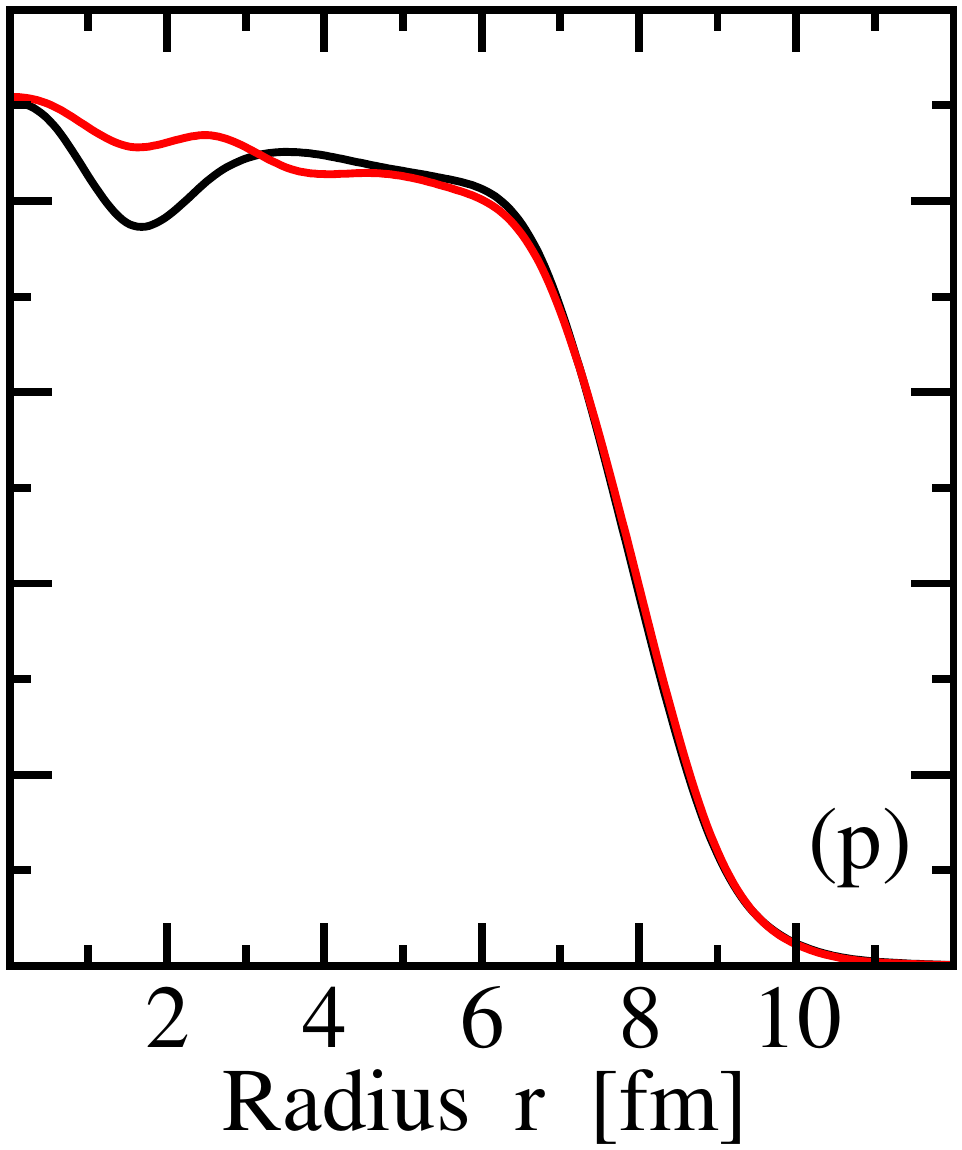}
\caption{The changes in proton (upper panels) and neutron (bottom panels) densities and
proton (second row of panels) and neutron (third row of panels) with increasing of proton 
and neutron numbers along the pathes discussed in the text. Note that in the columns the 
particle number (indicated above the top panel) either in proton or neutron subsystem is 
fixed. 
\label{density-potentials}
}
\end{figure*}
%%%%%%%%%%%%%%%%%%%%%%%%%%%%%%%%%%%%%%%%%%%%%%%%
\clearpage
%%%%%%%%%%%%%%%%%%%%%%%%%%%%%%%%%%%%%%%%%%%%%%%%
\begin{figure}[htb]
\centering
\includegraphics[width=8.4cm]{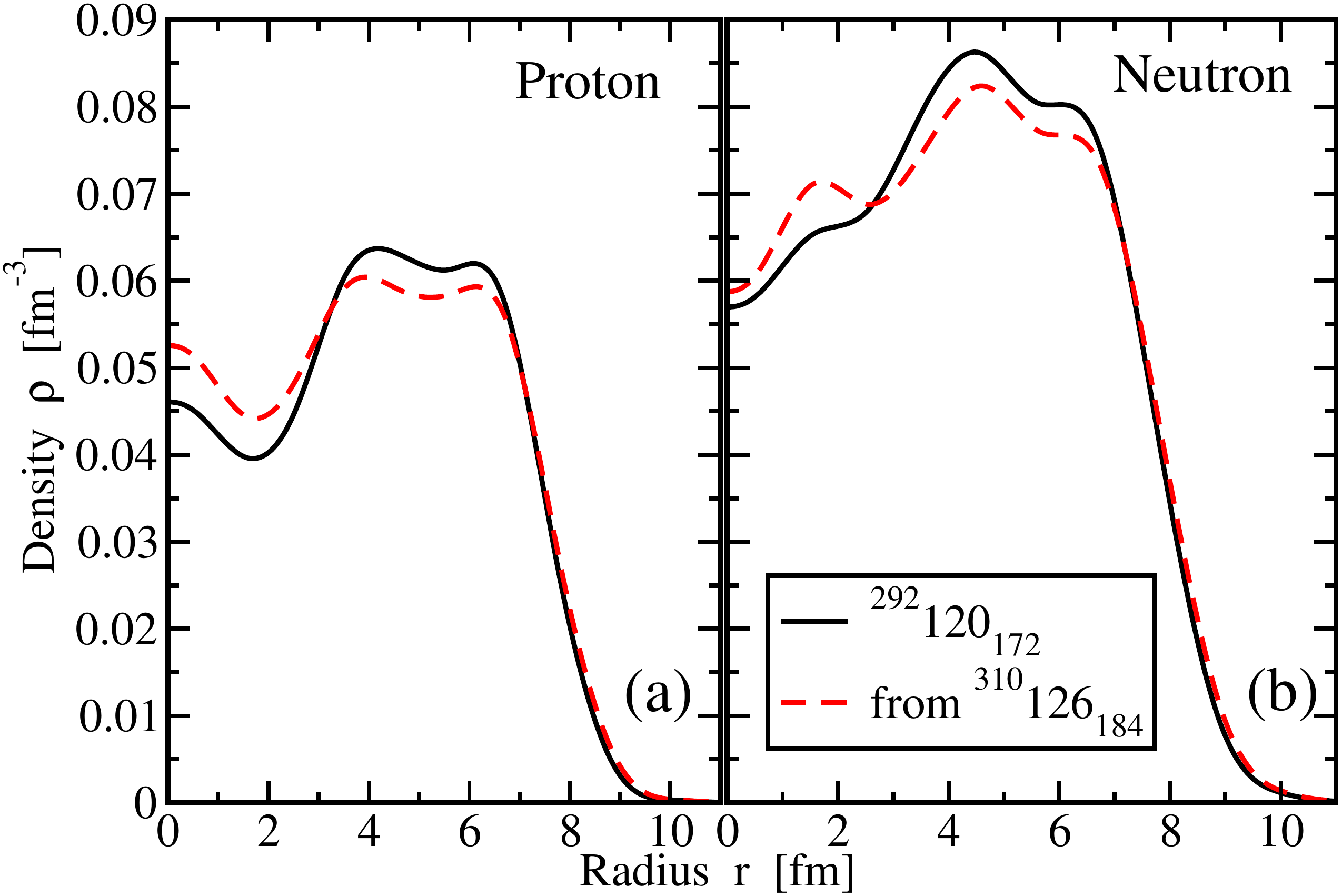}
\caption{The comparison of the densities in the $^{292}$120 nucleus obtained 
in a fully self-consistent way (black solid lines) with those derived by means
of the additivity rule (see Eq.\ (\ref{Eq:Additivity})) from the self-consistent densities 
of the $^{310}$126 nucleus (dashed red lines).
\label{BackCalc_120}
}
\end{figure}
%%%%%%%%%%%%%%%%%%%%%%%%%%%%%%%%%%%%%%%%%%%%%%%%

%%%%%%%%%%%%%%%%%%%%%%%%%%%%%%%%%%%%%%%%%%%%%%%%%%
\begin{figure}[htb]
\centering
\includegraphics[width=8.4cm]{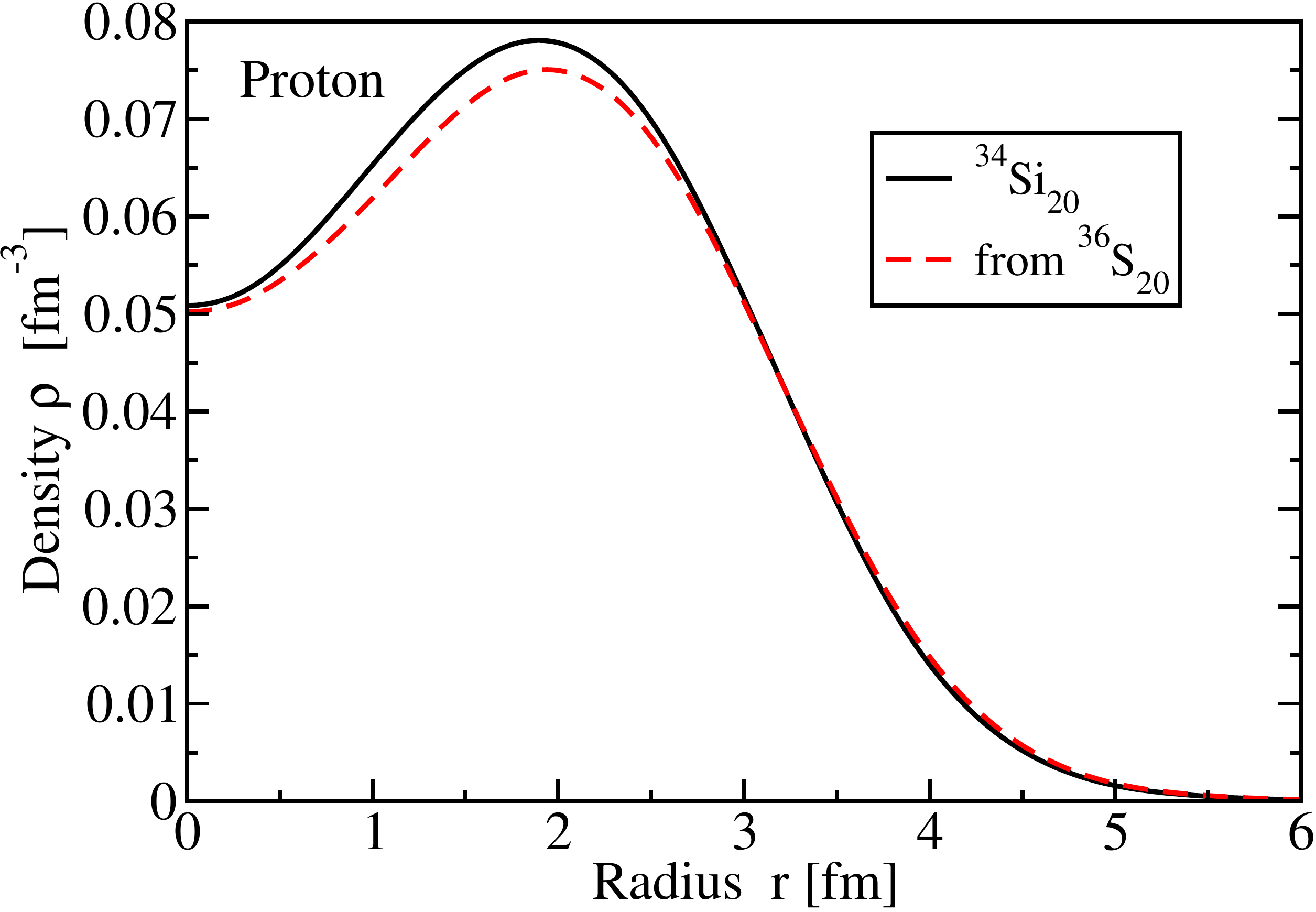}
\caption{The same as in Fig.\ \ref{BackCalc_120} but for the light nuclei. Black
line shows self-consistent proton densities of $^{34}$Si. Red dashed
line displays proton densities of this nucleus obtained by means
of the additivity rule from the self-consistent densities 
of the $^{36}$S nucleus.
\label{BackCalc_Si_Proton}
}
\end{figure}
%%%%%%%%%%%%%%%%%%%%%%%%%%%%%%%%%%%%%%%%%%%%%%%% 

   The additivity rule for densities is used here in order to verify whether the formation of 
the bubbles in the nuclei is predominantly due to single-particle degrees of freedom. This 
additivity rule is given as
\begin{eqnarray}
\rho_{tot}^B(r)  =    \rho_{tot}^A(r)  - \sum_{i} \rho_{sp-A}^i (r). 
\label{Eq:Additivity} 
\end{eqnarray}
Here, reference nucleus $A$ (either $^{310}$126 or $^{36}$S) is characterized by
flat bottom potentials while that in compared nucleus $B$ (either $^{292}$120 or 
$^{34}$Si) by wine bottle  potential(s). Single-particle density contributions 
$\rho_{sp-A}^i (r)$ of the single-particle states by which nuclei A and B differ
are defined in the reference nucleus $A$.

    The application of additivity rule is demonstrated in Figs.\ \ref{BackCalc_120}  
and \ref{BackCalc_Si_Proton}.  One can see that  starting from proton and neutron 
self-consistent densities  $\rho_{tot}^A(r)$ and respective single-particle densities
$\rho_{sp-A}^i (r)$ in the $^{310}$126 nucleus, the additivity rule reasonably well
predicts proton and neutron densities in the $^{292}$120 nucleus (see Fig.\ \ref{BackCalc_120}).  
The same is true for proton densities in the pair of the nuclei $^{34}$Si and $^{36}$S. 
Note that the level of the deviation of the densities obtained via additivity rule from self-consistent ones
in similar in both pairs of nuclei. This is due to the similarity of the relative change in total 
particle numbers between nuclei B and A in both pairs (5.9\% in the $^{34}$Si/$^{36}$S 
pair and 6.2\% in the $^{292}$120/$^{310}$126 pair).  Note that above discussed relative change 
in total particle number is comparable with the upper limit used in the analysis of additivity rule for 
relative quadrupole moments and effective alignments in Refs.\ \cite{SDDN.96,ALR.98,MADLN.07}.

  These results strongly point to the same mechanism of the formation of central depression of 
density distribution which is related to the single-particle degrees of freedom. In addition, in contrast 
to the results of Ref.\ \cite{SNR.17} they suggest that electrostatic repulsion does not play a dominant
role in the formation of bubble superheavy nuclei.

  Note that the addition or removal of particle(s) to the nucleonic configuration modifies via the 
polarization effects the total and single-particle radii (see Refs.\ \cite{RF.95,PAR.21}). For example, 
subsequent addition of neutrons leads to an increase of total charge radii \cite{RF.95,Pb-Hg-charge-radii-PRL.21,PAR.21} and 
proton single-particle radii \cite{APR.21} in the Pb isotopic chain. These polarization effects are 
minimized in the considered pairs of the nuclei. This is because the rms radii of proton/neutron 
matter distributions are very similar in compared nuclei (see Table \ref{Table-rms-radii}). It is 
reasonable to expect that with the increase of the difference of these radii in the pairs of nuclei under 
comparison the accuracy of the additivity principle for the single-particle densities will somewhat 
decrease. This is because the polarization effects will lead to a larger difference of the rms radii of 
proton and neutron singe-particle states in compared nuclei.

%%%%%%%%%%%%%%%%%%%%%%%%%%
\section{General observations}
\label{Gen-observations}
%%%%%%%%%%%%%%%%%%%%%%%%%%

\begin{figure*}[htb]
\centering
\includegraphics[width=6.4cm]{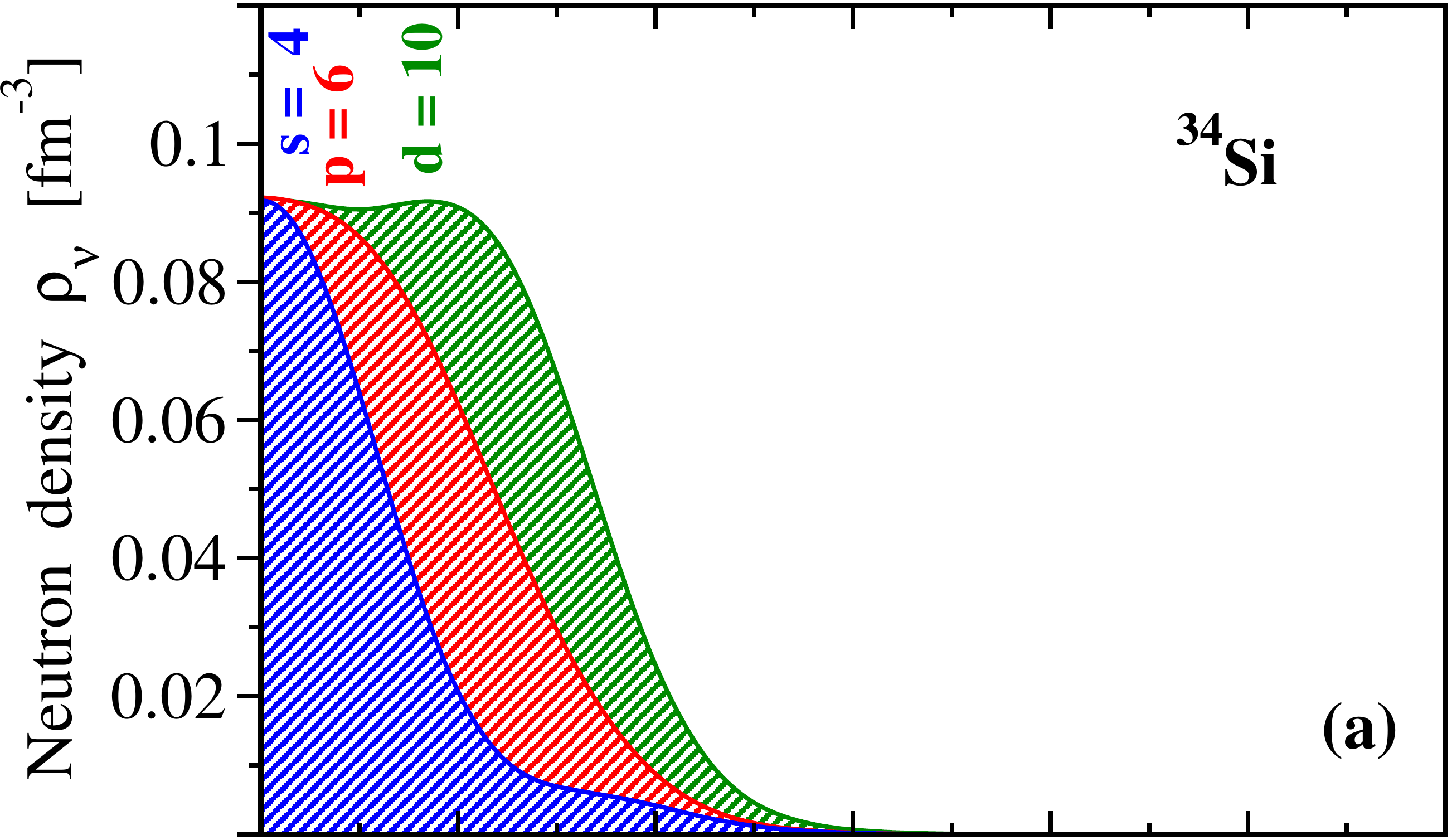}
\includegraphics[width=5.2cm]{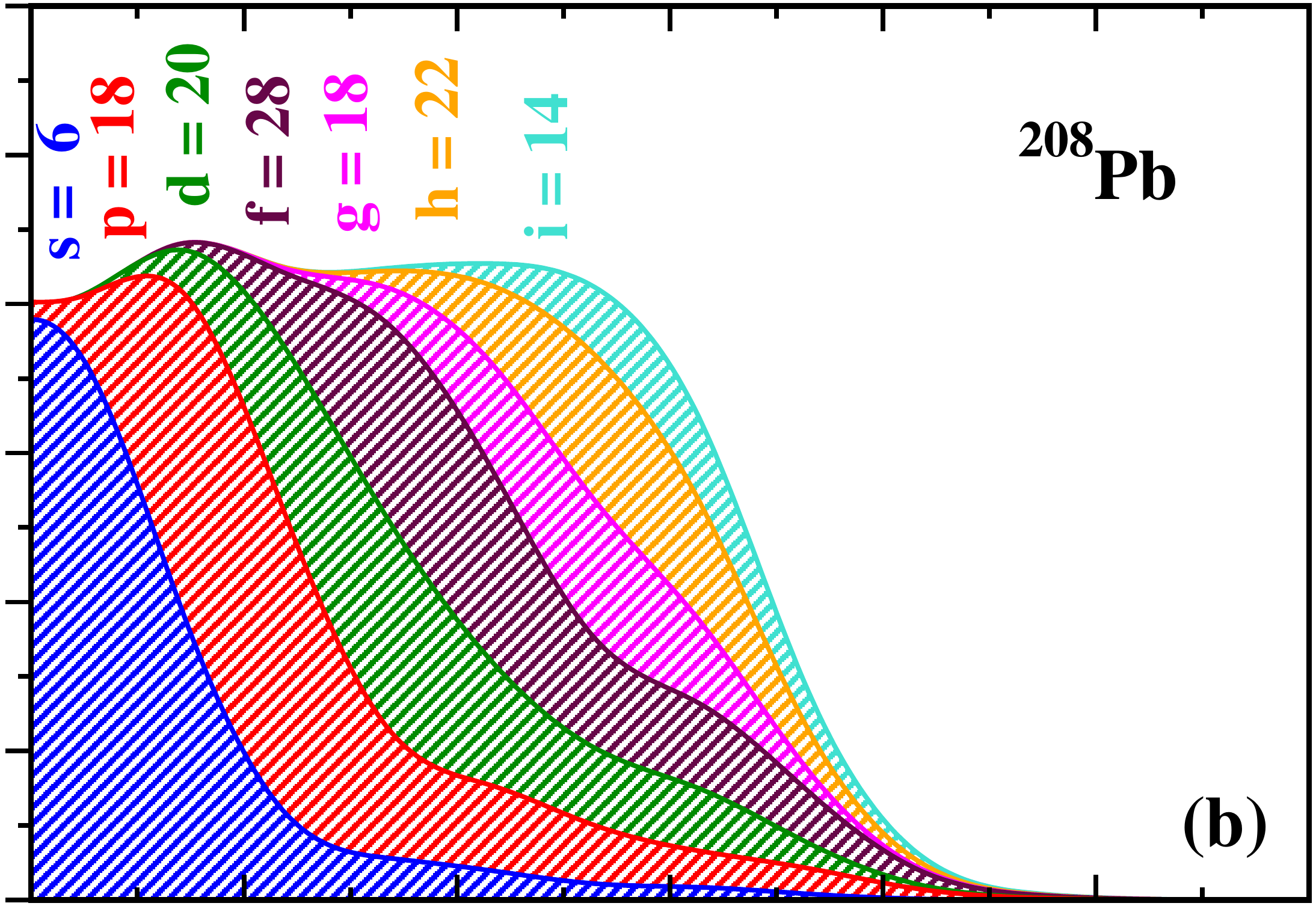}
\includegraphics[width=5.2cm]{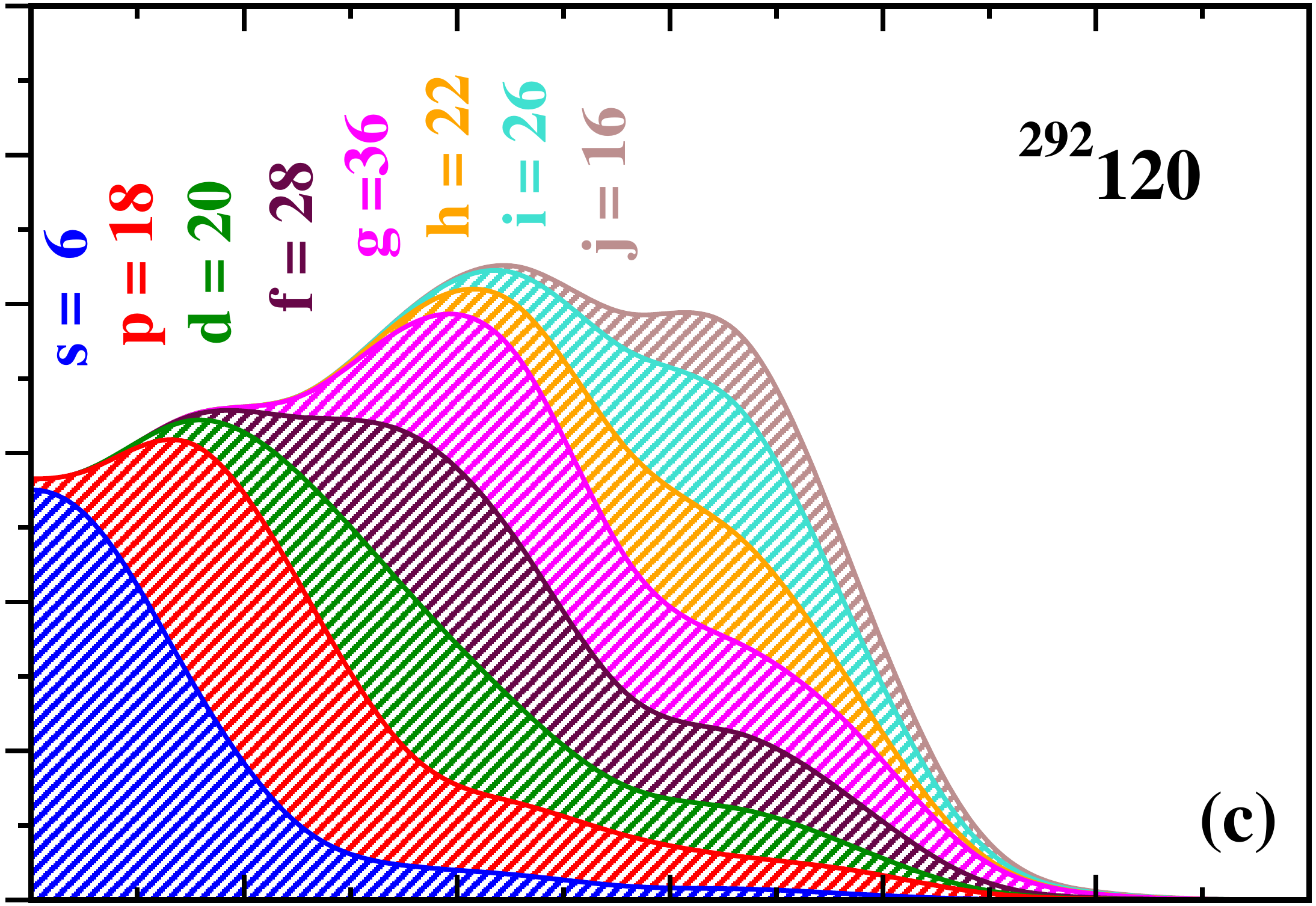}
\includegraphics[width=6.4cm]{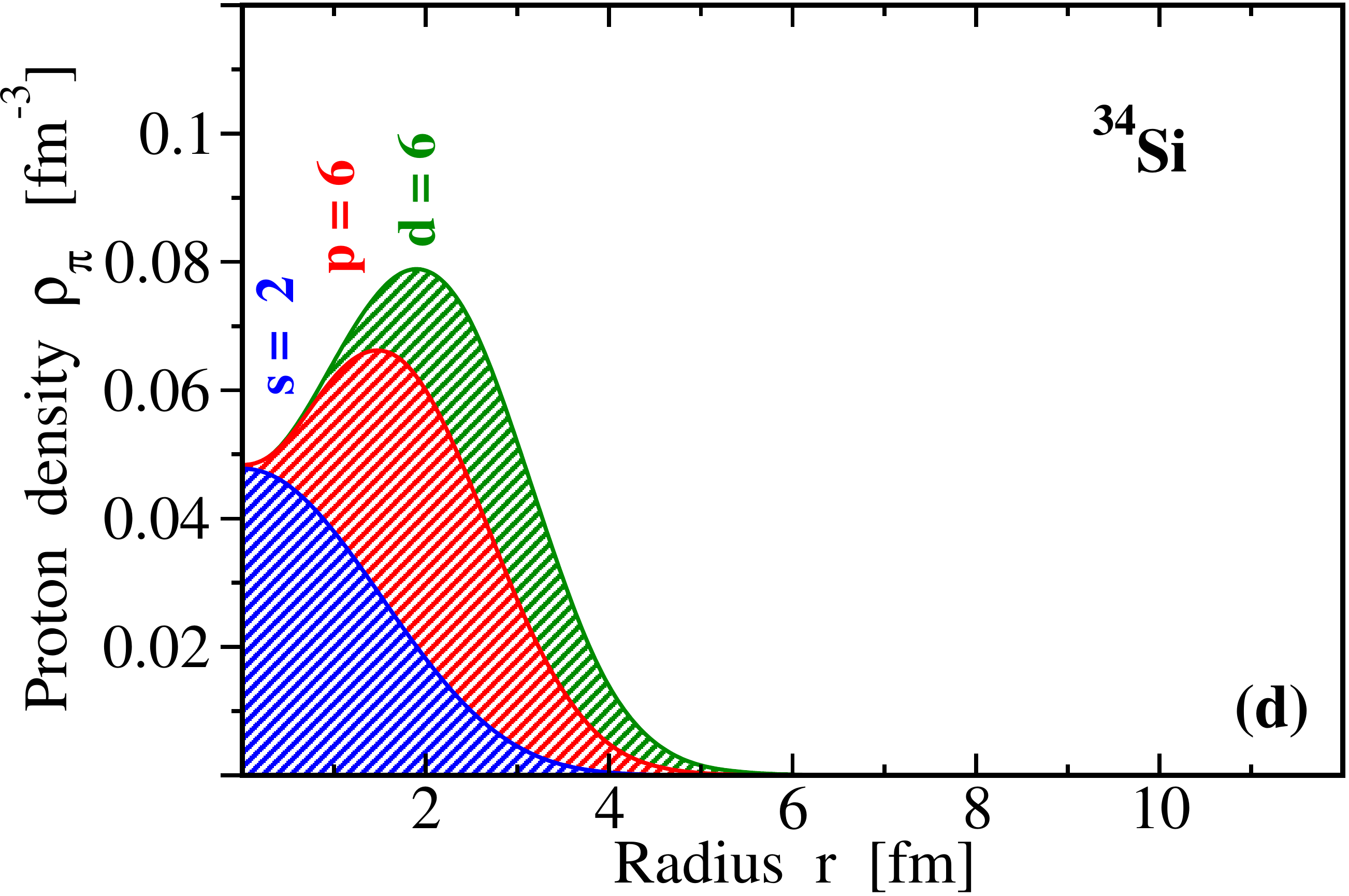}
\includegraphics[width=5.2cm]{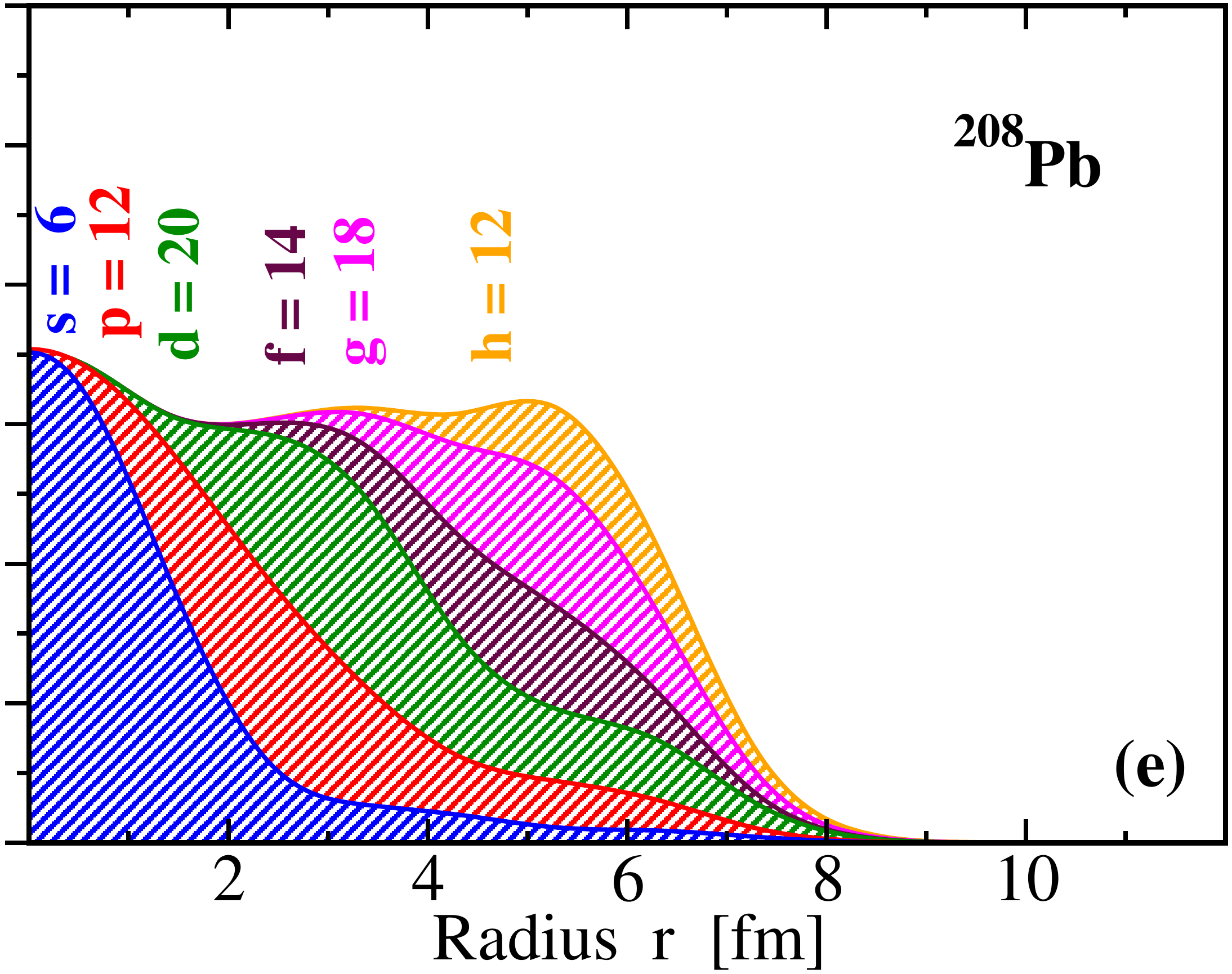}
\includegraphics[width=5.2cm]{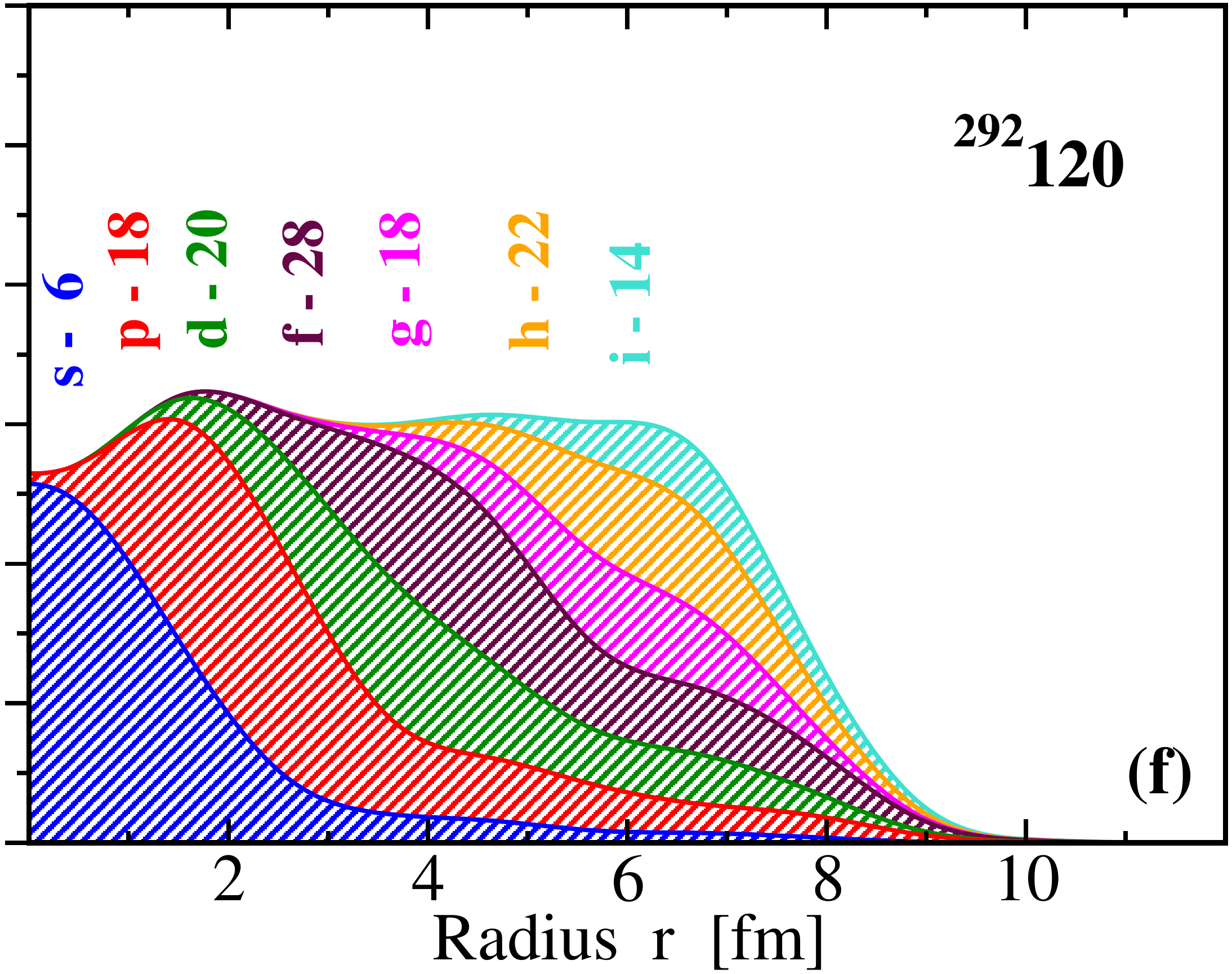}
\includegraphics[width=6.4cm]{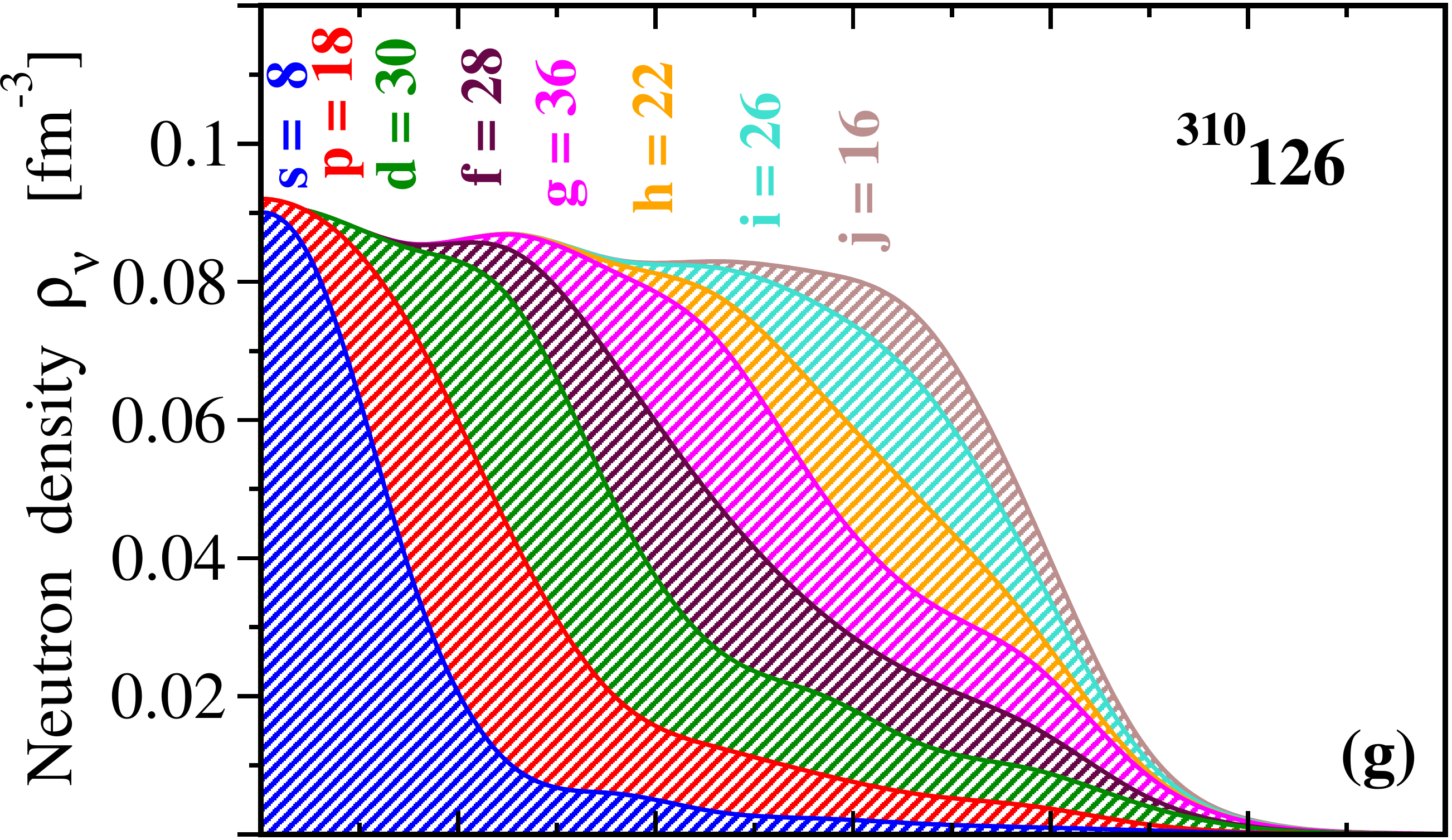}
\includegraphics[width=5.2cm]{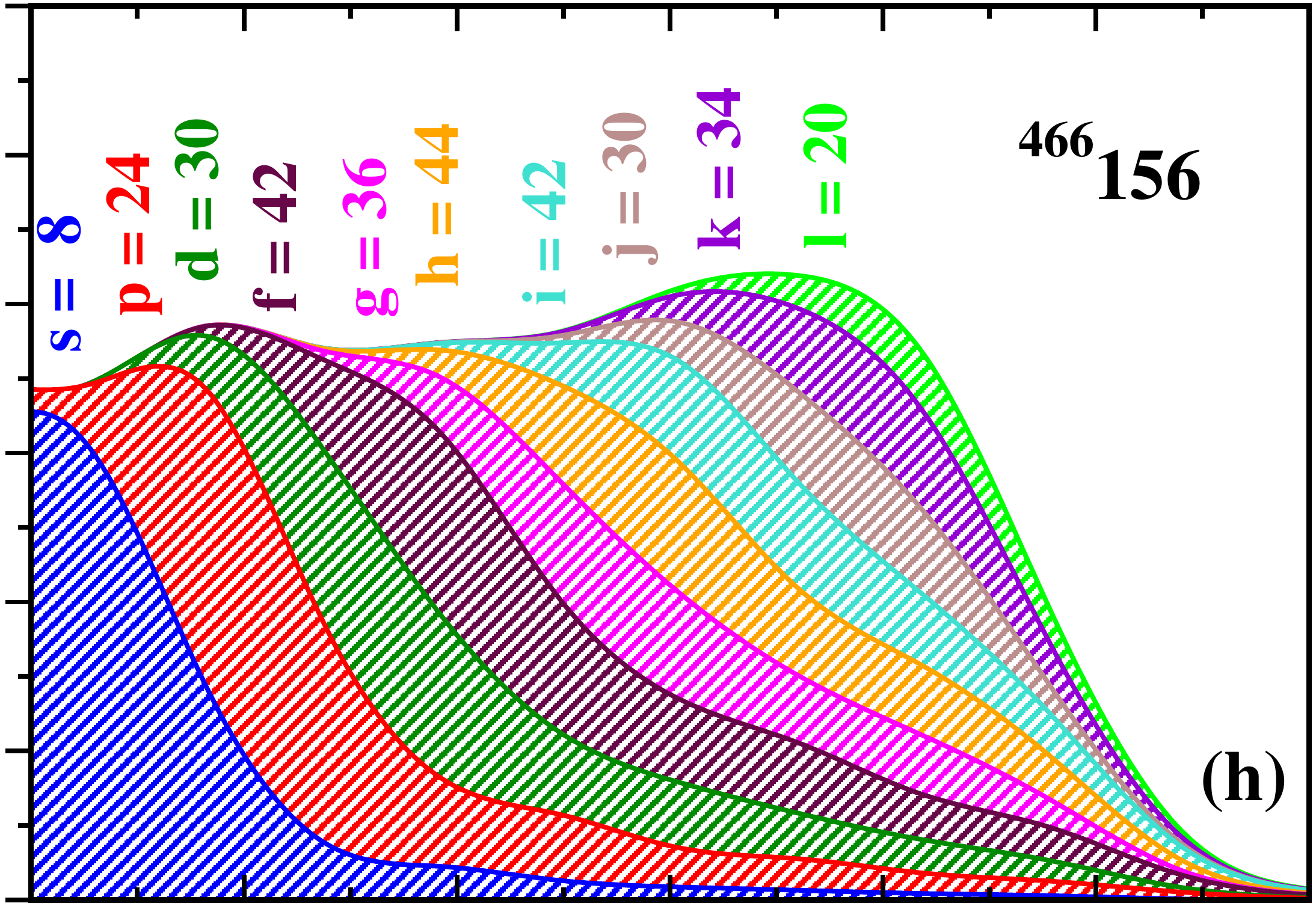}
\includegraphics[width=5.2cm]{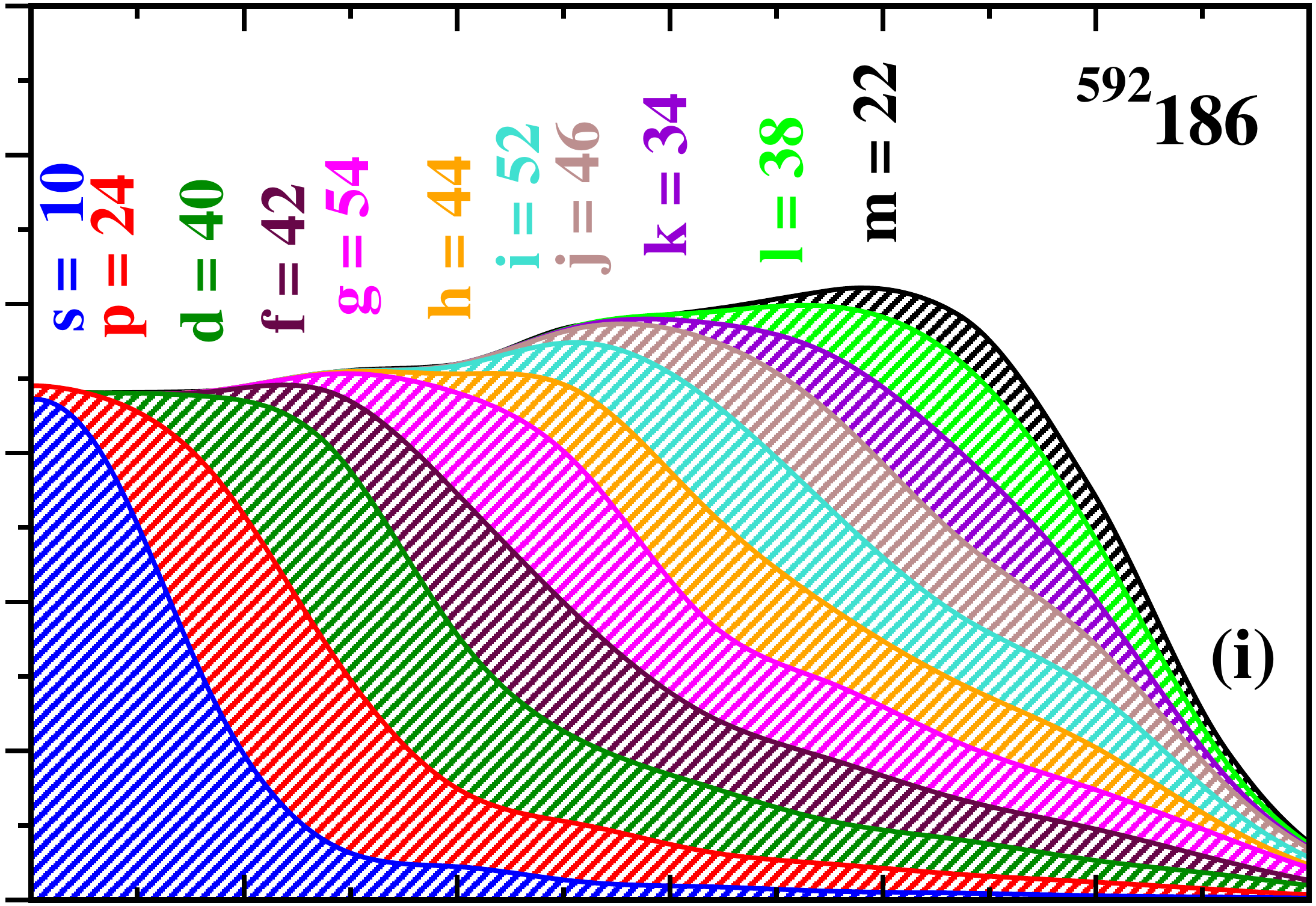}
\includegraphics[width=6.4cm]{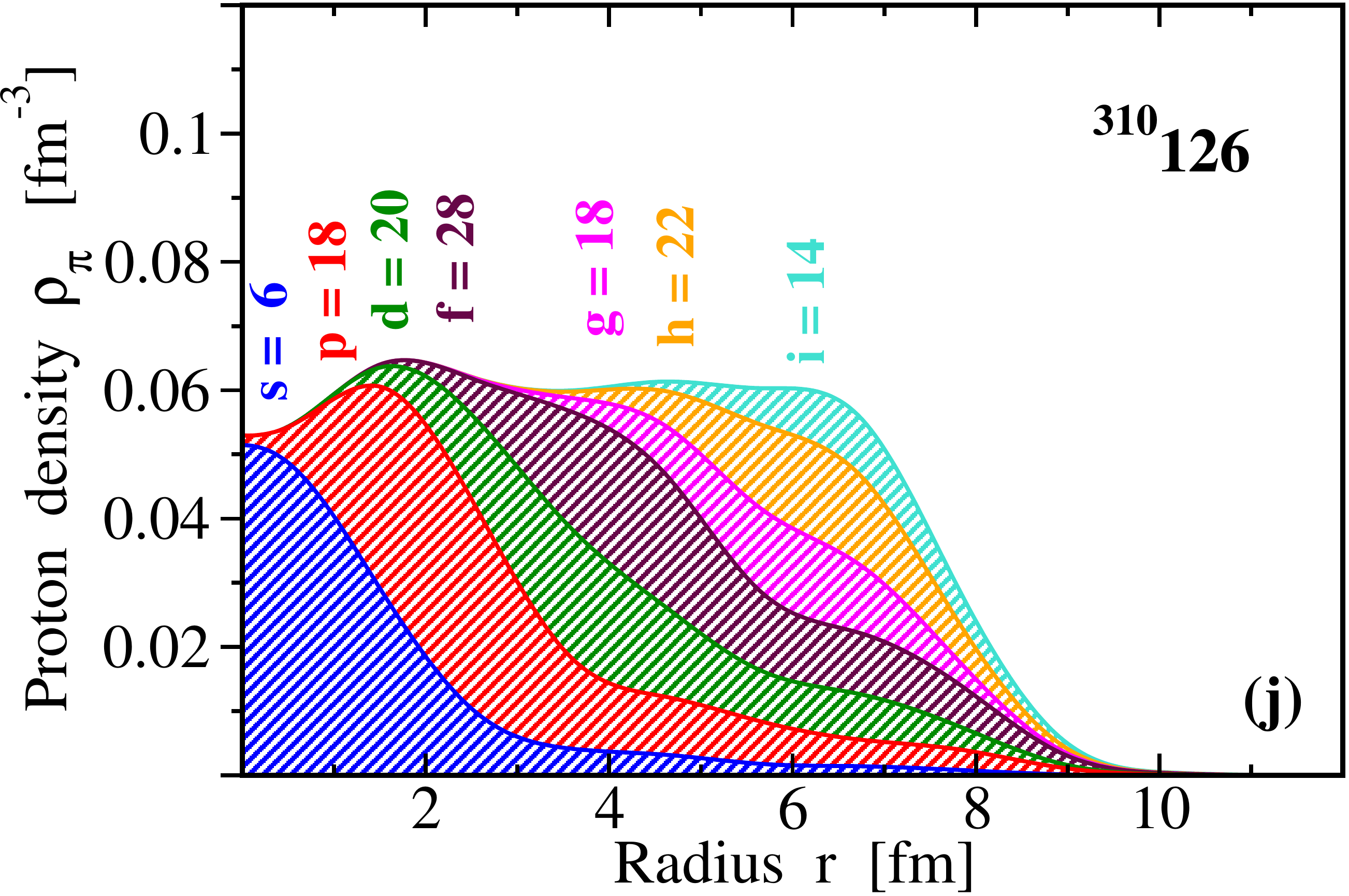} 
\includegraphics[width=5.2cm]{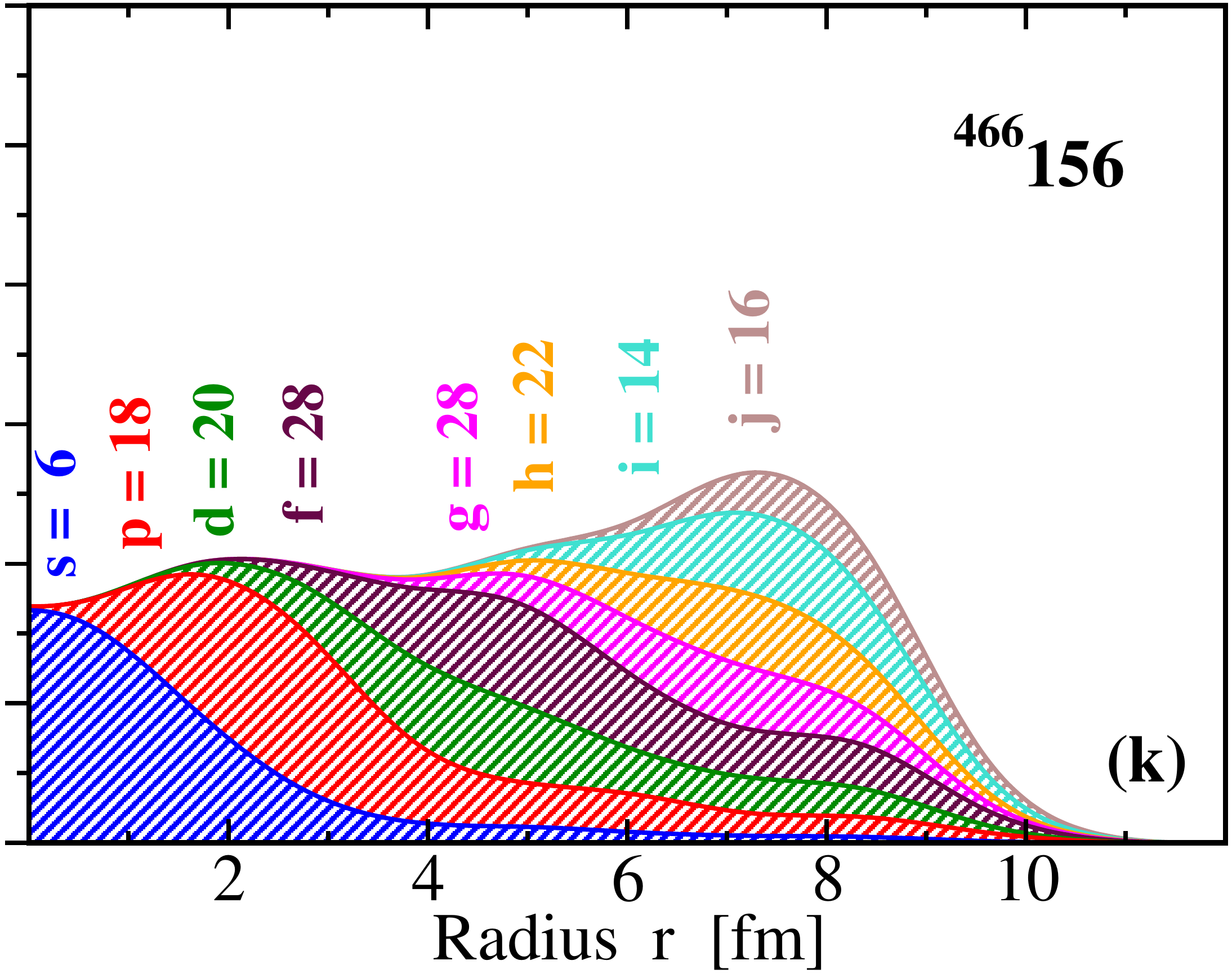}
\includegraphics[width=5.2cm]{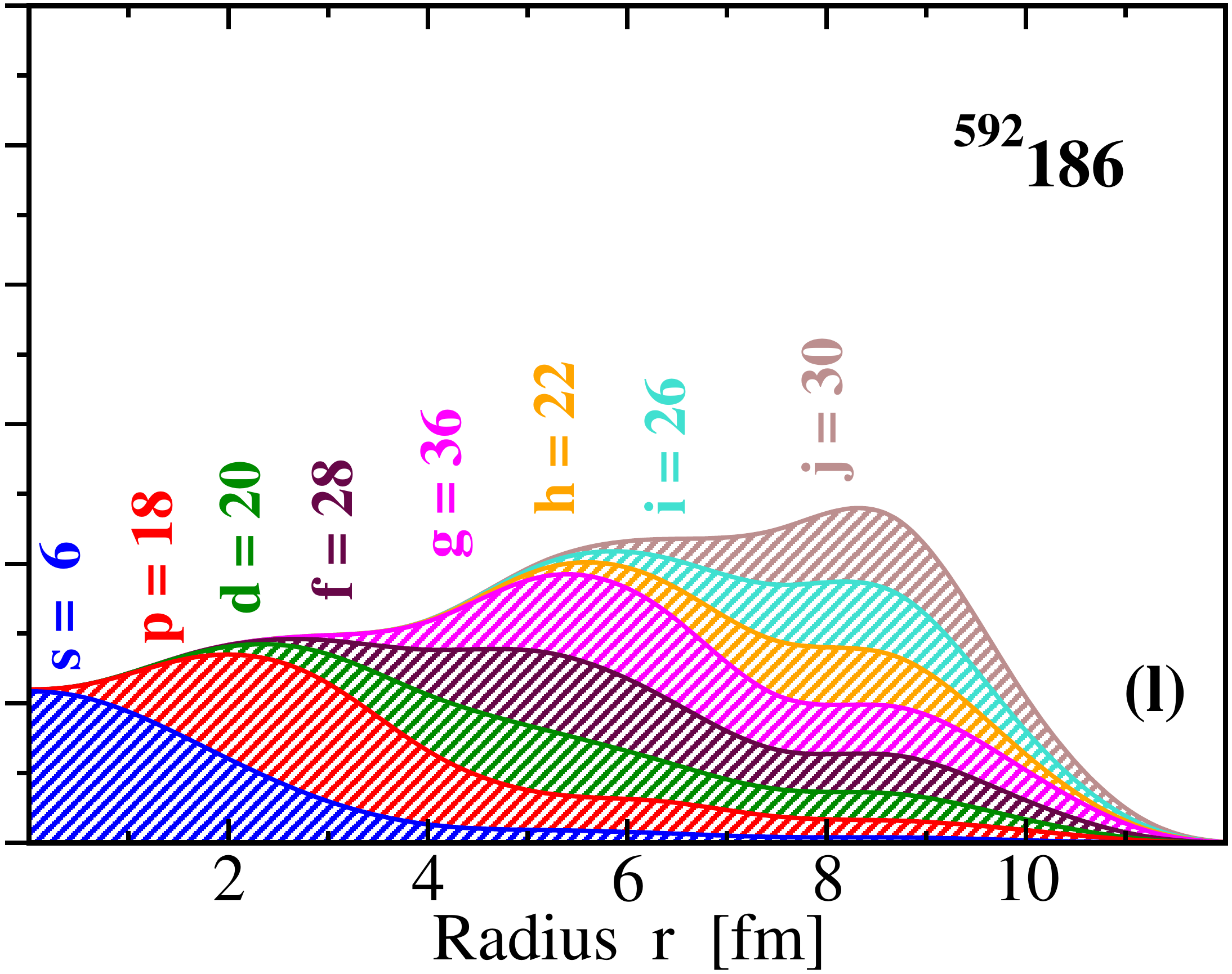}
\caption{The buildup of total density from the contributions of spherical subshells 
with given orbital angular momentum $l$. These contributions are shown by the shaded 
areas of different color and are indicated by the labels of the same color with the format 
"$l$=$num$". In these labels $l$ stands for the orbital angular momentum ($s$ for $l=0$, 
$p$ for $l=1$ and so on) and "$num$" shows the number of particles in such subshells. 
\label{density-buildup}
}
\end{figure*}

   In order to better understand the origin of the central depressions in  density 
distributions let us consider the contributions of different groups of the single-particle states 
with given orbital angular momentum $l$ to the total neutron and proton densities. They
are shown in Fig.\  \ref{density-buildup} for selected set of spherical nuclei across the nuclear 
chart.  These nuclei include a very light $^{34}$Si nucleus, doubly magic $^{208}$Pb nucleus, 
superheavy $^{292}$120 and $^{310}$126 nuclei and hyperheavy $^{466}$156 and 
$^{592}$186 nuclei (located in the centers of potential islands of stability of spherical 
hyperheavy nuclei (see Refs.\ \cite{AATG.19,AA.21}). The analysis of this figure leads to 
several important conclusions.

%%%%%%%%%%%%%%%%%%%%%%%%%%%%%%%%%%%%%%%%%%%%%%%%
%%%%%%%%%%%%%%%%%%%%%%%%%%%%%%%%%%%%%%%%%%%%%%%%%%
  
   First, let us consider the average density $\rho_{i}^{ave}$ in the respective $i$-th subsystem
for radial coordinate below the one at which the surface density reaches its maximum value. Neutron
average density $\rho_{\nu}^{ave}$ is located near saturation density $\rho_{\nu}^{sat} \approx 0.08$ 
fm$^{-3}$ for all these nuclei.  Although there is some trend of the decrease of this density with increasing 
proton number $Z$, it is significantly less pronounced than the one for proton average densities  
$\rho_{\pi}^{ave}$. The neutron average density tries to saturate at $\rho_{\nu}^{sat} \approx 0.08$ fm$^{-3}$ and 
this forces proton  subsystem to expand and follow the radial pattern of neutron density distribution.  
Because of the imbalance between proton and neutron numbers this can be achieved only by reducing 
overall proton density down to $\rho_{\pi}^{sat} \approx  0.06$  fm$^{-3}$ in the $^{208}$Pb, $^{292}$120 
and $^{310}$126 nuclei and down to $\rho_{\pi}^{sat} \approx  0.04$ fm$^{-3}$ in very neutron-rich 
hyperheavy $^{466}$156 and $^{592}$186 nuclei.  Similar features are seen in the nuclei of the isotopic 
chains with $Z=82, 106, 120$ and 126 (see Fig.\ 2 in Ref.\ \cite{AF.05-dep}) and in global survey of Ref.\ 
\cite{SKKJA.19}: neutron average densities stay close to the  $\rho_{\nu}^{sat} \approx 0.08$ fm$^{-3}$ 
while average proton densities decrease with increasing neutron number. These observations (in particular, 
the saturation of neutron density $\rho_{\pi}^{ave}$ near $\rho_{\nu}^{sat} \approx 0.08$ fm$^{-3}$) suggest 
that overall behavior of nuclear system is predominantly defined  by nuclear forces and  not by the Coulomb 
interaction.

 Second, the proton and neutron densities in the center of the nucleus, in its central and 
surface regions depend sensitively on the availability for occupation of the single-particle states with
respective radial properties. The densities at $r=0$ are built almost entirely  by the $s$ states. However, 
with increasing particle numbers additional $s$ states are not always available (see, for example, Fig.\ 
\ref{Single-particle-120-126} in the present paper and Figs. 5 and 8 in Ref.\ \cite{AATG.19}). 
The most striking example is the proton subsystem in which  only six $s$ states are available for 
occupation in the ground states of the nuclei with $Z\geq 82$ (see Figs.\ \ref{density-buildup}(e), (f), (j), 
(k), and (l)). Indeed, the transition from the $^{208}$Pb nucleus to $^{592}$186 one (which is equivalent 
to an addition of 104 protons to the $^{208}$Pb nucleus) does not provide any additional $s$ state. 
As a consequence, the density at $r=0$ falls down from $0.07$ fm$^{-3}$ in the $^{208}$Pb nucleus to 
$0.02$ fm$^{-3}$ in the $^{592}$186 nucleus [compare Figs.\ \ref{density-buildup}(e) and (l)]. The same 
features are also seen in neutron subsystem:  the transition from the $^{208}$Pb nucleus to the $^{292}$120 
one [both of them have six occupied $s$ states, see Figs.\ \ref{density-buildup}(b) and (c)] and from the 
$^{310}$126 nucleus to $^{466}$156 one [both of them have eight occupied $s$ states, see Figs.\ 
\ref{density-buildup}(g) and (h)] do not bring additional occupation of the $s$ states which results in 
the reduction of the density at $r=0$. 

%%%%%%%%%%%%%%%%%%%%%%%%%%%%%%%%%%%%%%%%%%%%%%%%  
\begin{figure}[htb] 
\centering
\includegraphics[width=8.4cm]{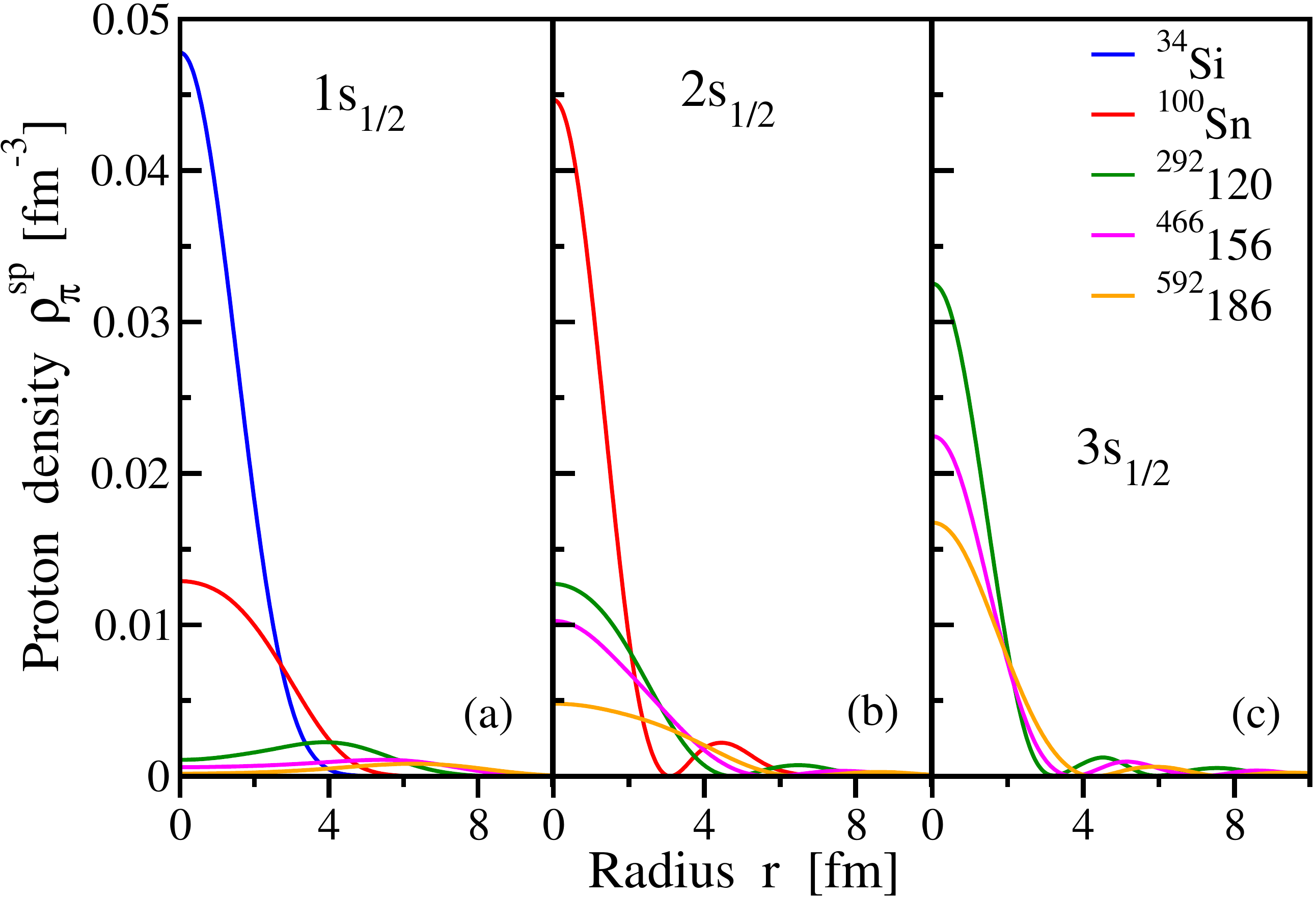}
\includegraphics[width=8.4cm]{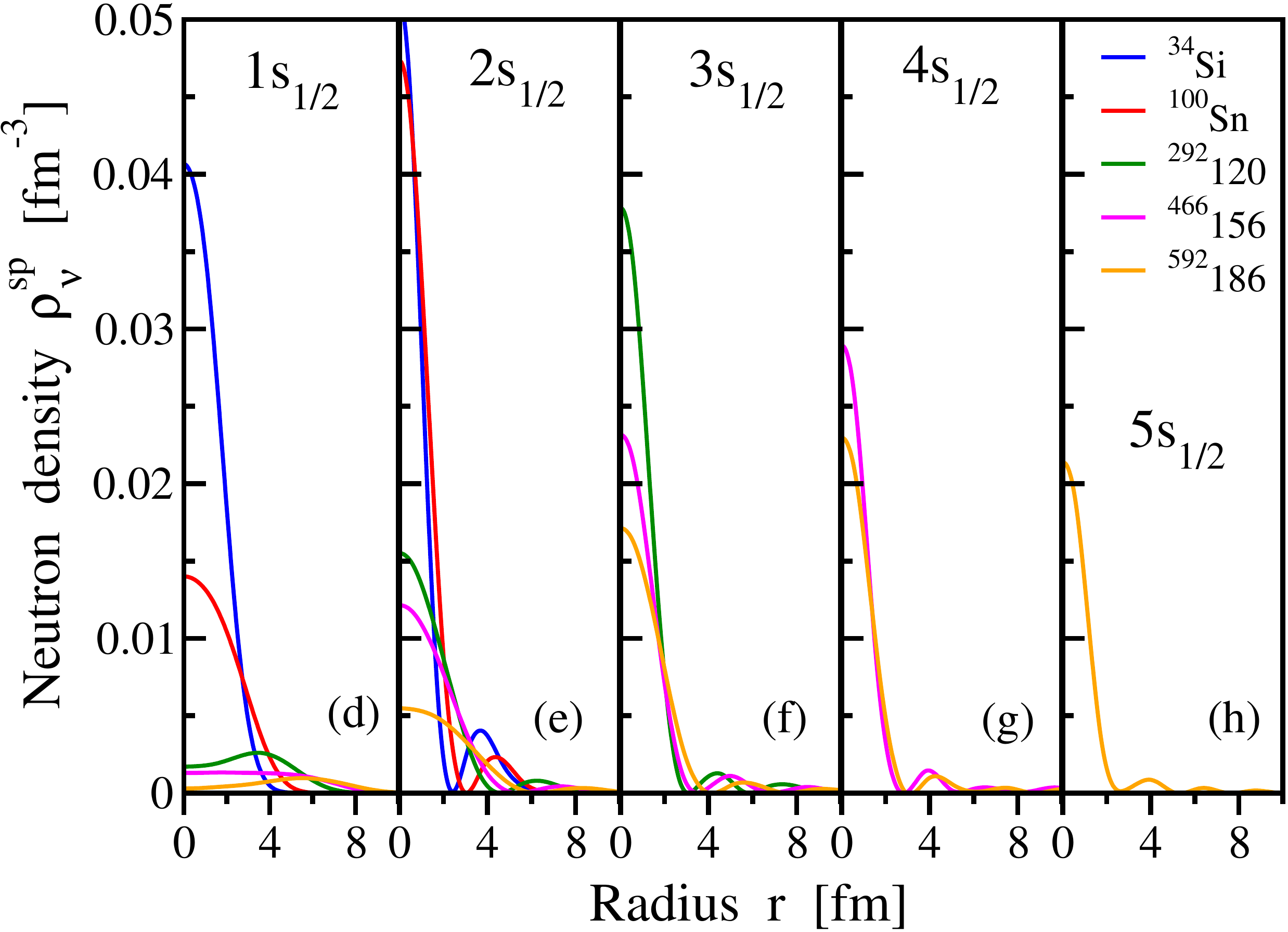}
\caption{Single-particle densities $\rho_i^{sp}$ of the $s$ states occupied in the
bubble nuclei ranging from linght $^{34}$Si up to hyperheavy nuclei.
\label{1s1_2_bubble}
}
\end{figure}
%%%%%%%%%%%%%%%%%%%%%

  This significant reduction of the density at the center of super- and hyperheavy nuclei is also
facilitated by two factors which affect the magnitude of the single-particle density of the $s$ states at $r=0$.
The first factor is the presence of classically forbidden regions in the nucleonic potentials which leads to a 
decrease of the density of the $1s_{1/2}$ and $2s_{1/2}$ states at $r=0$ [see Figs.\ \ref{1s1_2_bubble}(a), 
(b), (d), and (e)]. This decrease is especially drastic in the $^{292}120$, $^{466}$156 and $^{592}$186 nuclei
and for the $1s_{1/2}$ states. For example, the density at the center of the proton subsystem of the 
$^{592}$186 nucleus is built only by the $3s_{1/2}$ and $2s_{1/2}$ states and the contribution of the 
$1s_{1/2}$ state is almost zero. Note that the 77\% of the  total proton density at $r=0$ is built 
by only two $3s_{1/2}$ protons.   Another factor is the stretching out the radial profile of the density 
distribution of a given  single-particle state with increasing proton number or mass of the nucleus 
(see Fig.\ \ref{1s1_2_bubble}).  It leads to the decrease of the density of the $s$ states at $r=0$ because 
of the normalization condition of Eq.\ (\ref{norm-cond}).

  Third, the densities in the central $0<r<2$ fm regions of proton and neutron subsystems 
are  built by the occupation of the $s$, $p$ and $d$ states in all nuclei under study (see Fig.\ 
\ref{density-buildup}). Again the availability of such states for occupation plays a critical 
role.  For example, the number of occupied proton $s$, $p$ and $d$ states is 6, 18 and 
20, respectively, in the $^{292}$120, $^{310}$126, $^{466}$156 and $^{592}$186 nuclei 
[see Figs.\ \ref{density-buildup}(f), (j), (k) and (l)].  As a consequence, the process of the 
increase of the radius of the nucleus with increasing $Z$ (see Table \ref{Table-rms-radii}) 
leads to a reduction of the proton density in the central region with increasing $Z$.  A similar 
example is seen in the neutron subsystem of the $^{208}$Pb and $^{292}$120 nuclei in which 
the number of occupied $s$, $p$ and $d$ states is exactly the same (6, 18 and 20, respectively)
[see Figs.\ \ref{density-buildup}(b) and (c)]. Again the increase of mass number triggers 
the reduction of neutron density in the central region.

Fourth, the densities at higher radial coordinate $r$ and in the surface and near-surface regions 
are built  predominantly by the groups of medium and high-$l$ orbitals, respectively (see Figs.\ 
\ref{sp-neutron-SHE} and \ref{density-buildup}). However, the attribution  of the orbitals to these two 
groups depends on the nucleus and in many cases it is not unique. This is because the contribution of the 
groups of the orbitals with specific orbital angular momentum $l$ to the nucleonic density stretches 
over considerable range of radial coordinate (see Fig.\ \ref{density-buildup}). In addition, the groups of the
states with fixed $l$ are built from a number of the subshells with different principal quantum numbers
$n$ which differ significantly in the nodal structure of density distribution (the number of the peaks of
single-particle density is equal to $n$) and in the localization of density in radial coordinate (see Fig.\ 
\ref{sp-neutron-SHE}).

   The nucleonic density profiles in these regions depend also on the availability of  specific 
groups of the orbitals for occupation. To illustrate that let us compare the proton densities of the 
$^{466}$156 and $^{592}$186 nuclei [see Figs.\ \ref{density-buildup}(k) and (l)]. The number of the $s$, $p$, $d$ 
and $f$ states building the density in the $r <3.0$ fm region is the same in both nuclei.  However, the 
transition from the $^{466}$156 nucleus to the $^{592}$186 one leads to the increase of mass number 
which triggers the increase of the size of nucleus (see Table \ref{Table-rms-radii}) and as a consequence 
the lowering of the density in the $r <3.0$ fm region of the $^{592}$186 nucleus as compared with 
the $^{466}$156 one. This transition is also associated with the addition of  eight $g$,  eight $i$ 
and fourteen $j$ protons to the proton subsystem of the $^{466}$156 nucleus: these orbitals build density 
mostly in near-surface and surface regions of the $^{592}$186 nucleus [see  Fig.\ \ref{density-buildup}(l)].
However, the maximum density at the surface of the latter nucleus is smaller than that in the former one
because of the increase of the size of proton subsystem [compare Figs.\ \ref{density-buildup}(k) and (l)].
As a consequence of these self-consistent processes, the proton depletion factor $F_{\pi}$ (see
definition in Eq.\ (\ref{Depletion factor}) below) of  the $^{592}$186 nucleus is significantly larger than 
that of the $^{466}$156 one (see Table \ref{Table-depl}). Interestingly enough the neutron depletion factor 
$F_{\nu}$ shows opposite trend (see Table \ref{Table-depl}) and this is predominantly due to occupation 
of two additional $s$ neutrons leading to an increase of the density at $r=0$ in the  $^{592}$186 nucleus 
as compared with the $^{466}$156 one [see Figs.\ \ref{density-buildup}(k) and (l)].

%%%%%%%%%%%%%%%%%%%%%%%%%%%%
\section{Bubble indicators and their physical content}
\label{bubble-indicators}
%%%%%%%%%%%%%%%%%%%%%%%%%%%%

%%%%%%%%%%%%%%%%%%%%%%%%%%%%%%%%%%%%%
\begin{table}[h!]
\begin{center}
\caption{Depletion factor $F_i$ for proton and neutron subsystems
obtained in the calculations for indicated nuclei. }
\begin{tabular}{|c|c|c|} \hline
 Nuclei              & $F_{\pi}$ (\%)  & $F_{\nu}$  (\%) \\  \hline 	
$^{34}$Si          & 34.8                 & 0                        \\
$^{36}$S           & 0                      & 0                        \\
$^{40}$Ca         & 0                      & 0                        \\
$^{208}$Pb       & 0                      & 8.2                     \\
$^{292}$120     & 27.7                 & 34.0                   \\
$^{310}$126     & 16.7                 & 0                        \\
$^{466}$156     &  36.3                & 20.4                   \\
$^{592}$186     &  53.8                &16.0                     \\   \hline
\end{tabular}
\label{Table-depl}
\end{center}
\end{table}
%%%%%%%%%%%%%%%%%%%%%%%%%%%%%%%%%%%%%

    Two measures of the central depression in nucleonic densities are used 
in the literature. The first one called as depletion factor $F$ is defined by 
\cite{GGKNPSGV.09,YRWZD.10}
\begin{eqnarray}
F = \frac{\rho_{max} - \rho_{c}}{\rho_{max}}
\label{Depletion factor}
\end{eqnarray}
where $\rho_{c}$ and  $\rho_{max}$ represent the central $(r=0)$ and maximum densities, 
respectively. This is the simplest measure and numerical values of $F$ for the nuclei of
interest are shown in Table \ref{Table-depl}.  More complicated measure of central 
depression  has been introduced in Ref.\ \cite{SNR.17} and it is defined as 
\begin{eqnarray} 
\bar{\rho}_{t,c} = \frac{\rho_{t, av} - \rho_{t,c}}{\rho_{t,av}}
\label{Depl-fac-2}
\end{eqnarray} 
where $t= (\pi,\nu)$,  $\rho_{t,c}$  is the density at the center of respective subsystem 
and $\rho_{t,av} = N_t/(4/3\pi R_d^3)$ is the average density of the nucleus assuming 
a constant density up to diffraction radius $R_d$ \cite{FV.82}. The authors of 
Ref.\ \cite{SNR.17}  use this radius instead of rms radius since it is not affected by 
surface thickness.   

%%%%%%%%%%%%%%%%%%%%%%%%%%%%%%%%%%%%%
\begin{table}[h!]
\begin{center}
\caption{Depletion factors $F_i$ for proton and neutron subsystems
obtained in the calculations for indicated $N=Z$ nuclei. }
\begin{tabular}{|c|c|c|} \hline
 Nuclei              & $F_{\pi}$ (\%)  & $F_{\nu}$  (\%) \\  \hline 	
$^{56}$Ni          & 0                 & 0                        \\
$^{100}$Sn           & 24.31                      & 21.84                        \\
$^{164}$Pb         & 0                      & 0                        \\
$^{240}$120       & 26.12                     & 20.88                     \\
$^{252}$126     & 15.91                & 17.39                  \\
$^{312}$156     & 34.45                 &28.23                       \\
$^{372}$186     & 58.25                & 51.36                  \\ \hline 
\end{tabular}
\label{Table-depl-mirror}
\end{center}
\end{table}
%%%%%%%%%%%%%%%%%%%%%%%%%%%%%%%%%%%%%

    Both indicators are strongly affected by the single-particle degrees of freedom. For example, $\rho_{c}$ in both 
definitions is determined almost entirely by the $s$ states and their availability for occupation 
across the nuclear chart. Second ingredient entering into Eqs.\  (\ref{Depletion factor}) 
and (\ref{Depl-fac-2}) is also not free from single-particle degrees of freedom. Let us first 
consider depletion factor $F$. In the cases when the density in center is larger than the 
one at the surface [see Figs.\ \ref{Total_den}(a), (b), (c) and (f)] then 
$\rho_{max} = \rho_{c}$ and $F=0$ (see Table \ref{Table-depl}). In other cases,  
$\rho_{max}$ is defined by the single-particle states which build maximum density in 
the region near the surface [see Figs.\ \ref{Total_den}(a), (d), (e), (f), (g) and (h), Fig.\ 
\ref{density-buildup} and discussion of the second part of Sec.\ \ref{Gen-observations}]. 
$\rho_{t,av}$ used in Eq.\ (\ref{Depl-fac-2}) also depends on the underlying single-particle
structure and availability of the single-particle states for occupation despite the fact that it 
averages densities up to diffraction radius $R_d$.

%%%%%%%%%%%%%%%
\begin{figure}[htb]
\centering
\includegraphics[width=8.4cm]{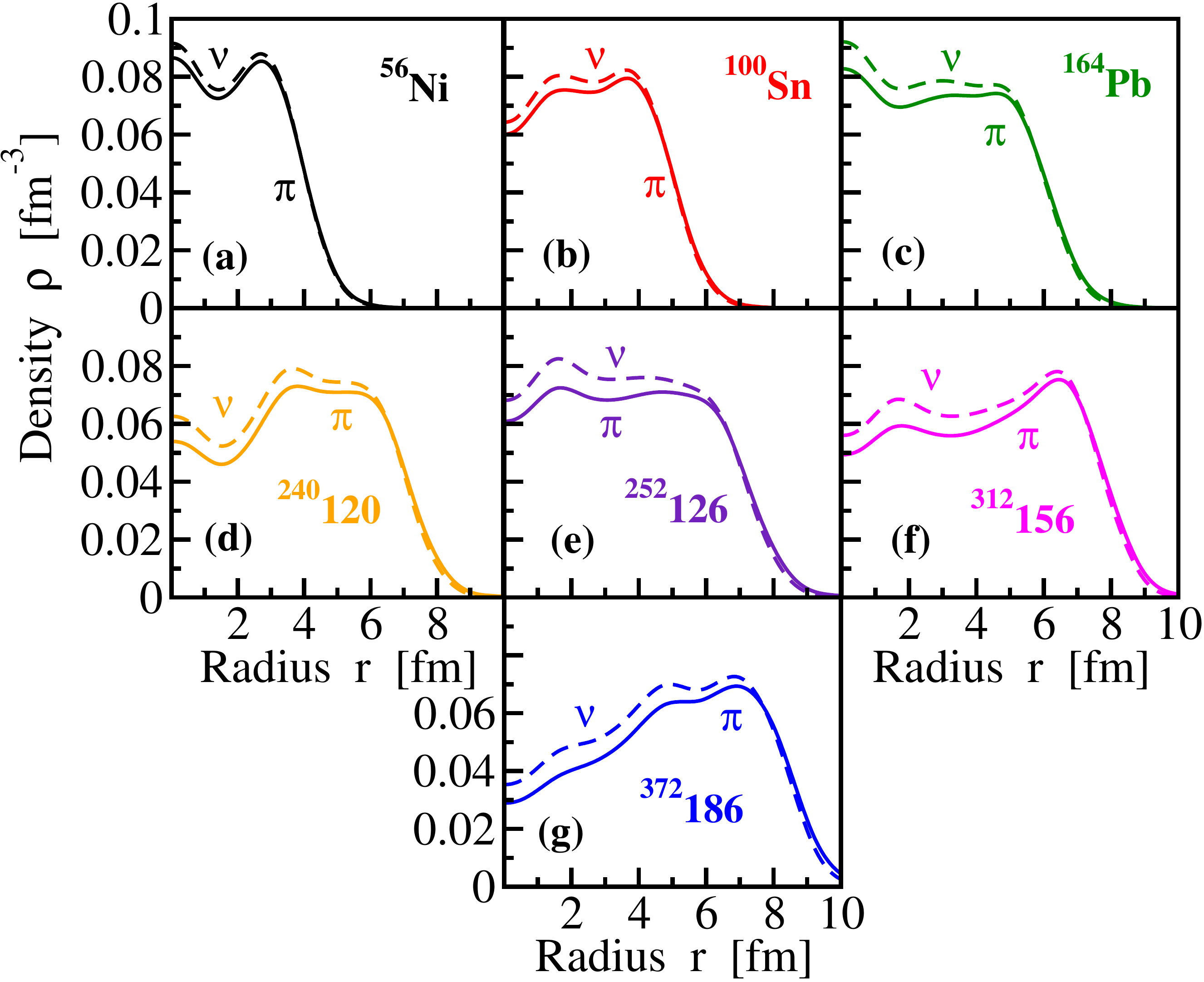}
\caption{Proton and neutron densities of selected $N=Z$ nuclei.
\label{Mirror_nuclei_no_pair_density}
}
\end{figure}
%%%%%%%%%%%%%%%%%%%%%%%%%%%%%%%%%%%%%%%%%%%%%%%% 

    These facts strongly suggest that both indicators cannot be reliable measures of 
bulk properties (such as those related to the Coulomb interaction). This is especially true because in wine bottle
nucleonic potentials the densities of the $s$ states, their magnitudes at $r=0$ and their radial 
profiles,  are affected strongly by classically forbidden regions of the potentials (see Secs.\ 
\ref{single-particle-degrees} and  \ref{Gen-observations}).  Thus, the conclusions of Ref.\ 
\cite{SNR.17}  that the central depression in superheavy nuclei is firmly driven by the 
electrostatic repulsion should be treated with extreme caution since they are based on the 
bubble indicator of Eq.\ (\ref{Depl-fac-2}).

%%%%%%%%%%%%%%%%%%%%%%%%%%%%%%%%%%%%%%%%%%%%%%%%%%
\begin{figure}[htb]
\centering
\includegraphics[width=8.4cm]{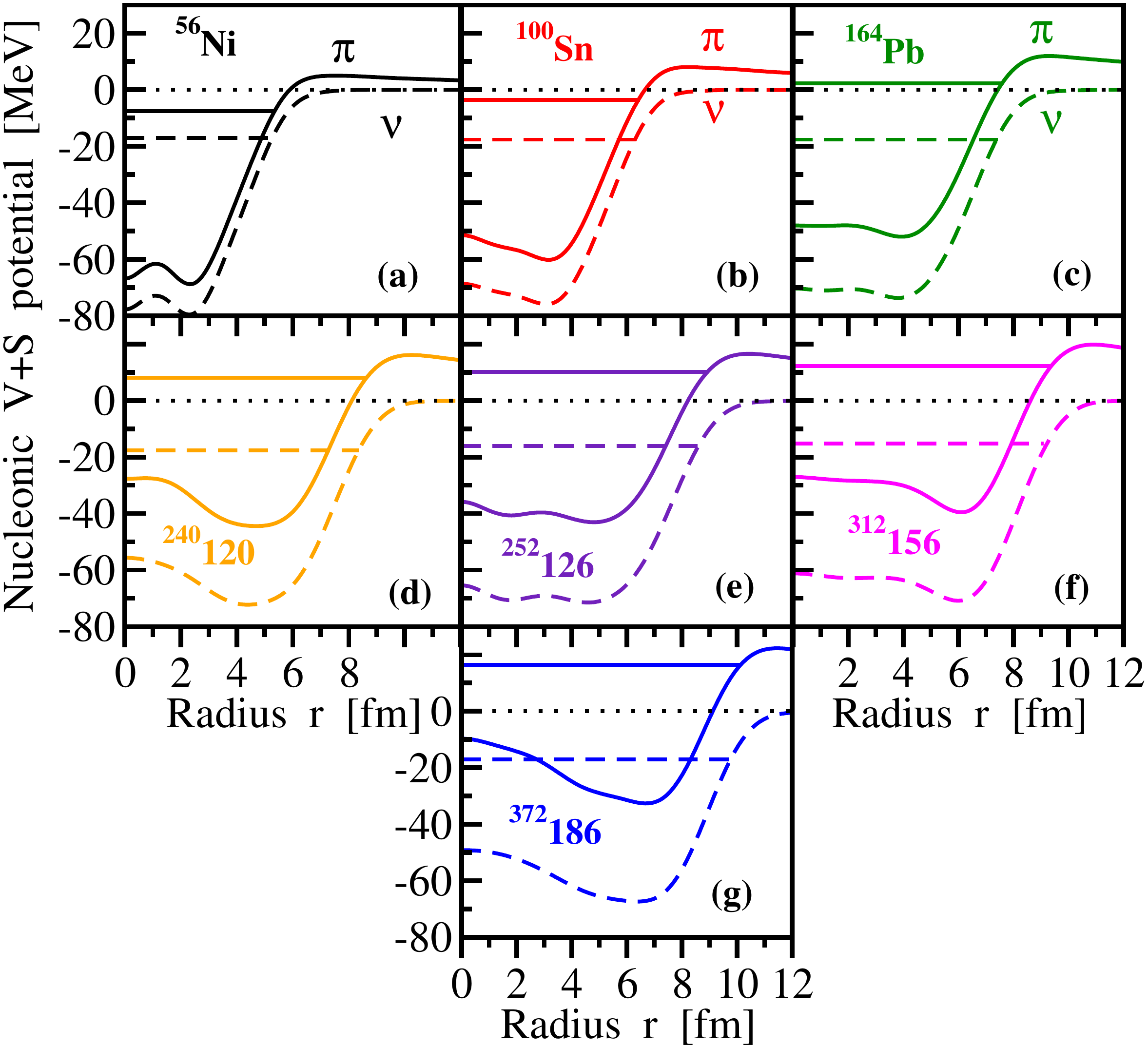}
\caption{Proton (solid lines) and neutron (dashed lines) nucleonic potentials in indicated
$N=Z$ nuclei. Horizontal solid and dashed lines show proton and neutron Fermi levels,
respectively.
\label{Mirror_nuclei_no_pair_Poten}
}
\end{figure}
%%%%%%%%%%%%%%%%%%%%%%%%%%%%%%%%%%%%%%%%%%%%%%%%

    In nuclei the protons feel combined nuclear and Coulomb potentials  which 
lead to proton single-particle states with specific density distributions over radial coordinate. 
In addition, there is a nuclear interaction between protons and neutrons which further 
complicates the situation. As a consequence, there is no straightforward procedure of the 
separation of nuclear and Coulomb interaction effects on the central depression in density 
distributions. On the other hand, in the light of the conclusions of Ref.\ \cite{SNR.17} it is 
important to estimate possible magnitude of the Coulomb interaction effects on these 
depressions. From out point of view, the only possible way to get that is by comparing 
proton and neutron depressions in symmetric $N=Z$ nuclei with the same nucleonic 
configurations in proton and neutron subsystems.

     Proton and neutron densities of selected set of the nuclei are shown in 
Fig.\ \ref{Mirror_nuclei_no_pair_density}. Most of these nuclei belong  to isotopic chains 
discussed above, but we also added $^{56}$Ni and $^{100}$Sn\footnote{The 
majority  of these nuclei are proton unbound (see Fig.\ \ref{Mirror_nuclei_no_pair_Poten}) 
and there is no local minimum at spherical shape in deformation energy curves of hyperheavy 
nuclei. The lowest in energy solutions of the hyperheavy $^{312}$156 and $^{372}$186 
nuclei in the axial RHB calculations correspond to toroidal shapes. Thus, spherical solutions 
in these two hyperheavy nuclei are used here as theoretical benchmarks 
for the investigation of the impact of single-particle degrees of freedom 
and Coulomb interaction on the formation/suppression of the bubble structures in hyperheavy
nuclei.}. One can see in Fig.\ \ref{Mirror_nuclei_no_pair_density} 
that in a given nucleus the proton densities closely follow the radial profiles of the neutron 
densities but with somewhat reduced absolute magnitudes. This is due to  the fact that 
Coulomb interaction (by means of electrostatic repulsion) somewhat  increases the radius 
of the proton density  as compared with the neutron one.  This corresponds to transfer of 
protons from sub-surface region into surface one.

     The resulting depletion factors $F_i$ are shown in Table \ref{Table-depl-mirror}.
One can see that on average they are larger in the proton  subsystem as compared with 
the neutron one by only approximately 20\%. This suggests that the Coulomb interaction 
plays only a secondary role in the formation of the depletions in the central  density distribution. 
It is interesting that the depletion factors are similar in the medium mass $^{100}$Sn and 
superheavy $^{240}$120 nuclei. This again supports the notion that single-particle degrees 
of freedom are dominant in creation of the bubbles and Coulomb interaction plays only a 
secondary role.
\vspace{-0.4cm}
%%%%%%%%%%%%%%%%%%%%%%%%%%%%%%%%%%%%%%
\section{The factors affecting the availability of the low-$l$ states for occupation}
\label{factors_1} 
%%%%%%%%%%%%%%%%%%%%%%%%%%%%%%%%%%%%%%

   The densities of the occupied single-particle states represent the 
basic building blocks of the total densities. To build a flat density distribution one 
should have a balanced combination of the occupied states which build the 
density in the center of the nuclei and in their middle and surface parts. However, the 
question of whether such a balanced combination of single-particle states is 
available for occupation in super- and hyperheavy nuclei has not even been raised 
so far in the literature.
 
 %%%%%%%%%%%%%%%%%%%%%%%%%%%%%%%%%%%%%%%%%%%%%%%%  
\begin{figure}[htb] 
\centering
\includegraphics[width=4.75cm]{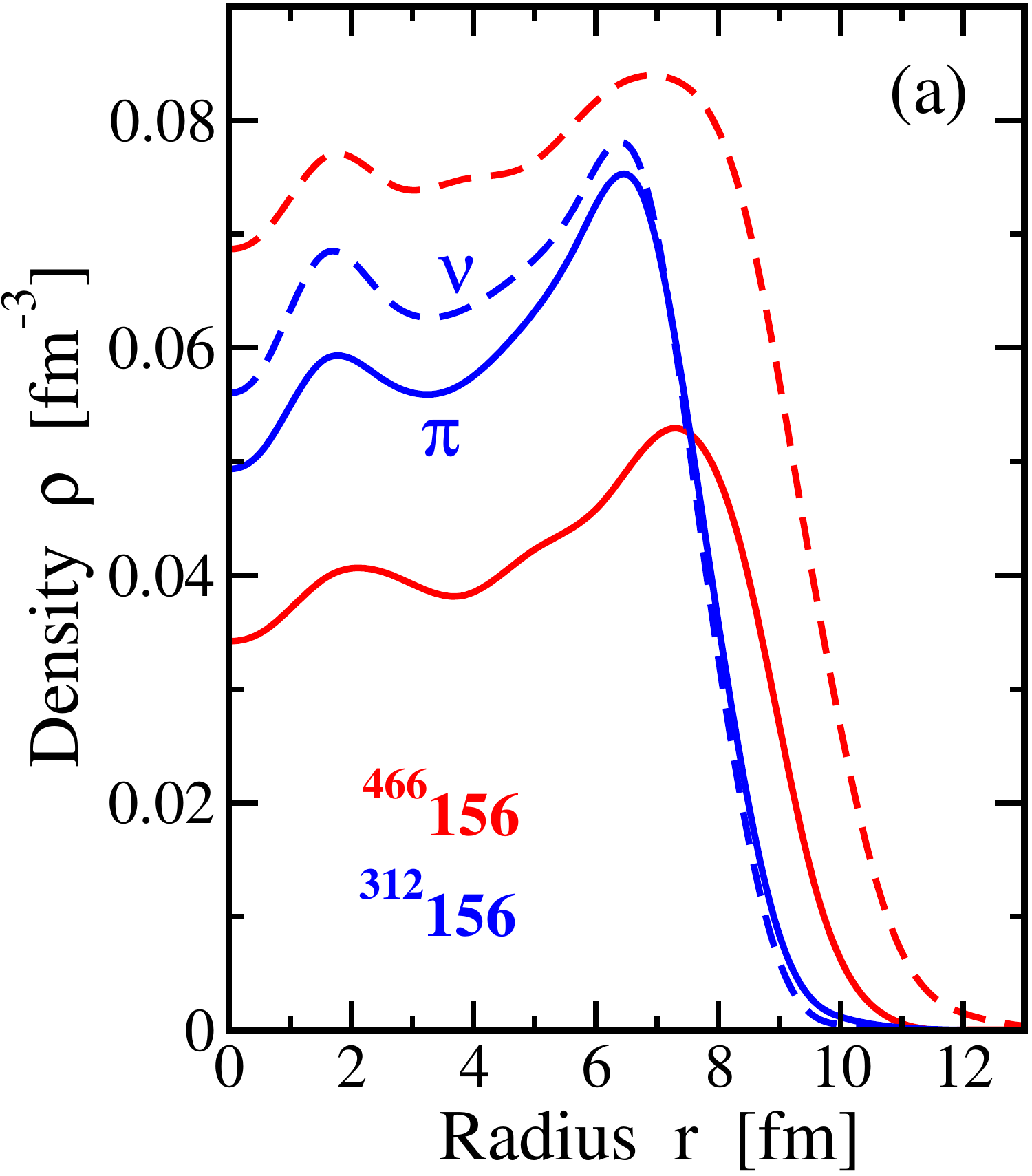}
\includegraphics[width=3.72cm]{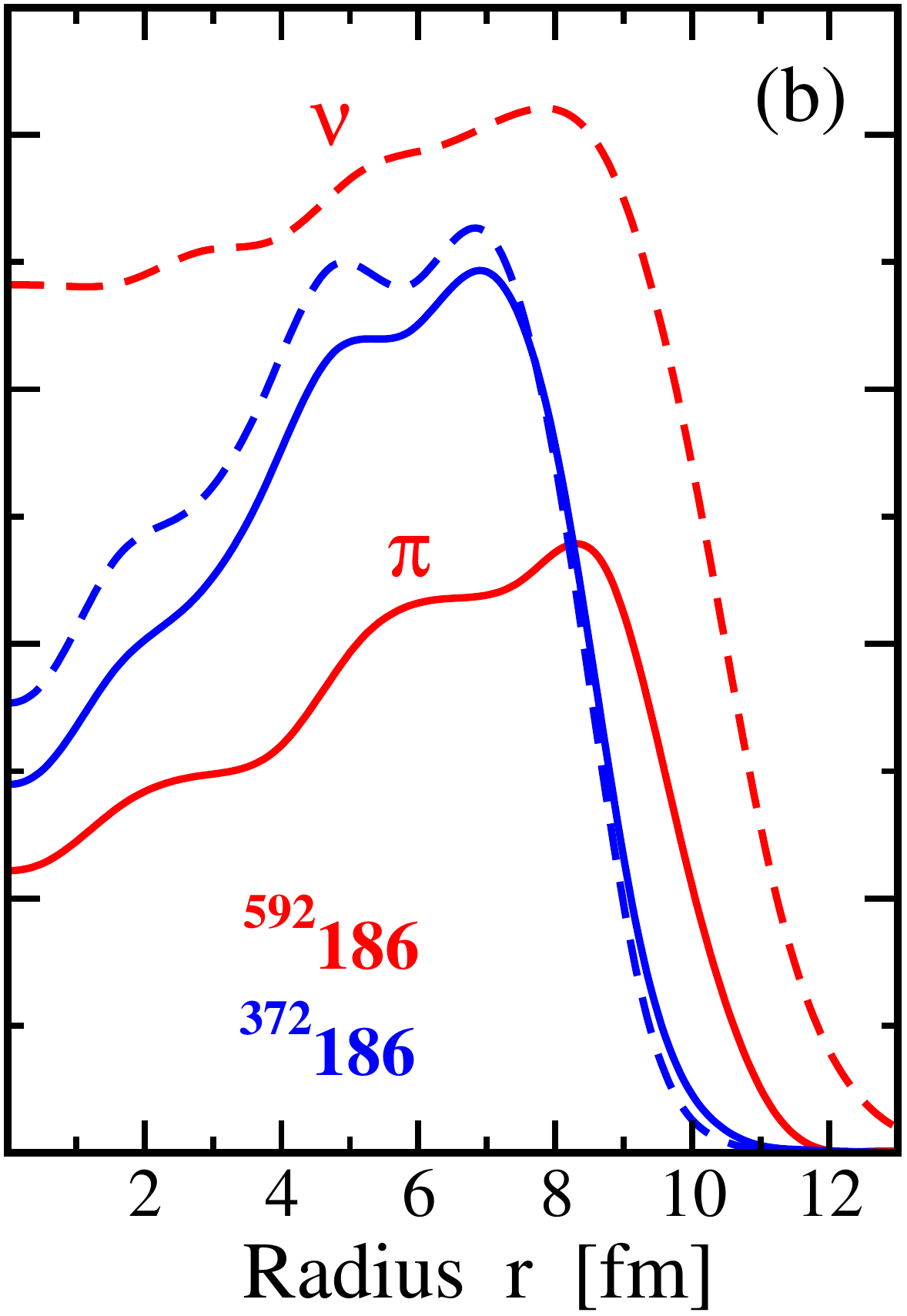}
\caption{Proton (solid lines) and neutron (dashed lines) densities of indicated nuclei.
\label{Density_466_312_156}
}
\end{figure}
%%%%%%%%%%%%%%%%%%%%%%%%%%%%%%%%%%%%%%%%%%%%%%%%  
\newpage
 %%%%%%%%%%%%%%%%%%%%%%%%%%%%%%%%%%%%%
\begin{table*}[ht]
	\begin{center}
		  \caption{The single-particle states of given principal quantum number $N$ and orbital
                                 angular momentum $l$ in the hyperheavy $^{592}$186 nucleus. This nucleus
                                 has 186 protons and 406 neutrons. The particle numbers are given in the
                                 format ${\bf n_{pos}}/p_{occ}/n_{occ}$. ${\bf n_{pos}}$ (in bold)  is total number of
                                 the states with given values of $N,l$ in spherical harmonic oscillator potential. 
                                 $p_{occ}$ and  $n_{occ}$ display the number of the occupied states with 
                                 given values of $N,l$ in the proton and neutron subsystems of the  
                                 $^{592}$186   nucleus, respectively.                           
                                 } 
		\begin{tabular}{|c|c|c|c|c|c|c|c|c|c|c|c|} \hline
			$l\diagdown N$ & 0 & 1 & 2 & 3 & 4 & 5 & 6 & 7 & 8 & 9 & 10   \\ \hline 	
			0 & \textbf{2}/2/2  &  &\textbf{ 2}/2/2 & &\textbf{2}/2/2 &&\textbf{ 2}/0/2&&\textbf{2}/0/2&&\textbf{2}/0/0 \\ \hline
			1 &   & \textbf{6}/6/6 &  &\textbf{ 6}/6/6 & &\textbf{6}/6/6 & &\textbf{6}/0/6 &&\textbf{6}/0/0 & \\ \hline
			2 &   &  & \textbf{10}/10/10 & &\textbf{10}/10/10&&\textbf{10}/0/10&&\textbf{10}/0/10&&\textbf{10}/0/0 \\ \hline
			3 &  &  & & \textbf{14}/14/14&&\textbf{14}/14/14& &\textbf{14}/0/14&&\textbf{14}/0/0& \\ \hline
			4 &  &  &  & &\textbf{18}/18/18 && \textbf{18}/18/18&&\textbf{18}/0/18&&\textbf{18}/0/0 \\ \hline
			5 &  &  & & & &\textbf{22}/22/22& &\textbf{22}/0/22&&\textbf{22}/0/0& \\ \hline
			 6 & &  & & &&& \textbf{26}/26/26&&\textbf{26}/0/26&&\textbf{26}/0/0 \\ \hline
			 7 &  &  & & & & & &\textbf{30}/30/30& &\textbf{30}/0/16& \\ \hline
			 8 & &  &  & & && &&\textbf{34}/0/34&&\textbf{34}/0/0 \\ \hline
			 9 & &  &  & & && &&&\textbf{38}/0/38& \\ \hline
			10 & &  &  & & && &&&&\textbf{42}/0/22 \\ \hline
		    Total &\textbf{2}/2/2&\textbf{6}/6/6&\textbf{12}/12/12&\textbf{20}/20/20&\textbf{30}/30/30/&\textbf{42}/42/42&\textbf{56}/44/56&\textbf{72}/30/72&\textbf{90}/0/90&\textbf{110}/0/54&\textbf{132}/0/22 \\ \hline
		\end{tabular}
		\label{Filling}
	\end{center}
\end{table*}
%%%%%%%%%%%%%%%%%%%%%%%%%%%%%%%%%%%%%%%%%%%%%%%%%%%%%%
   A specific feature of a realistic nuclear potential is the fact that within 
a shell with a given principal quantum number $N$ the states with highest possible 
orbital angular momentum $l$ are the lowest in energy while those with lowest 
$l$  (such as the $s$ states in even-$N$ shells and the $p$ states in odd-$N$ shells)
are typically located at the highest or near highest energies in the shell (see,
for example, Fig. 6.3 in Ref.\ \cite{NilRag-book}). Thus, with the filling of a 
specific $N$ shell the density is first built at the surface, then in the middle
part of the nucleus and only then in the central region and at $r=0$.

   The detailed analysis of the occupation of different groups of the $(N,l)$ 
states in the $^{592}$186 nucleus reveals that only high-$l$ subshells are occupied in 
the high-$N$  shells (see Table \ref{Filling}). Let us consider proton subsystem. All 
$N=5$ states are occupied in it (see Table \ref{Filling}). However, only $l=6$ and $l=4$ 
states are occupied in the $N=6$ shell and only $l=7$ states in the $N=7$ 
shell\footnote{Similar pattern of the occupation is seen in the neutron subsystem in which
the last fully occupied shell has $N=8$. Only $l=9$ and  $l=7$ subshells are occupied
in the $N=9$ shell and only $l=10$ subshell in the $N=10$ shell  (see Table \ref{Filling}).} 
(see Table \ref{Filling}). This imbalance between the occupation of the high-$l$ and low-$l$ 
subshells is definitely responsible for a preferential buildup of the density in the surface 
region and as a consequence of the formation of pronounced bubble in this nucleus 
[see Fig.\ \ref{Density_466_312_156}(b)]. This feature becomes even more pronounced 
for symmetric $N=Z$ $^{372}$186 nucleus which has the same nucleonic configurations 
in proton and neutron subsystems [see Fig.\ \ref{Density_466_312_156}(b)].

   Let us consider how the neutron system of the $^{592}$186 nucleus is built from 
the one in the $^{372}$186 nucleus.  The former nucleus has 220 extra neutrons
which according to the Table \ref{Filling} are placed into the $s$ states (4 neutrons),
$p$ states (6 neutrons), $d$ states (20 neutrons) and the rest into  higher $l$ states.
The presence of these low-$l$ states allows to increase the density in the central
region and to build significantly flatter neutron density distribution (compare dashed 
red  curves for the $^{372}$186 and $^{592}$186 nuclei in Fig.\ \ref{Density_466_312_156}(b)]).  
Detailed analysis reveals that similar features are also active in the $Z=156$ isotopes (see 
Fig.\  \ref{Density_466_312_156}(a)]).

   These two examples clearly indicate that although the effects of the Coulomb interaction 
are increased in hyperheavy nuclei as compared with lighter ones they alone cannot explain 
the density profiles seen in  Fig.\ \ref{Density_466_312_156}. It turns out that the unavailability 
of the low-$l$ states for occupation plays an extremely important role in the formation
of the bubble structures in such nuclei.

   It is important to evaluate which factors affect the availability of the 
low-$l$ single-particle states for occupation. Since earlier studies it became clear
that the bubble can have a profound impact on relative energies of the low- and 
high-$l$ states. For example, it was shown in Ref.\ \cite{Wong.73} 
(see discussion of Figs. 30 and 31 in this paper) within schematic shell model 
approach that low-$l$ (high-$l$) states rise in energy (go down in energy) with 
increasing inner bubble radius $R_1$. However, this approach is unrealistic 
since it assumes zero density inside the bubble for $r<R_1$. Such a scenario is 
not realized in the nuclei under study and thus the effect of the bubble is 
substantially overestimated in Ref.\ \cite{Wong.73}. Similar effect (see Fig. 8 in Ref.\ 
\cite{DP.97}) is seen also in the calculations of Ref.\ \cite{DP.97} in a phenomenological 
shell approach which allows partial filling of the hole.  Non-relativistic Hartree-Fock 
calculations of the $^{200}$Hg nucleus show that for realistic shapes of the bubble 
the effect is significantly smaller: only the $s$ states rise in energy by a few MeV with 
increasing of the bubble size while the energies of other states remains almost constant 
(see Fig.\ 4 in Ref.\ \cite{CS.73}). The Hartree-Fock-Bogoliubov calculations with
the D1S force also show that the energies of low-$l$ states (high-$l$) states rapidly 
rise (gradually decrease) in energy with the increase of the size of the bubble
(see Figs. 18 and 19 in Ref.\ \cite{DBGD.03}).

   All these results suggest that in some situations the presence of the 
bubble in the density can lead to unavailability of the low-$l$ states for occupation
at given particle number as compared with the case of flat density distribution in the 
subsurface region of nucleus.  One should keep in mind that constraining 
bubble potential $F(r)$ is usually added to hamiltonian $H$ in order to evaluate the 
evolution of single-particle levels with bubble size/shape by minimization of 
$H+\lambda F(r)$ (see, for example, Refs.\ \cite{CS.73,DBGD.03}). Here $\lambda$ 
is bubble parameter.  Unfortunately, the same shape of the bubble is assumed for proton and 
neutron subsystems in the calculations of Refs.\ \cite{Wong.73,CS.73,DP.97,DBGD.03}
and this contradicts to calculated total densities seen in Figs.\ \ref{Total_den}, 
\ref{Total_density_3_nuclei} and \ref{Density_466_312_156}.  Note that the bubbles 
are different in proton and neutron subsystems even  in the $N=Z$ nuclei (see 
Fig.\ \ref{Mirror_nuclei_no_pair_density}). In addition, the results of the calculations  
depend on the assumption about the form on $F(r)$.

\begin{figure*}[t] 
\centering
%\hspace{3.0}\includegraphics[width=6.52cm]{VS_Proton_120E_around.eps}
\hspace{3.0pt}\includegraphics[width=6.52cm]{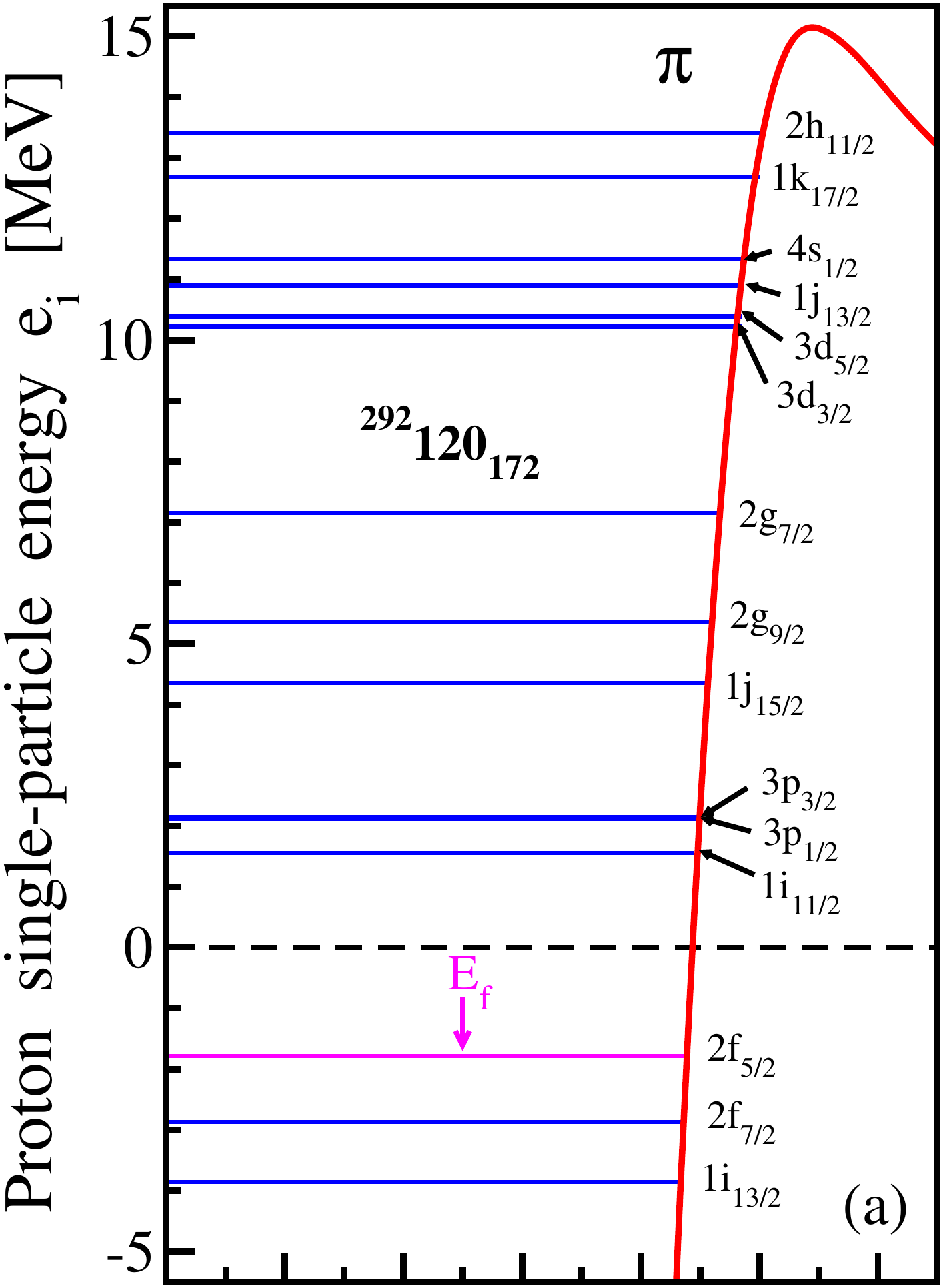}
\includegraphics[width=5.4cm]{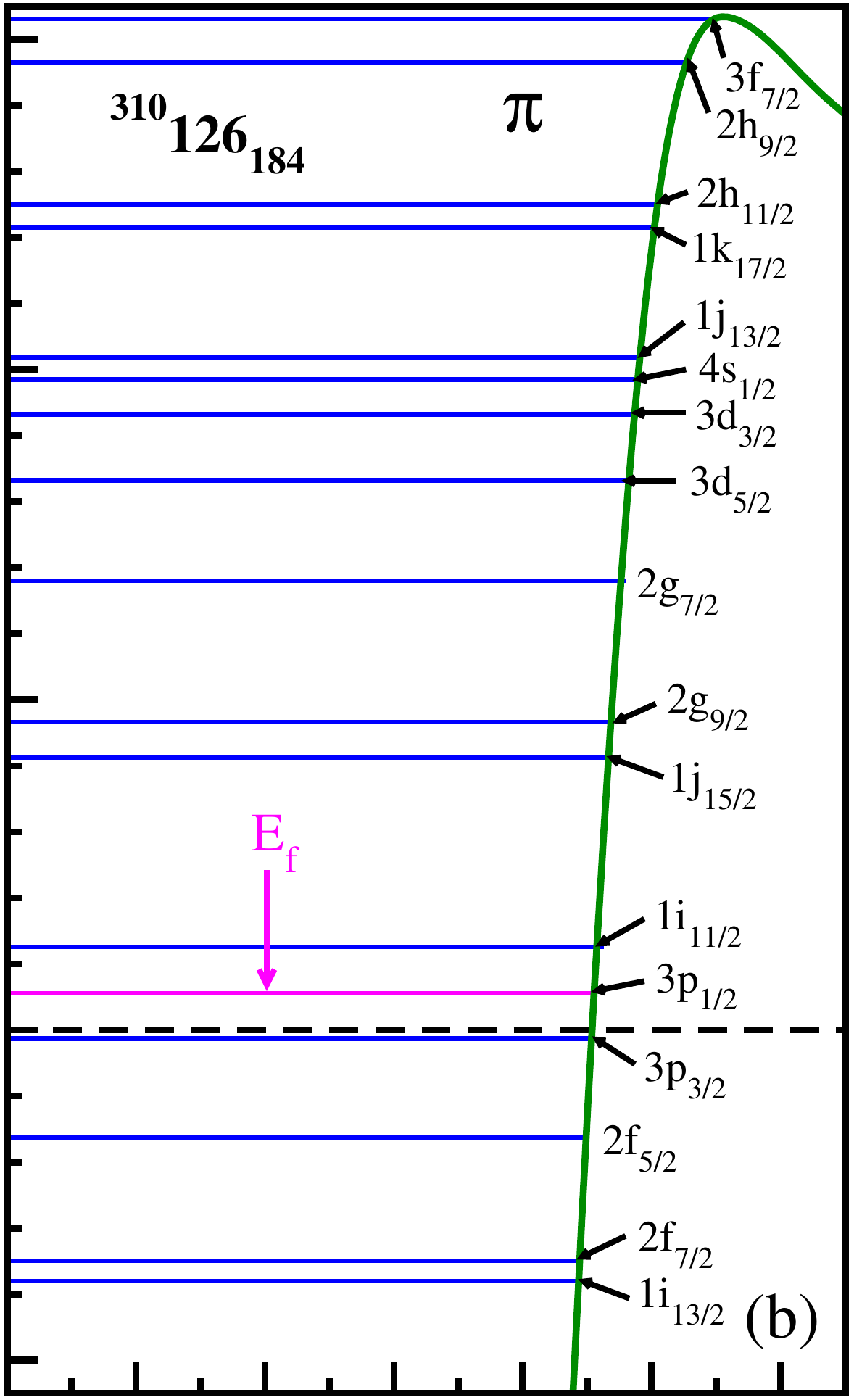}
\includegraphics[width=6.64cm]{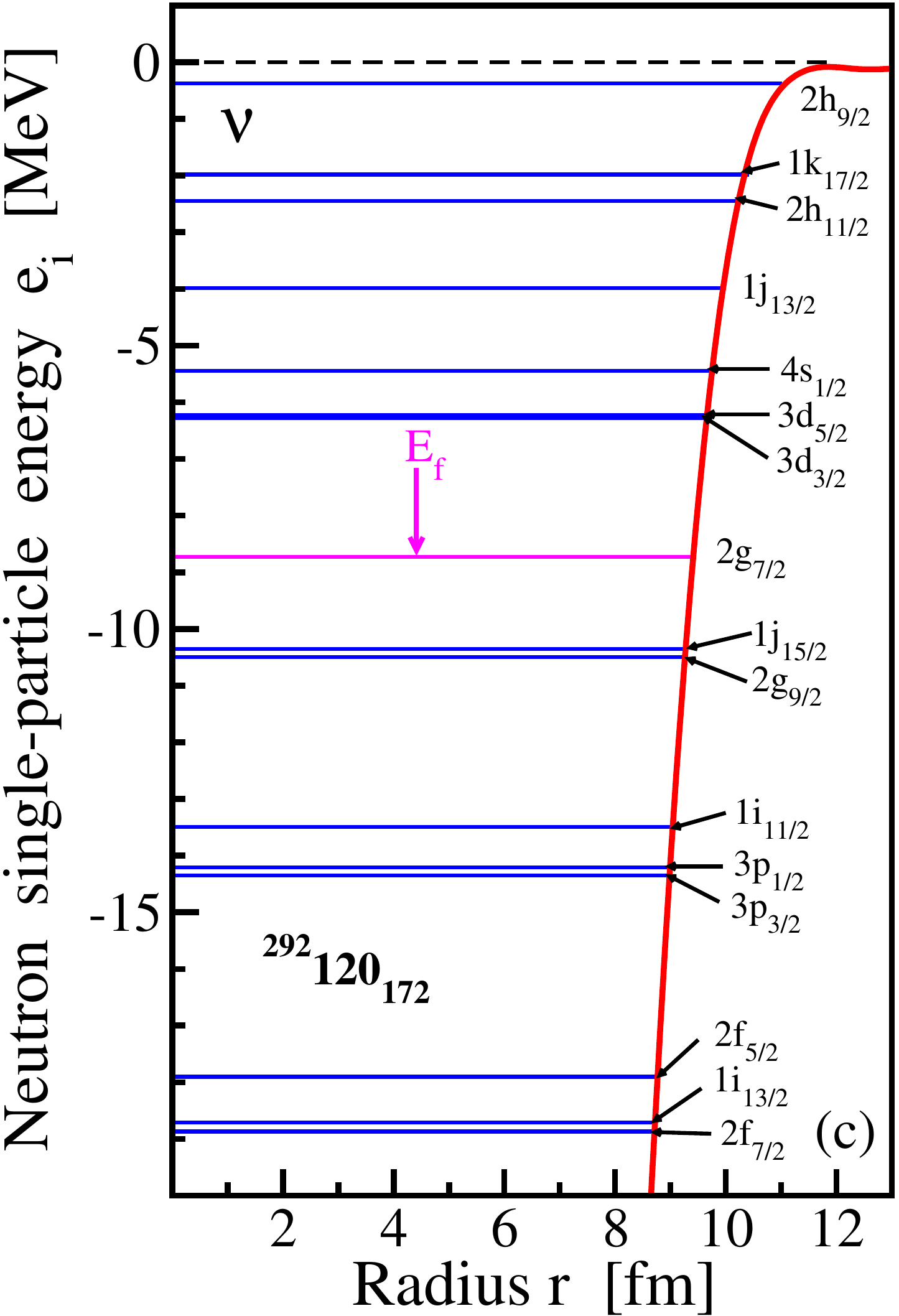}
\includegraphics[width=5.4cm]{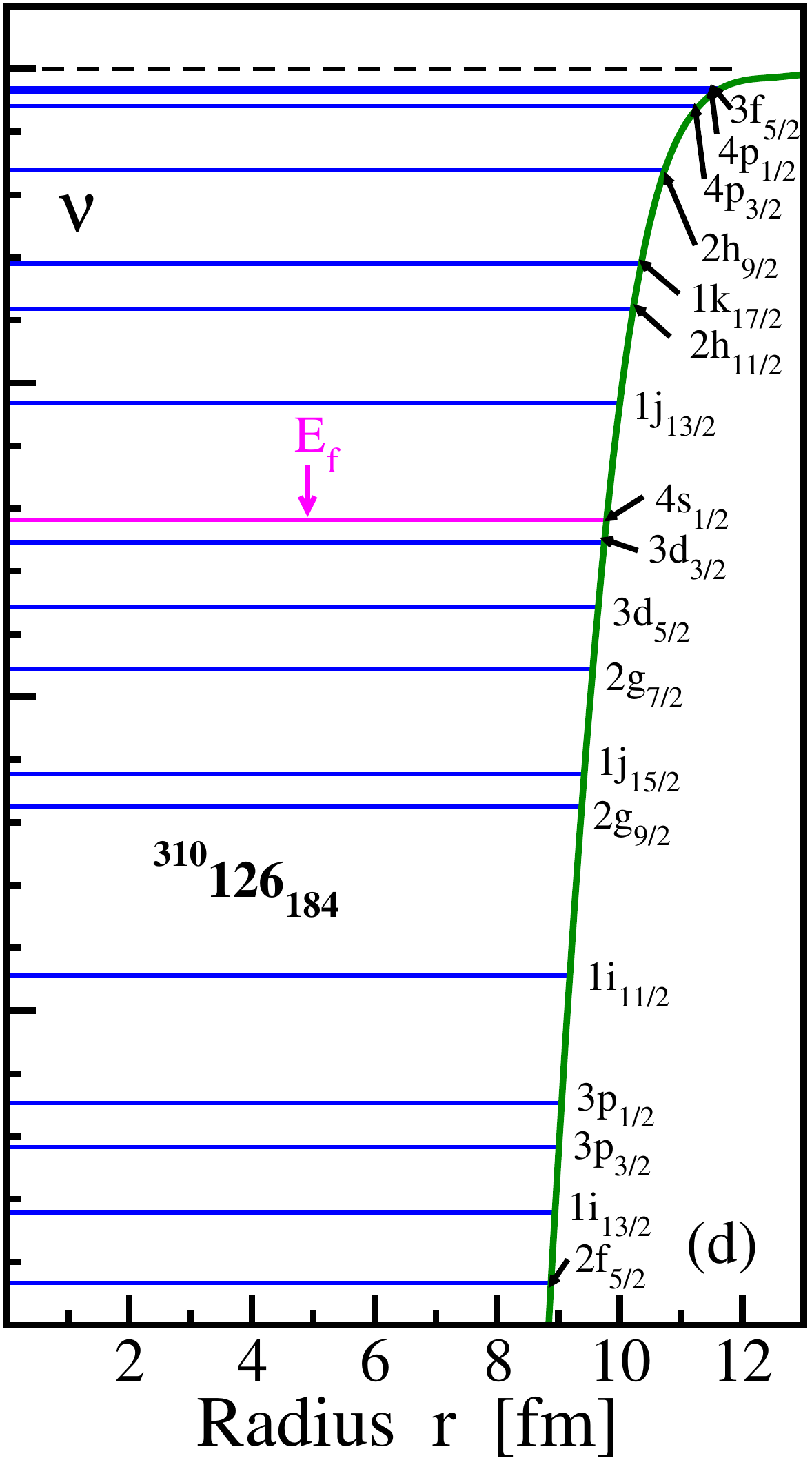}
\caption{The same as in Fig.\ \ref{Single-particle-120-126} but for the 
single-proton states located between the Fermi level and the top of 
the Coulomb barrier and for the neutron single-particle states located 
below continuum  threshold. The energy range on vertical axis is 
the same in all panels.
\label{Single-particle-120-126_around}
}
\end{figure*}

   Thus, the results of the calculations discussed above should be taken 
with some grain of salt and alternative methods of the analysis of the impact of the 
bubble on the single-particle structure should be considered.  The comparison 
of the single-particle spectra in the pair of the nuclei with and without bubble 
structures provide such an alternative. The best example of such a comparison is 
provided by the pair of the $^{292}$120 and $^{310}$126 nuclei (see Fig.\ 
\ref{Single-particle-120-126_around}) since these two nuclei  have very similar 
rms radii of proton and neutron matter distributions (see Table \ref{Table-rms-radii}).

 The sequence of the proton states  from the vicinity of the Fermi level up to the top 
of the Coulomb potential in the flat density nucleus  $^{310}$126 is $2f_{5/2}$, 
$3p_{3/2}$, $3p_{1/2}$, $1i_{11/2}$, $1j_{15/2}$, $2g_{9/2}$, $2g_{7/2}$, $3d_{5/2}$,  $3d_{3/2}$, 
$4s_{1/2}$, $1j_{13/2}$, $1k_{17/2}$, $2h_{11/2}$, $2h_{9/2}$ and $3f_{7/2}$  (see  
Fig.\ \ref{Single-particle-120-126_around}(b)). Almost the same sequence with a pair 
of exceptions discussed below are seen in the proton subsystem of the bubble nucleus 
$^{292}$120 (see  Fig.\ \ref{Single-particle-120-126_around}(a)). The energies of the 
$l>2$ proton states are typically located within 1 MeV in both nuclei. So the bubble 
does not  produce a significant impact on their energies. On the contrary, it has a more 
pronounced impact on the energies of the low-$l$ $s$ and $p$ states: it moves 
them from below their high-$l$ neighbours ($1i_{11/2}$ and $1j_{13/2}$, respectively) 
in flat density  $^{310}$126 nucleus to above them in  bubble nucleus 
$^{292}$120 [compare Figs.\ \ref{Single-particle-120-126_around}(a) and (b)].
It is interesting that the effect of the bubble is less pronounced in the neutron
subsystem: the sequence of the states is the same in both nuclei 
[compare Figs.\ \ref{Single-particle-120-126_around}(c) and (d)].
This is due to the fact that wine bottle potential is less pronounced in the neutron 
subsystem of the $^{292}$120 nucleus than in the proton one [compare 
Figs.\ \ref{Single-particle-120-126}(a) and (c)]. In contrast, the bubble is
more pronounced in neutron densities (see Fig.\ \ref{Total_density_3_nuclei}). 
Thus, one can conclude that these are the modifications in the potentials (and 
not in densities) which govern the behavior of the single-particle states.

   Although the presence of the bubble somewhat increases the energies of the 
$s$ and $p$ states in the proton subsystem and affects the availability of  these 
states for occupation as a function of proton number,  this effect in superheavy 
nuclei is not that drastic. It will only shift in proton number the position in the 
$(Z,N)$ plane at which bubble structures are either enhanced or suppressed.

\begin{figure}[htb]
\centering
\includegraphics[width=8.3cm]{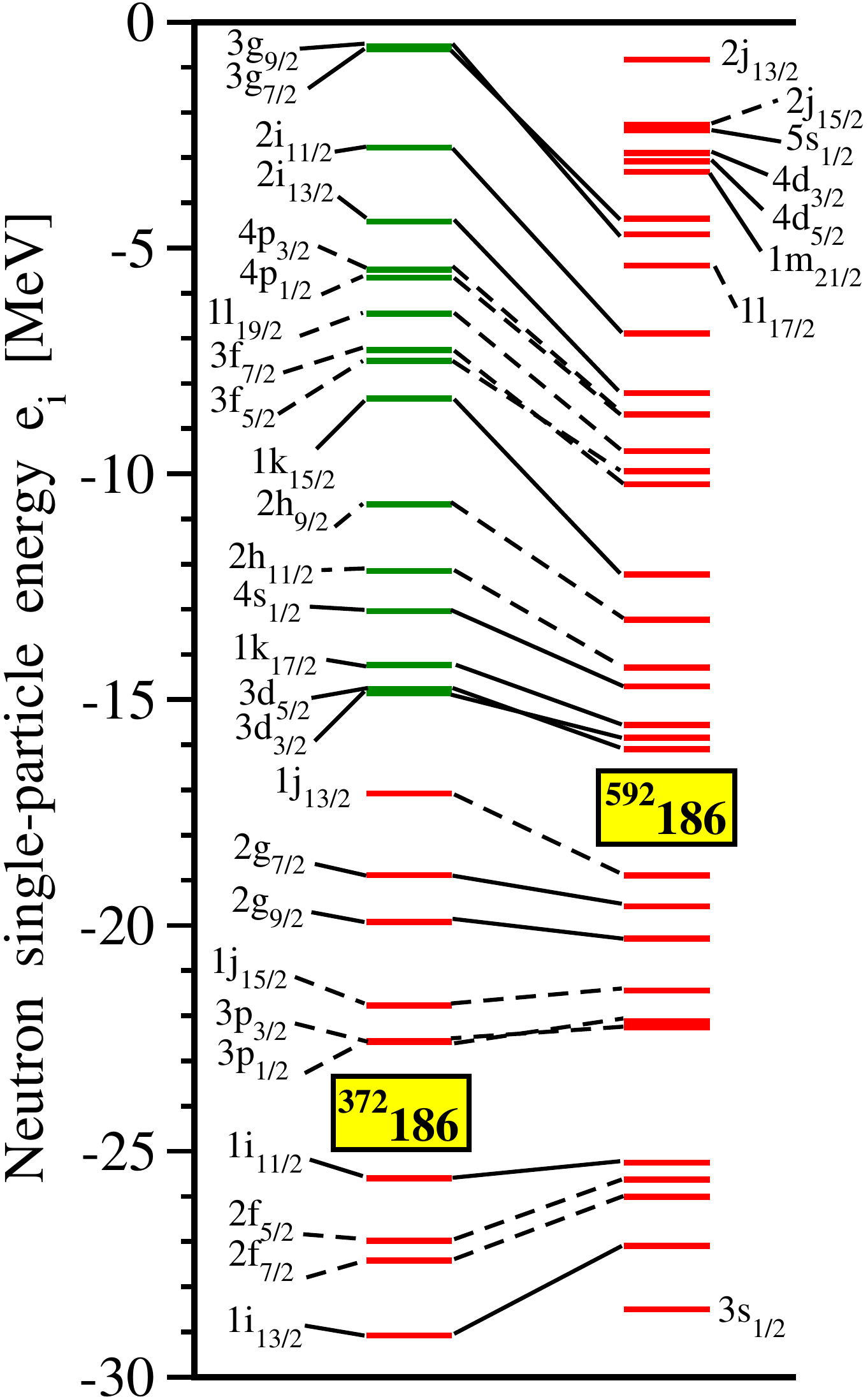}
\caption{Neutron single-particle states in indicated nuclei. Red  and green horizontal lines 
correspond to occupied and unoccupied states, respectively. Solid and dashed black lines 
are used to connect/indicate the states of positive and negative parities, respectively.
\label{sp-neutron-Z186}
}
\end{figure}

    Such kind of comparison as the one discussed above for the $^{292}$120/$^{310}$126
pair of superheavy nuclei is not feasible for  hyperheavy nuclei. This is because  it is impossible  
to find a pair of hyperheavy nuclei with and without bubble structures located close  enough in nuclear 
chart so that their sizes are comparable. However, it is still interesting to see how the formation 
or suppression of the bubble structures affects the single-particle structure. For that we compare
single-particle structures of the $N=Z$ $^{372}$186  and  $^{592}$186 nuclei. Both nuclei have
the same proton configuration. However, the latter nucleus is created from the former one by
the addition of the 220 neutrons including four $s$ neutrons, six $p$ neutrons and twenty $d$ 
neutrons (compare Figs.\ \ref{density-buildup}(i) and (l) and see Fig.\ \ref{sp-neutron-Z186}) 
which leads to a substantial/some  suppression of the bubble structure in neutron/proton 
subsystems (see Fig.\ \ref{Density_466_312_156}(b)  and compare Tables \ref{Table-depl-mirror} 
and \ref{Table-depl}).

\begin{figure}[htb]
\centering
\includegraphics[width=8.4cm]{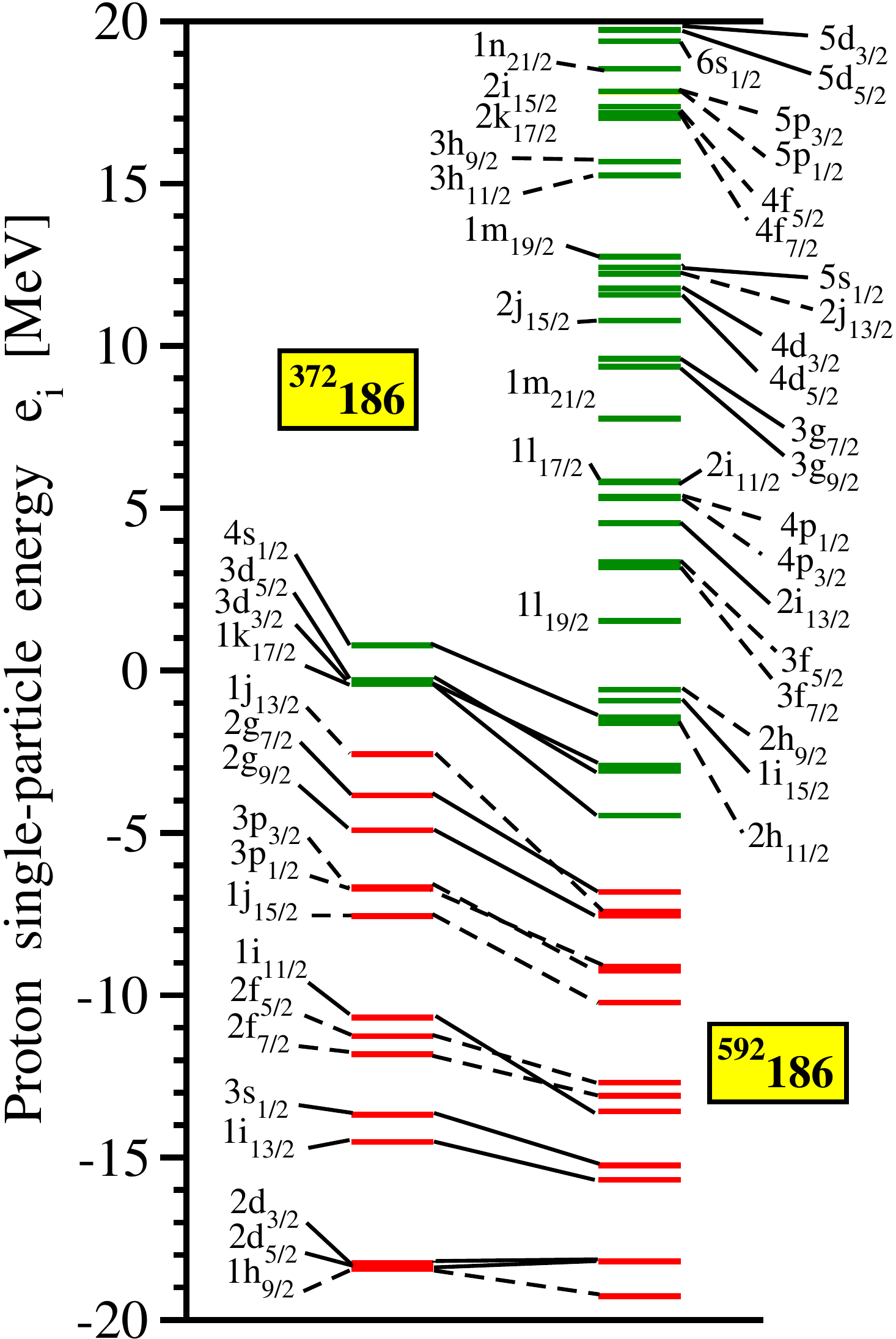}
\caption{The same as in Fig.\ \ref{sp-neutron-Z186} but for the proton single-particle states.  
Note that in order to make a comparison easier the states of the $^{372}$186 nucleus are 
shifted down by 19 MeV.  Only the states below the top of the Coulomb barrier are shown in 
both nuclei.
\label{sp-proton-Z186}
}
\end{figure}

    Neutron single-particle energies of these two nuclei are compared in Fig.\ \ref{sp-neutron-Z186}.
The sequence of the states is basically the same in both nuclei with the exception of the fact that the
order  of the states is inverted in the spin-orbit doublets built on the low-$l$ orbitals.
The neutron spectra are more compressed  in the $^{592}$186  nucleus as compared with the $^{372}$186 
one: this is due to larger radii of the neutron density and potential in the former nucleus. In the proton 
subsystem, there are some changes in the sequence of the single-particle states in two nuclei under study.
It is caused by a substantial reduction of the spin-orbit splitting of the high-$j$ orbitals (such as $1i_{13/2}$ 
and $1i_{11/2}$  or $1j_{15/2}$ and $1j_{13/2}$, see Fig.\ \ref{sp-proton-Z186}) on transition from $N=186$ 
to $N=406$ $Z=186$ isotope.  Similar effect is also seen in the neutron subsystem but it does not
affect the sequence  of the states in two nuclei under study (see Fig.\ \ref{sp-neutron-Z186}).

    These figures also illustrate the relative rarity of the $s$ states in hyperheavy nuclei. For 
example, the neutron $3s_{1/2}$, $4s_{1/2}$ and $5s_{1/2}$ states in the $^{592}$186
nucleus are located at the energies $\approx -28.5$ MeV, $\approx -14$ MeV and $\approx 2.5$ 
MeV, respectively (see the right column of Fig.\ \ref{sp-neutron-Z186}).  Thus, starting from the 
system with occupied  $3s_{1/2}$ state one should add 108 neutrons to occupy the $4s_{1/2}$ 
state and starting from the system with occupied  $4s_{1/2}$ state one should add 142 neutrons 
to occupy the $5s_{1/2}$ state. This rough estimate is obtained under the assumption that the 
occupation of the states does not change the sequence of the states shown in the right column 
of Fig.\ \ref{sp-neutron-Z186}.  Similar estimates could be obtained from the analysis of the 
proton single-particle states shown in the right column of Fig.\ \ref{sp-proton-Z186}.

%%%%%%%%%%%%%%%%%%%%%%%%%%%%%%%%%%%%%%%%%%%%%%%%  
%%%%%%%%%%%%%%%%%%%%%%%%%%%%%%%%%%%%%%%%%%%%%%%%  
    This analysis suggests that similar to superheavy nuclei, the availability (as a function
of particle number) of the low-$l$ (in particular, the $s$ states) states for occupation is not that 
drastically affected by the transition from flat to bubble density distributions in hyperheavy 
$^{592}$186 and  $^{372}$186 nuclei. It will only somewhat shift in proton and neutron numbers 
the position in the $(Z,N)$ plane at which  bubble structures are either enhanced or suppressed.

  It is well known that the spin-orbit interaction is modified in bubble nuclei (see Refs.\ 
\cite{DKW.73,DP.97,DP.98,DBGD.03,KLRL.17}). The spin-orbit potential in the CDFT is
given by \cite{VALR.05}
\begin{eqnarray}
V_{ls} = \frac{m}{m_{eff}} (V-S) ,
\end{eqnarray}
where $m$ is the mass of nucleon and $m_{eff}$ is its effective mass. This potential
in the case of spherical symmetry produces a spin-orbit term of the following form 
\cite{VALR.05}
\begin{eqnarray} 
V_{s.o} = \frac{1}{2m^2}\left( \frac{1}{r} \frac{\partial}{\partial r} V_{ls}(r)\right) \vec{l}\cdot\vec{s}  .
\end{eqnarray} 
%

%%%%%%%%%%%%%%%%%%%%%%%%%%%%%%%%%%%%%%%%%
%%%%%%%%%%%%%%%%%%%%%%%%%%%%%%%%%%%%%%%%%%

\begin{figure}[htb]
\centering
\includegraphics[width=8.4cm]{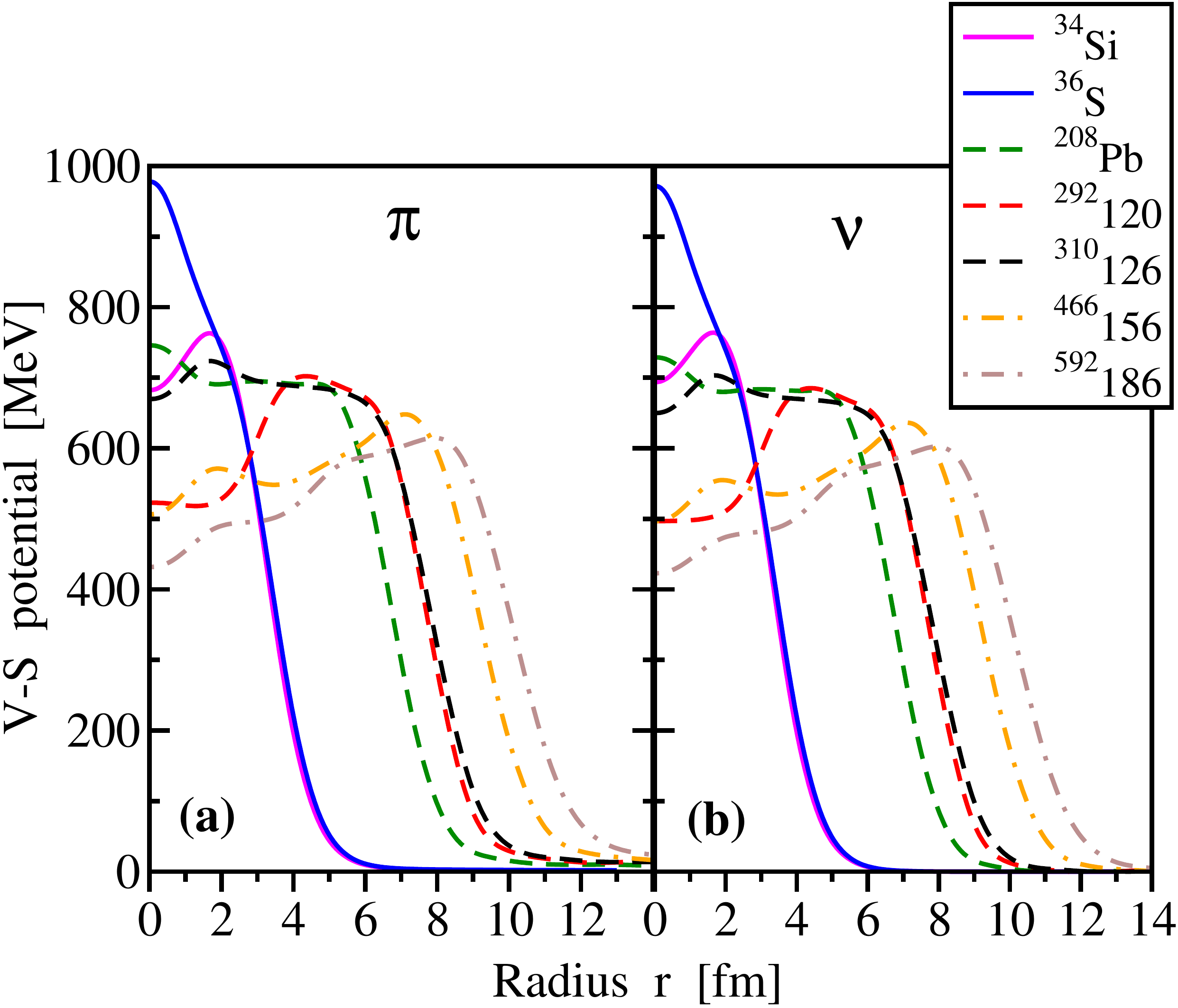}
\caption{Proton and neutron  $(V - S)$  potentials for the nuclei under study.
\label{V_M_S}
}
\end{figure}

As  a consequence, the spin-orbit splitting of the $(l\pm 1/2)$ states in the spin-orbit 
doublet with orbital angular momentum $l$ depends on the derivative of the difference
of vector and scalar potentials $(V-S)$.  These differences for the nuclei under study  
are shown in Fig.\ \ref{V_M_S}. They range from $\approx$ 1000 MeV in the center 
of the  $^{36}$S  nucleus to $\approx 450$ MeV in the center of the $^{592}$186 
nucleus. However, for most of the nuclei the $(V-S)$ values are in the vicinity of 
$600-700$ MeV in the subsurface. 

%%%%%%%%%%%%%%%%%%%%%%%%%%%%%%%%%%%%%%%%%
%%%%%%%%%%%%%%%%%%%%%%%%%%%%%%%%%%%%%%%%%%
    In the nuclei with flat density distributions such as $^{208}$Pb and $^{310}$126, 
the $(V-S)$ potential is almost flat in the subsurface region. Thus, this part of the nucleus 
contributes  only marginally to the spin-orbit splittings which are almost entirely defined by the 
decrease of the $(V-S)$ potential in the surface region. In contrast, in the bubble nuclei the 
$(V-S)$ potential increases with increasing $r$ in the subsurface region. Thus, this region 
contributes to the spin-orbit splittings  but with the sign opposite to the one produced in the surface 
region where the $(V-S)$ potential decreases with increasing $r$. This mechanism is 
responsible for the modifications of the spin-orbit splittings such as the reduction 
and/or inversion of spin-orbit splittings of the low-$l$ spin-orbit doublets in the bubble
nuclei discussed above.
\clearpage
%%%%%%%%%%%%%%%%%%%%%%%%%%%%%%%%%%%%%%%%%
\section{Potential impact of deformation on the balance of the single-particle and 
Coulomb interaction contributions to the bubble structures}
\label{factors_2} 
%%%%%%%%%%%%%%%%%%%%%%%%%%%%%%%%%%%%%%%%%

%%%%%%%%%%%%%%%%%%%%%%%%%%%%%%%%%%%%%%%%%%%%%%%%
%%%%%%%%%%%%%%%%%%%%%%%%%%%%%%%%%%%%%%%%%%%%%%%%

   It is an interesting question on how the balance of the contributions 
of  the single-particle degrees of freedom and Coulomb interaction to the formation 
of bubble structures changes on the transition from spherical to deformed nuclei. 

   The emergence of the deformation has two important consequences 
for the single-particle structure. First, the deformation leads to a more even 
distribution of the deformed single-particle states emerging from the high-$j$ and 
low-$j$ spherical subshells  as compared with the one of the single-particle states 
at spherical shape (see, for example, the Nilsson diagrams in Figs. 3 and 4  of 
Ref.\ \cite{CAFE.77} and in Figs.\ 1-3 of Ref.\ \cite{DABRS.15}). Second, the wave 
functions of deformed single-particle states contain the contributions from different 
spherical $j$-subshells and the mixture of such contributions increases with increasing 
deformation.  Both these factors effectively reduce the contribution of the single-particle 
states into the formation of the bubble structures.  Thus,  in a given nucleus the density 
profile of a deformed solution is flatter  than that of a spherical one (see Fig. 5 in Ref.\ 
\cite{AF.05-dep} and Fig. 10 in Ref.\ \cite{PXS.05}).
 
\begin{figure}[htb]
\centering
\includegraphics[width=8.4cm]{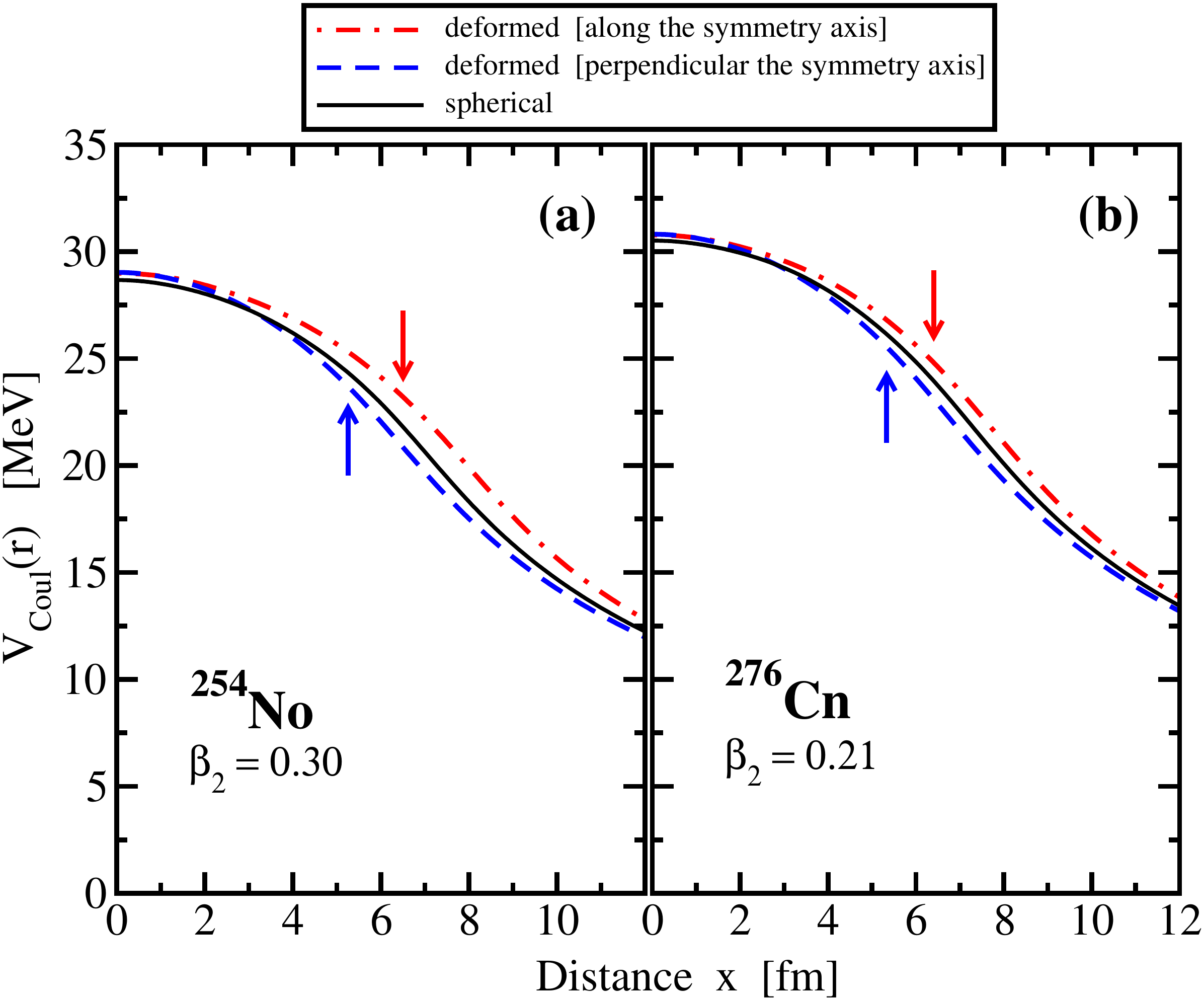}
\caption{Coulomb potentials in deformed ground state and excited 
spherical solution of the $^{254}$No and $^{276}$Cn nuclei obtained in the RHB 
calculations. They are shown as a function of radial coordinate for spherical shape 
(solid black line) and as a function of distance along (red dod-dashed line) and 
perpendicular (blue dashed line) the symmetry axis for deformed shape. Equilibrium 
deformations of deformed states are indicated. The density profiles of these nuclei 
(not shown) are very similar to those  displayed in Fig. 5 of Ref.\ \cite{AF.05-dep}.
Note that the results of the calculations without pairing are very close to those 
shown in the figure.
%The arrows indicate the distances from the center of the nucleus at which the surface 
%region appears in the direction along (red arrows) and perpendicular (blue arrow) the 
%symmetry axis. 
\label{V-coul-def}
}
\end{figure}

  In contrast, the transition from spherical to deformed shapes has a relatively small 
impact on the Coulomb potential. This is illustrated here by the examples of the $^{254}$No 
and $^{276}$Cn nuclei the bubble structures in the proton and neutron densities of which 
have been studied earlier (see discussion of Fig.\ 5 in Ref.\ \cite{AF.05-dep}). The 
Coulomb potential in axially deformed nuclei depends on $z$ and $r_{\bot}$. Here $z$ is 
the distance from the center of the  nucleus along the symmetry axis and $r_{\bot}$ is the 
distance in radial direction. For simplicity, the Coulomb potential $V_{Coul}^{def}$ in these 
deformed  nuclei is considered  as a function of the distances along ($r_{\bot}=0$) and 
perpendicular ($z=0$) the symmetry axis (see Fig.\ \ref{V-coul-def}). For comparison, the Coulomb
potential $V_{Coul}^{spher}(x)$ of spherical solution is also presented in this figure. One can see 
that 
$V_{Coul}^{def}(z=x, r_{\bot}=0) >  V_{Coul}^{def}(z=0, r_{\bot}=x)$ and the splitting between these two 
branches increases with deformation but it is rather modest. Note that the average of 
these two branches is very close to $V_{Coul}^{spher}(x)$. 
 
  As discussed in Sec.\ \ref{SHE-Coulomb}, the impact of the Coulomb potential on 
the formation of wine bottle potential and bubble structures is defined by the difference 
$V_{Coul}(x=0) - V_{Coul}(x_{surf})$, where $x_{surf}$ is the coordinate in a given 
direction at which the density is maximal in near surface region (the point at which 
the surface region starts).  Red and blue arrows in Fig.\ \ref{V-coul-def} indicate the 
distances $x_{surf}$ from the center of the nucleus in the direction along  and 
perpendicular the symmetry axis, respectively.  $x_{surf}$ of the spherical 
solution is located approximately at the middle point between these two arrows. 
There are some differences in the values of $V_{Coul}$ and $x_{surf}$ of deformed 
and spherical solutions. Despite that the impact of the Coulomb interaction 
on the formation of the bubble structure in deformed nuclei integrated over the 
volume of the nucleus is expected to be close to that of spherical solution. This is because 
two factors, namely,  (i) 
$V_{Coul}^{def} (z=x, r_{\bot}=0) \approx  V_{Coul}^{def}(z=0, r_{\bot}=x) \approx
V_{Coul}^{spher}(x)$ for $x<3$ fm 
and (ii) 
$V_{Coul}^{def} (z=x, r_{\bot}=0) > V_{Coul}^{spher}(x) > V_{Coul}^{def}(z=0, r_{\bot}=x)$
for $x<x_{surf}$.

   These observations have two consequences. First, with increasing
deformation the relative impact of the Coulomb interaction on the formation of 
the bubble structures in deformed nuclei  increases since the one due 
single-particle degrees of freedom decreases. Second, the bubble structures 
in superheavy deformed nuclei are relatively small (see Fig. 5 in Ref.\ 
\cite{AF.05-dep}, Figs. 6, 7 and 10 in Ref.\ \cite{PXS.05} and Fig. 2 in 
Ref.\ \cite{SNR.17}): this is especially true for the nuclei located far away 
from double  shell closures.  In many cases, they are  smaller that those in 
deformed light nuclei (see Ref.\ \cite{SA.14}). Let us ignore the impact of the 
single-particle degrees of freedom and attribute the effect of the bubble 
creation entirely to the Coulomb interaction. Then the fact that the bubble structures
are either absent or relatively small in deformed superheavy nuclei allows
to conclude that the effect of the Coulomb interaction on the formation of the 
bubble structures in such nuclei is rather modest.  This conclusion is expected
to be valid also for spherical superheavy  nuclei since the impact of the Coulomb
interaction on the formation of the bubble structures only weakly depend on the 
deformation.
\clearpage
%%%%%%%%%%%%%%%%%%%%%%%%%%%%%%%%%%%%%%%%%%%%%%%%
%%%%%%%%F%%%%%%%%%%%%%%%%%%%%%%%%%%%%%%%%%%%%%%%

%%%%%%%%%%%%%%%%%%%%%%%%%%
\section{Conclusions}
\label{concl}
%%%%%%%%%%%%%%%%%%%%%%%%%%

 The detailed investigation of  microscopic mechanisms leading to the
formation of bubble structures in the nuclei with main emphasis on the role
of the single-particle degrees of freedom and Coulomb interaction has been
performed in the framework of covariant density functional theory. Many of 
existing publications such as Refs.\ \cite{DBGD.03,SNR.17} emphasize 
dominant role of the Coulomb interaction in the creation of the bubble
structures in super- and hyperheavy nuclei.  However, our detailed analysis 
paints much more complicated picture in which single-particle degrees
of freedom  play a significant role which overshadows the role of the 
Coulomb interaction in superheavy nuclei. The main results can be 
summarized as follows:    

\begin{itemize}

\item
   There is a central classically forbidden region at the bottom of the wine 
bottle potentials the size of which depends on the nucleus. The presence of this region 
leads to a substantial reduction of the densities of the $1s_{1/2}$ states and somewhat
smaller reduction in the densities of the $2s_{1/2}$ states for radial coordinate $r=0$ 
and near it as compared with the case of flat bottom potential.  The densities of 
the $l=1$ and $l=2$ states located at the  bottom of the wine bottle potential can also 
be pushed away from $r=0$. This represents a new never discussed before 
microscopic mechanism of the creation of bubble structures in nuclei. It is responsible 
for a significant reduction of the nucleonic densities at $r=0$ in hyperheavy 
nuclei.

\item
  Microscopic mechanisms of the formation of the wine bottle nucleonic potentials 
have been investigated in detail. It was shown that the formation of the bubble 
structure in the densities of the subsystem A (proton or neutron) of the nucleus 
leads to a significant enhancement  of wine  bottle features of the potential in 
other subsystem B (neutron or proton). The microscopic origin of this feature lies 
in the isovector character of nuclear interaction which tries to keep proton and 
neutron densities alike. The formation of bubble structure in the densities of the 
subsystem A with increasing particle number proceeds by the occupation of the 
states in the vicinity of the Fermi  level and it has only minor impact on radial 
profile of the bottom of the nucleonic potential in this subsystem. For a  fixed 
number of the particles in the subsystem B, the formation of the 
bubble structure  in its densities, driven by the formation of  the bubble structures 
in the subsystem A, can be achieved only by a significant enhancement of wine 
bottle features of its potential.

\item
    The bubbles in nucleonic total densities also depend on  the availability of 
low-$l$ single-particle states for occupation since their densities  represent the basic 
building blocks of total densities. However, such states (in particular, the $s$ states) 
appear less frequently as
compared with medium and high $l$ states with increasing principal quantum 
number $N$. This is a typical feature of realistic nucleonic potential: within 
a shell with a given principal quantum number $N$ the states with highest possible 
orbital angular momentum $l$ are the lowest in energy while those with lowest 
$l$  (such as the $s$ states in even-$N$ shells and the $p$ states in odd-$N$ 
shells) are typically located at the highest or near highest energies in the shell. 
Thus, with filling of a specific $N$ shell the density is first built at the surface,
then in the middle part of the nucleus and only then in the central region and 
at $r=0$. The balanced  distribution of the occupation of low, medium and high-$l$ 
states is required for building flat density distribution. However, this balance is 
substantially  broken in hyperheavy nuclei and the density is built predominantly in 
near surface region by the high-$l$ states.

\item
    Existing bubble indicators [see Eqs.\ (\ref{Depletion factor}) and (\ref{Depl-fac-2})] 
are  strongly affected by single-particle properties. In particular, the central density 
$\rho_{c}$  is defined almost entirely by occupied $s$ states. Thus, they cannot be 
reliable measures of  bulk properties (such as a Coulomb interaction).  This is especially 
true for the nuclei characterized  by wine bottle nucleonic potentials since the densities 
of the lowest $s$ states at  $r=0$  are strongly affected by classically forbidden regions
of the potentials.

\item
   An additivity rule for the densities has been proposed for the first time. It was verified 
on the pairs of the $^{34}$Si/$^{36}$S and $^{292}$120/$^{310}$126 nuclei: the first 
nucleus in the pair has bubble structure while second one is characterized by flat 
density distributions. The additivity rule works with comparable accuracy in both
pairs of the nuclei.  This strongly suggests the same mechanism of the formation 
of the central depression in lighter nucleus of the pair which is related to emptying 
of specific  low-$l$ singe-particle orbitals.

\item
  The global evolution of the densities is governed also by saturation mechanisms. The 
analysis of the densities shown in the present paper and in Refs.\ \cite{AF.05-dep,SKKJA.19} 
clearly reveals that average neutron densities $\rho^{ave}_{\nu}$ in subsurface region of the nuclei 
try to stabilize near saturation density $\rho_{sat} \approx 0.08$ fm$^{-3}$.  In contrast, average proton 
densities  $\rho^{ave}_{\pi}$ in subsurface region can be significantly below this value especially 
in neutron-rich nuclei. This strongly suggests that Coulomb interaction effects are secondary 
to  nuclear interaction ones in absolute majority of the nuclei.

\item 
    Self-consistency effects are characterized by a very complex nature of 
the impact of the nuclear densities on the nucleonic potentials. For example, the removal 
of two protons from $^{36}$S leads to a creation of bubble structure in proton densities
of $^{34}$Si. However,  it has a substantially larger impact on neutron potential of $^{34}$Si (which becomes
wine bottle one) than on the proton one (which becomes flat bottom one). Similar effects have 
been seen before in Ref.\ \cite{AF.05-dep}:  particle-hole excitations in neutron subsystem 
led to substantial changes in neutron densities but this process results in larger changes
in the proton potential as compared with the neutron one.

\end{itemize}

   One can see in some publications the statements that the Coulomb interaction
is at the origin of the systematic deviations from a uniform charge distributions since
the system can lower its (positive) electrostatic energy by forming bubble structures
(see, for example, Refs.\ \cite{DBGD.03,SNR.17}). In extreme, the lowest Coulomb energy would be 
reached if all the protons were located in a thin layer at the nuclear surface. However, 
in nuclei this trend is counteracted by the quantum nature of the single-particle states: 
only specific single-particle states with specific density profiles can be occupied with 
increasing proton and neutron numbers. In addition, there is a nuclear interaction between protons 
and neutrons which further complicates the situation.  The pattern of the saturation of neutron 
density at $\rho_{sat} \approx 0.08$ fm$^{-3}$ counteracts the frequent
argument that the neutron density follows to a certain extent the trend produced by
the protons as a result of the strongly attractive neutron–proton interaction.   As a 
consequence, the formation of bubble structures depends on the competition of several 
factors and there is no simple indicator which would clearly allow to separate nuclear and 
Coulomb interaction effects on the central depression in density distributions. However,
our detailed analysis of different aspects of bubble physics strongly suggests that the 
formation of bubble structures in superheavy nuclei is dominated by  single-particle 
effects. This is in contrast to the conclusions of Ref.\  \cite{SNR.17}  that the central 
depression in superheavy nuclei is firmly driven by the  electrostatic repulsion. The
role of the Coulomb interaction increases in hyperheavy nuclei but even for such 
systems we do not find strong arguments that the formation of  bubble structures  
is dominated by the Coulomb interaction.

%%%%%%%%%%%%%%%%
\section{Acknowledgement}
%%%%%%%%%%%%%%%%

  This material is based upon work supported by the U.S.
Department of Energy, Office of Science, Office of Nuclear
Physics under Award No. DE-SC0013037.

\bibliography{references-34-PRC-bubble-nuclei-rev}

\end{document}